\documentclass[a4paper,10pt]{article}
\pdfoutput=1

\usepackage{jheppub}
\usepackage[T1]{fontenc}
\usepackage{natbib}
\usepackage{float}
\usepackage{multirow}
\usepackage{graphicx}
\usepackage{subcaption}
\usepackage{booktabs,array}
\usepackage{multirow}
\usepackage{bbm}
\usepackage{cleveref}
\usepackage{amssymb}
\usepackage{mathtools}
\usepackage{hyperref}
\usepackage{cleveref}
\usepackage{xcolor}
\usepackage{textcomp}
\usepackage{multicol}
\usepackage{slashed}
\usepackage{comment}
\usepackage{empheq}
\usepackage{easy-todo}

\def\eps{\epsilon}
\renewcommand{\i}{\,\mathrm{i}\,}
\renewcommand{\d}{\mathrm{d}}

\newcommand{\p}[1]{(\ref{#1})}
\newcommand \vev [1] {\langle{#1}\rangle}
\newcommand{\pa}{\partial}

\title{
    Two-Loop Five-Point Two-Mass Planar Integrals and
    Double Lagrangian Insertions in a Wilson Loop 
}

\preprint{
  \begin{tabular}{l}
   CERN-TH-2024-136 \\
   LAPTH-043/24 \\
   ZU-TH 40/24 \\
  \end{tabular}
}

\author[1,2]{Samuel Abreu,}

\affiliation[1]{Theoretical Physics Department, CERN, Geneva, Switzerland}
\affiliation[2]{Higgs Centre for Theoretical Physics, School of Physics and Astronomy,\\
The University of Edinburgh, Edinburgh EH9 3FD, Scotland, UK}

\author[3]{Dmitry Chicherin,}
\affiliation[3]{LAPTh, Universit\'e Savoie Mont Blanc, CNRS, B.P.~110, F-74941 Annecy-le-Vieux, France}

\author[4]{Vasily Sotnikov,}
\affiliation[4]{Physik-Institut, University of Zurich, Winterthurerstrasse 190, 8057 Zurich, Switzerland}

\author[1]{Simone Zoia}

\emailAdd{samuel.abreu@cern.ch}
\emailAdd{chicherin@lapth.cnrs.fr}
\emailAdd{vasily.sotnikov@physik.uzh.ch}
\emailAdd{simone.zoia@cern.ch}

\abstract{
We consider the complete set of planar two-loop five-point Feynman integrals with two off-shell external legs.
These integrals are
relevant, for instance, for the calculation of the second-order QCD corrections to the 
production of two heavy vector bosons in association with a jet or a photon at a 
hadron collider.
We construct pure bases for these integrals and reconstruct their analytic differential equations in canonical form through numerical sampling over finite fields.  
The newly identified symbol alphabet, one of the most complex to date, 
provides valuable data for bootstrap methods.  
We then apply our results to initiate the study of double Lagrangian insertions in a four-cusp Wilson loop in planar maximally supersymmetric Yang-Mills theory, 
computing it through two loops.
We observe that it is finite, conformally invariant in four dimensions, and of uniform transcendentality.  
Furthermore, we provide numerical evidence for its positivity within the amplituhedron region through two loops.  
}

\begin{document}

\maketitle

\section{Introduction}
\label{sec:intro}

Scattering amplitudes are essential components that bridge the formal aspects
of quantum field theory with the observables that can be measured at particle
colliders. Beyond the leading order in perturbation theory, 
these amplitudes are computed as sums of
Feynman integrals which originate from the exchange of virtual particles. Indeed,
understanding and being able to compute Feynman integrals is crucial in
providing precise theoretical predictions for high-energy particle colliders such as 
the Large Hadron Collider (LHC). Besides their phenomenological importance, 
Feynman integrals are also interesting objects in their own right
as their mathematical structure is very rich, with physical
constraints being blended into complicated multi-valued functions.
The calculation of Feynman integrals is thus central to the advancement of
our understanding of quantum field theory,
leading to both practical and theoretical advancements in particle physics.

While the functions which Feynman integrals evaluate to are generally not
known (see ref.~\cite{Bourjaily:2022bwx} for a recent review), many physically relevant
Feynman integrals can be expressed as $\mathbb{Q}$-linear combinations of
Chen's iterated integrals~\cite{Chen:1977oja} over logarithmic differential
one-forms, known as symbol letters. These so-called pure 
integrals~\cite{Arkani-Hamed:2010pyv}
satisfy differential equations in canonical form~\cite{Henn:2013pwa} and
possess several properties that make them particularly valuable for representing
the transcendental functions involved in scattering amplitudes
\cite{Abreu:2022mfk}. They provide compact and physically insightful
representations of scattering amplitudes and are often suitable for efficient
numerical evaluation. The intriguing relationship between symbols, leading
singularities, and Landau singularities has gained considerable attention
recently, with new methods having being developed to predict sets of 
symbol letters~\cite{Abreu:2017mtm,Arkani-Hamed:2017ahv,Chen:2023kgw,
Fevola:2023kaw,Fevola:2023fzn,Jiang:2023qnl,Dlapa:2023cvx,He:2023umf,Jiang:2024eaj,
Helmer:2024wax,Caron-Huot:2024brh}, known as the alphabet. 
Additionally, certain constraints on the iterated integrals can sometimes be
derived \cite{Hannesdottir:2021kpd,Hannesdottir:2024cnn}, which can
be leveraged into bootstrap approaches to systematically construct and verify the
structure of scattering amplitudes. These provide a powerful method to explore and
understand the underlying physics without relying solely on explicit
calculations.

It is a priori not known if a set of Feynman integrals contributing to a given scattering processes can be represented in this form.
While several methods have been devised to facilitate finding such representations~\cite{Lee:2014ioa,Meyer:2017joq,Prausa:2017ltv,Gituliar:2017vzm,Lee:2020zfb,Henn:2020lye,Chen:2020uyk,Chen:2022lzr,Dlapa:2020cwj,Dlapa:2021qsl},
in the case of multi-scale Feynman integrals this still remains a very challenging task.
At two loops, among the most complex cases considered are 
all integrals for five-point one-mass scattering \cite{Abreu:2020jxa,Abreu:2021smk,
Abreu:2023rco}, some integrals for six-point massless scattering 
\cite{Henn:2024ngj},
as well as selected integral topologies relevant for $t \bar{t}j$ and $t \bar{t} H$ in hadron collisions \cite{FebresCordero:2023pww,Badger:2022hno,Badger:2024fgb} 
(although in the latter cases not all integrals were found to be pure).
The first five-point integral topology with two external masses has been considered in 
ref.~\cite{Jiang:2024eaj}. These calculations all follow a similar pattern: one finds
a good basis for the space of Feynman integrals under consideration, and then obtains
the corresponding differential equations. 
In some cases the solutions in terms of iterated integrals can be efficiently
evaluated numerically~\cite{Chicherin:2020oor,Chicherin:2021dyp,Abreu:2023rco}.

In this work we consider the complete set of planar two-loop five-point integrals with 
two off-shell external legs. 
These integrals are relevant, for instance, 
for the yet unknown second-order corrections in the 
strong-coupling constant for the production of two heavy vector bosons production 
in association with a jet or a photon,
or for the third-order QCD corrections to the production of two heavy vector bosons.
We find that it is possible to construct sets of so-called master integrals 
whose differential equations are in canonical form \cite{Henn:2013pwa}, that is where the dependence on
the dimensional regulator factorises and only logarithmic one-forms with 
algebraic arguments appear.
We employ finite-field techniques \cite{vonManteuffel:2014ixa,Peraro:2016wsq,Peraro:2019svx} to 
reconstruct the differential equations.
The logarithmic forms, or symbol letters, mostly but not completely match the 
results predicted by the method of ref.~\cite{Jiang:2024eaj}.
Even if incomplete, the partial knowledge of the symbol alphabet greatly
simplifies the task of reconstructing the analytic differential equation.
We find that the complexity of the alphabet increases substantially compared 
to the case where a single external leg is off-shell~\cite{Abreu:2020jxa}.

The Feynman integrals that we calculate in this work are also interesting for studies
of more formal aspects of QFT. We are motivated by the renowned duality between 
the vacuum expectation values of the null polygonal Wilson loops and scattering 
amplitudes in maximally super-symmetric Yang-Mills theory (MSYM)~\cite{Alday:2007hr,
Drummond:2007aua,Brandhuber:2007yx}. This duality suggests the definition of a class
of finite gauge-invariant multi-scale observables in MSYM which are closely related 
to the scattering amplitudes and their integrands~\cite{Arkani-Hamed:2012zlh}. 
We consider the correlations functions of the null Wilson loop and several 
Lagrangians, normalised by the vacuum expectation value of the null Wilson 
loop. This ratio of correlation functions is finite and is expected to possess 
a number of intriguing properties, as we discuss in this paper. 
We refer to it as Lagrangian insertions in the null Wilson loop.

The case of a single Lagrangian insertion in the four-cusp Wilson loop 
has been extensively studied in the literature.
It is calculated at strong coupling~\cite{Alday:2011ga} and at weak coupling up to
three-loop order~\cite{Alday:2012hy,Alday:2013ip,Henn:2019swt}. The single 
Lagrangian insertion in multi-cusp Wilson loop has been studied 
in \cite{Chicherin:2022bov}, where hidden symmetries and dualities with pure 
Yang-Mills amplitudes have been revealed. 
The single Lagrangian insertion in the five-cusp Wilson loop has been calculated 
in the two-loop approximation in \cite{Chicherin:2022zxo}. A recent inspiring 
development in the perturbative study of the single Lagrangian insertion in the 
four-cusp Wilson loop originated from its geometric description. 
The four-dimensional loop integrands of scattering amplitudes are completely specified by 
the amplituhedron~\cite{Arkani-Hamed:2013jha}. However, the implications of the 
amplituhedron construction are less transparent for integrated amplitudes, which 
require infrared regularisation. Unlike amplitudes, the Lagrangian insertion in the Wilson
loop is well-defined in four space-time dimensions, and the geometric constructions for the 
four-dimensional loop integrands can be promoted to the integrated loop corrections. 
A decomposition of the loop corrections into negative geometries has been studied 
in refs.~\cite{Arkani-Hamed:2021iya,Brown:2023mqi}, and certain geometries have been 
solved and resummed to all loop orders.
The negative geometry decomposition has been also extended to the ABJM 
theory~\cite{Henn:2023pkc,He:2023exb,Lagares:2024epo}. 

Given the numerous beautiful properties observed for the single Lagrangian insertion, 
it is natural to wonder if these also hold for double Lagrangian insertions. 
In this work we initiate the study of double Lagrangian insertions. 
We consider the double Lagrangian insertion in the four-cusp null Wilson loop, 
which is the simplest null polygonal contour, and calculate it for the first time 
through two loops. As compared to the single Lagrangian insertion, the kinematic space 
of the double Lagrangian insertion is multidimensional already for the four-cusp contour 
considered in this paper. The perturbative two-loop calculation we perform here
supports the expectation that the double Lagrangian insertion is finite,
conformally invariant in four dimensions, and has uniform transcendentality.
The fact that these properties emerge from our calculation give a strong check of
the correctness of the integrals we compute in this paper.

Another remarkable geometric observation focuses on the positivity of the integrated 
loop corrections of the single Lagrangian insertion. More precisely, relying on 
the available perturbative data, it was observed in 
refs.~\cite{Arkani-Hamed:2021iya,Chicherin:2022bov} that the loop corrections of 
the single Lagrangian insertion have uniform sign in a certain subregion of the Euclidean 
kinematic region which is predicted by the amplituhedron construction. 
Recently, ref.~\cite{Henn:2024qwe} provided evidence that the positivity of the 
loop corrections could be extended to a much stronger statement of complete 
monotonicity, which restricts the sign of the derivatives of all orders in 
kinematic variables. In this work we provide numerical evidence that the integrated 
loop corrections of the double Lagrangian insertion have uniform sign inside the
amplituhedron region, at least up to two loops.

The paper is organised as follows. In \cref{sec:kinematics} we discuss our 
notation and conventions for the integrals considered in this paper. 
In \cref{sec:bases}, we discuss the pure bases of master integrals and the corresponding
analytic differential equations, including the determination of the alphabet
relevant for the description at two loops of planar scattering processes involving 
five particles, out of which two are off shell. 
We also provide sample numerical evaluations and discuss the checks we performed.
In \cref{sec:WilsonLoop} we present the calculation of the double Lagrangian insertion in a 
(quadrilateral) Wilson loop at the two-loop order.
We summarise our results and discuss
some possible next steps in \cref{sec:conclusions}.
We conclude with two appendices. 
In \cref{app:5ptSectors} we list our master integrals for all the five-point sectors of the two-loop families.
In \cref{sec:app_kin} we define a number of kinematic regions relevant in this work.
Our ancillary files are available at~\cite{ancillary}.

\section{Kinematics and Definitions}
\label{sec:kinematics}

\subsection{Kinematics}
\label{sec:kin}
We compute all two-loop planar integrals required to describe scattering 
processes involving five external legs, out of which three are massless
and two are massive. All internal propagators are massless.
The momenta of the massless external
legs are denoted $p_1$, $p_2$ and $p_3$, and the momenta of the two
massive legs are denoted $p_4$ and $p_5$. We take the
momenta to be outgoing, and they satisfy the momentum-conservation relation
\begin{equation}
p_1+p_2+p_3+p_4+p_5 = 0\,.
\end{equation}
With these kinematics, there are seven independent Mandelstam variables,
which we choose to be
\begin{equation}
\label{eq:sdef}
X \coloneqq \bigl\{ s_{12}, s_{23}, s_{34}, s_{45}, 
s_{15}, s_4, s_5 \bigr\} \,,
\end{equation}
with $s_{ij} \coloneqq (p_i+p_j)^2$ and $s_i \coloneqq p_i^2$.
We work in dimensional regularisation, with $d=4-2 \eps$ spacetime dimensions and four-dimensional external momenta.

The kinematics of these processes are also described by
Gram determinants. We define them~as
\begin{align}\begin{split}
& {\rm Gram}\bigl(\{a_1,\ldots,a_m \}, \{b_1,\ldots,b_m \} \bigr) \coloneqq {\rm det}\left( 2\, a_i\cdot b_j \right)\bigl|_{i,j=1,\ldots,m} \,, \\
& {\rm Gram} (a_1,\ldots,a_m ) \coloneqq {\rm Gram}\bigl(\{a_1,\ldots,a_m \}, \{a_1,\ldots,a_m \} \bigr)  \,.
\end{split}\end{align}
In particular, we will need
\begin{align}\label{eq:grams}
& \Delta_3(p_i,p_j) \coloneqq {\rm Gram}\bigl(p_i,p_j \bigr)
=-\lambda(s_i,s_j,s_{i,j})\,, \\
& \Delta_5 \coloneqq {\rm Gram}\bigl(p_1,p_2,p_3,p_4 \bigr) \,,
\end{align}
where $\lambda(x,y,z)$ is the K\"all\'en function,
\begin{align}
\lambda(x,y,z) \coloneqq x^2 + y^2 + z^2 - 2 x y - 2 y z - 2 z x \,.
\end{align}
The Gram determinant $\Delta_5$ is related to the five-particle pseudo-scalar invariant via
\begin{align}
\label{eq:delta5}
\Delta_5 = {\rm tr}(\slashed{p}_1 \slashed{p}_2 \slashed{p}_3 \slashed{p}_4 \gamma_5)^2 \,.
\end{align}

We also use Gram determinants to express the $(-2\eps)$-dimensional components 
of the loop momenta, conventionally denoted $\mu_{ij}$, in terms of scalar products of external and loop momenta:
\begin{align} \label{eq:muij}
\mu_{ij} & \coloneqq k_i^{[-2\eps]} \cdot k_j^{[-2\eps]} \,, \nonumber \\
& = \frac{ {\rm Gram}\bigl(\{k_i, p_1,p_2,p_3,p_4\} , \{k_j, p_1,p_2,p_3,p_4\} 
\bigr) }{2 \, \Delta_5} \,,
\end{align}
where $k_i \eqqcolon k_i^{[4]} + k_i^{[-2\eps]}$, with $ k_i^{[-2\eps]} \cdot p_j = 0 =  k_i^{[-2\eps]} \cdot k_j^{[4]}$.
These objects play an important role in the construction of compact pure integral bases (see \cref{sec:bases}).

\subsection{Integral Families}
\label{sec:families}

\begin{figure}[t]
    \centering
    \begin{subfigure}[c]{0.3\linewidth}
        \centering
        \includegraphics[width=\textwidth]{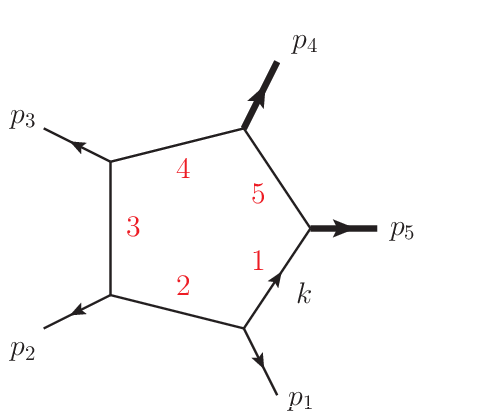}
        \caption{Pa}
        \label{fig:Pa}
    \end{subfigure}
    \begin{subfigure}[c]{0.3\linewidth}
        \centering
        \includegraphics[width=\textwidth]{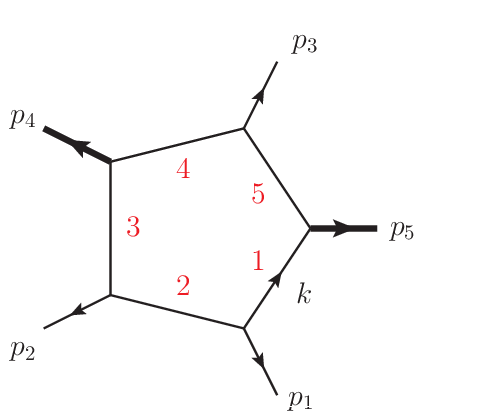}
        \caption{Pb}
        \label{fig:Pb}
    \end{subfigure}

    \caption{Independent set of one-loop ``pentagon'' topologies.}
    \label{fig:pentagons}
\end{figure}

\begin{figure}[t]
    \centering
    \begin{subfigure}[c]{0.32\linewidth}
        \centering
        \includegraphics[width=\textwidth]{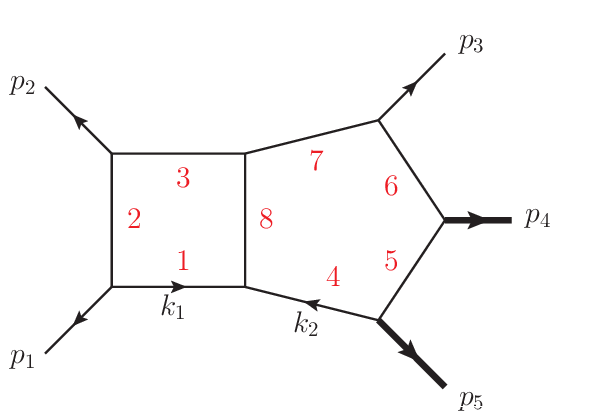}
        \caption{PBmmz}
        \label{fig:PBmmz}
    \end{subfigure}
    \begin{subfigure}[c]{0.32\linewidth}
        \centering
        \includegraphics[width=\textwidth]{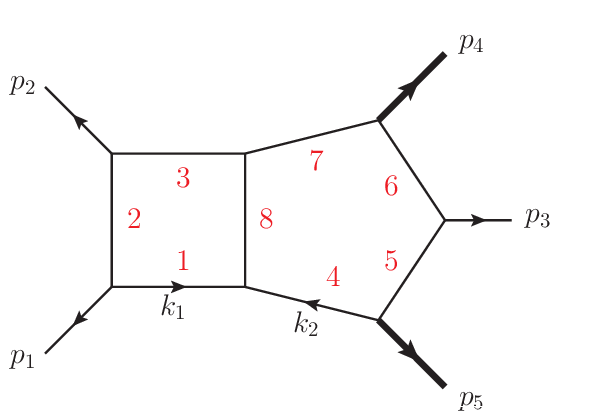}
        \caption{PBmzm}
        \label{fig:PBmzm}
    \end{subfigure}
    \begin{subfigure}[c]{0.32\linewidth}
        \centering
        \includegraphics[width=\textwidth]{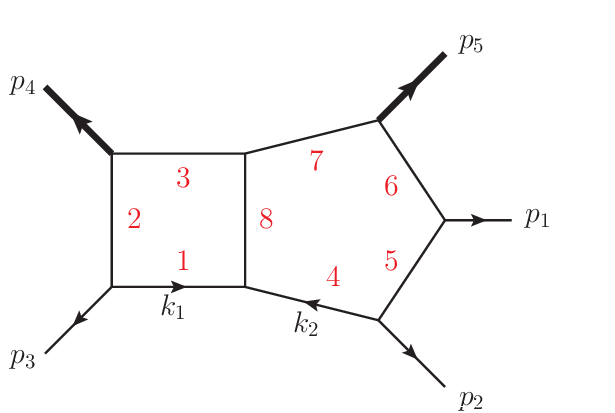}
        \caption{PBmzz}
        \label{fig:PBmzz}
    \end{subfigure}
    
    \begin{subfigure}[c]{0.32\linewidth}
        \centering
        \includegraphics[width=\textwidth]{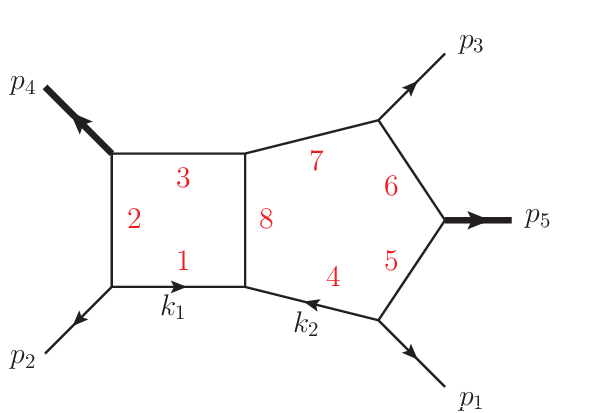}
        \caption{PBzmz}
        \label{fig:PBzmz}
    \end{subfigure}
     \begin{subfigure}[c]{0.32\linewidth}
        \centering
        \includegraphics[width=\textwidth]{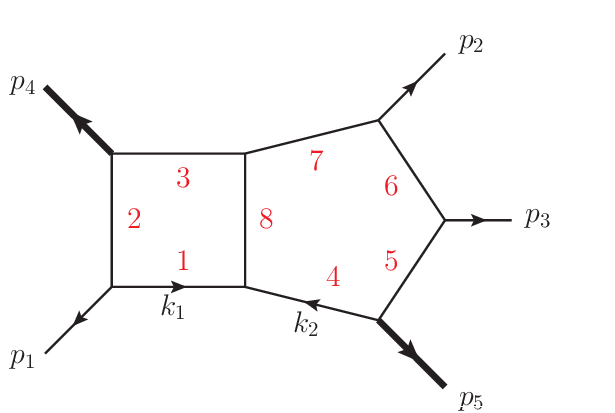}
        \caption{PBzzm}
        \label{fig:PBzzm}
    \end{subfigure}
     \begin{subfigure}[c]{0.32\linewidth}
        \centering
        \includegraphics[width=\textwidth]{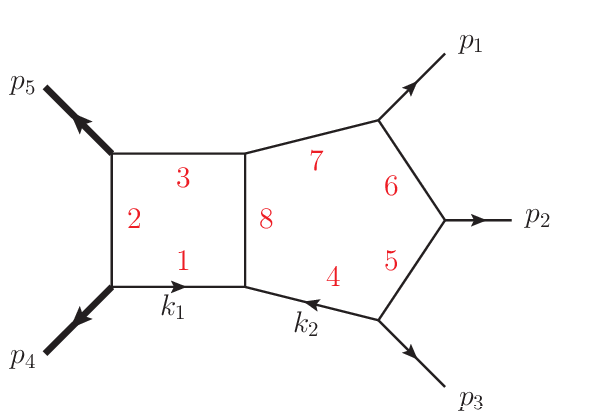}
        \caption{PBzzz}
        \label{fig:PBzzz}
    \end{subfigure}

    \caption{Independent set of two-loop ``pentagon-box'' topologies.}
    \label{fig:pentaboxes}
\end{figure}

The set of integrals we will compute can be organised into two
one-loop families and six two-loop families, distinguished
by the relative position of the two massive legs.
They are depicted in \cref{fig:pentagons,fig:pentaboxes},
together with our convention for the routing of the loop
momenta and the naming of each family.
Each diagram in \cref{fig:pentagons,fig:pentaboxes}
is associated with a set of master integrals.
For instance, to the diagram of \cref{fig:Pa}
we associate a vector space corresponding to integrals of the form
\begin{align}\begin{split}\label{eq:PaExp}
    I_{\text{Pa}}(\vec\nu)=&\,
    \int\frac{\d^{4-2\epsilon}k\,e^{\epsilon\gamma_E}}{\i\pi^{2-\epsilon}} \,
    \frac{1}
    {\rho^{\nu_1}_1\rho^{\nu_2}_2\rho^{\nu_3}_3\rho^{\nu_4}_4
    \rho^{\nu_5}_5}\\
    =&\,\int\mathcal{D}^{4-2\epsilon}k \,
    \frac{1}{(k^2)^{\nu_1}}\frac{1}{[(k+p_1)^2]^{\nu_2}}
    \frac{1}{[(k+p_1+p_2)^2]^{\nu_3}}\frac{1}{[(k-p_4-p_5)^2]^{\nu_4}}
    \frac{1}{[(k-p_5)^2]^{\nu_5}}
\end{split}\end{align}
for integer $\nu_i$. We omit Feynman's prescription for the propagators.
Each element in this vector space corresponds
to a set of exponents $\vec{\nu}$, and in this paper we compute a basis of this
space. In \cref{eq:PaExp}, we introduced the integration measure
in dimensional regularisation
\begin{equation}
    \mathcal{D}^{4-2\epsilon}k\coloneqq
    \frac{\d^{4-2\epsilon}k\,e^{\epsilon\gamma_E}}{\i\pi^{2-\epsilon}}\,,
\end{equation}
which also defines the normalisation of our integrals. 
The inverse propagators $\rho_i$ can
be read off the diagram in \cref{fig:Pa}, where we included a (red)
index for each propagator. The corresponding expression
for the second one-loop family, denoted $I_{\text{Pb}}(\vec\nu)$,
can be easily obtained from \cref{fig:Pb}.

At two loops, the diagrams in \cref{fig:pentaboxes} are not sufficient
to fully specify our conventions, as we must also define the 
so-called irreducible scalar products (ISPs). 
To each diagram in \cref{fig:pentaboxes} we associate a vector 
space corresponding to integrals of the form
\begin{equation}\label{eq:PBExp}
    I_{F}(\vec\nu)
    =\int\mathcal{D}^{4-2\epsilon}k_1\,\mathcal{D}^{4-2\epsilon}k_2 \,
    \frac{\rho^{-\nu_{9}}_{9}\rho^{-\nu_{10}}_{10}\rho^{-\nu_{11}}_{11}}
    {\rho^{\nu_1}_1\rho^{\nu_2}_2\rho^{\nu_3}_3\rho^{\nu_4}_4
    \rho^{\nu_5}_5\rho^{\nu_6}_6\rho^{\nu_7}_7\rho^{\nu_8}_8}\,,
\end{equation}
with $F\in\{\text{PBmmz, PBmzm, PBmzz, PBzmz, PBzzm, PBzzz}\}$ and
for integer $\nu_i$ such that $\nu_i\leq0$ for $i=9,10,11$.
The complete set of definitions for all the families considered
in this paper can be found in our ancillary files~\cite{ancillary}.

The dimension $\dim(F)$ of the vector space associated 
with each family $F$ in \cref{fig:pentagons,fig:pentaboxes}
corresponds to the number of master integrals we must compute for 
each of them. We collected these numbers in \cref{tab:masterCount}.
We determined them by generating systems of integration-by-parts (IBP) relations~\cite{Tkachov:1981wb,Chetyrkin:1981qh} with \textsc{LiteRed}~\cite{Lee:2012cn,Lee:2013mka} and \textsc{NeatIBP}~\cite{Wu:2023upw}, and solving them with the Laporta
algorithm~\cite{Laporta:2001dd} within the finite-field framework \textsc{FiniteFlow}~\cite{Peraro:2019svx,Peraro:2019okx}.

\begin{table}
\begin{center}
  \begin{tabular}{lcc@{\hskip 4em}lcc}
\toprule
Family & $\dim(F)$  &  $\dim(\mathcal{A}_F)$ &  Family & $\dim(F)$ & $\dim(\mathcal{A}_F)$ \\
\midrule
Pa & 16 &    43 &  PBmzz & 105 & 80  \\
Pb & 15 &    39 &  PBzmz & 104 & 96 \\
PBmmz & 94 & 85 &  PBzzm & 104 & 82  \\
PBmzm & 87 & 52 &  PBzzz & 127 & 104\\
\bottomrule
\end{tabular}
\end{center}
\caption{The number of master integrals $\dim(F)$, and the dimension of the alphabet $\dim(\mathcal{A}_F)$ of each family $F$.}
\label{tab:masterCount}
\end{table}

We close the discussion of integral families with two comments.
First, we do not consider here two-loop families that are
products of one-loop integrals, as they can be trivially obtained
from the integrals computed in this paper. 
Second, all planar integral topologies 
for instance ``triangle-hexagon'', are reducible to
the integrals from pentagon-box topologies computed here.

\section{Pure Bases and Canonical Differential Equations}
\label{sec:bases}

A pure basis and canonical differential equations for $\text{PBzzz}$ were already obtained in ref.~\cite{Jiang:2024eaj}.
In this section, we discuss the construction of {pure} bases \cite{Arkani-Hamed:2010pyv} of master integrals (MIs) for all families.
Let $\vec{I}_F$ be the list of all MIs for the family $F$.
We say that $\vec{I}_F$ is pure if it satisfies a system of differential equations (DEs) in the {canonical form}~\cite{Henn:2013pwa}
\begin{align} \label{eq:DEsCanonical1}
\d \vec{I}_F(X, \eps) = \eps \, \d {A}_F (X) \cdot \vec{I}_F(X, \eps) \,,
\end{align}
where the {connection matrix} ${A}_F(X)$ is given by a $\mathbb{Q}$-linear combination of logarithms,
\begin{align} \label{eq:DEsCanonical2}
{A}_F(X) =  \sum_i a^{(F)}_i \, \log\left( W_i(X) \right) \,.
\end{align}
Here, $\d$ is the total differential with respect to the kinematic invariants
in $X$ (see \cref{eq:sdef}),
\begin{align}
\d = \sum_{x \in X} \d x \, \frac{\partial}{\partial x} \,,
\end{align}
$a^{(F)}_i$ is a matrix of rational numbers, and the $W_i(X)$'s are algebraic functions of $X$ called {letters}.
The ensemble of all letters, called the {alphabet} and denoted 
$\mathcal{A}_F$ below, encodes the singularity structure of the MIs.
We devote \cref{sec:analyticDEs} to the identification of the letters and the expression of the connection matrices in terms of them, and focus here on the problem of constructing a basis $\vec{I}_F$ such that the dimensional regulator $\eps$ factorises as in \cref{eq:DEsCanonical1}.

Given a list of candidate MIs $\vec{I}_F(X, \eps)$, we differentiate it using \textsc{LiteIBP}~\cite{Peraro:2019svx,Peraro:2019okx}, and rewrite the derivatives in terms of $\vec{I}_F(X, \eps)$ by solving IBP relations.
We generate the required IBP relations using \textsc{LiteRed}~\cite{Lee:2012cn,Lee:2013mka} and \textsc{NeatIBP}~\cite{Wu:2023upw}.
The latter provides optimised systems of IBP relations by solving syzygy equations~\cite{Gluza:2010ws}, allowing for a faster and less memory-consuming solution.
We solve the IBPs via the Laporta algorithm~\cite{Laporta:2001dd} within the finite-field framework \textsc{FiniteFlow}~\cite{Peraro:2019svx,Peraro:2019okx}.
This allows us to side-step the intermediate expression swell which plagues multi-variable computations by replacing the symbolic manipulations with numerical evaluations over finite fields~\cite{vonManteuffel:2014ixa,Peraro:2016wsq}.
In other words, all rational coefficients are evaluated numerically for random integer values of all variables $X$ and $\eps$ modulo some (large) prime number.

The complete factorisation of $\eps$ requires the introduction of several square roots. 
A number of approaches have been proposed to include them in the finite-field framework, e.g.\ by sampling over phase-space points where the arguments of the roots are perfect squares in the finite field (see e.g.~\cite{Abreu:2020jxa}).
These approaches are however inconvenient when there are many distinct square roots.
Following ref.~\cite{Peraro:2019svx}, we prefer to reconstruct the DEs for what we call ``pre-canonical'' bases $\vec{I}'_F$, i.e., bases whose MIs are pure up to overall normalisation by square-root factors.
In other words, their DEs take the form
\begin{align}
\frac{\partial \vec{I}'_F(X, \eps) }{\partial x}= \left[ A_{F,x}^{(0)}(X) + \eps \, A_{F,x}^{(1)}(X) \right] \cdot \vec{I}'_F(X, \eps) \,,
\end{align}
for all $x \in X$, where $A_{F,x}^{(0)}(X)$ and $A_{F,x}^{(1)}(X)$ are matrices of rational functions, and $A_{F,x}^{(0)}(X)$ is diagonal, with non-zero entries only in correspondence with those MIs which require a square-root normalisation.
The finite-field sampling of the matrices $A_{F,x}^{(k)}(X)$ can then proceed via the standard algorithms.
We perform the functional reconstruction by following the strategy described in e.g.~refs.~\cite{Abreu:2018zmy,Badger:2021imn}, based on fitting linear relations among the rational coefficients and matching factors against an ansatz over univariate phase-space slices.
We also set $s_{12} = 1$ and restore its dependence via dimensional analysis.
With these techniques and the optimised IBP relations generated by \textsc{NeatIBP}, the functional reconstruction of the DEs for the pre-canonical bases is fairly simple.

Finally, we achieve the factorisation of $\eps$ with a basis transformation
\begin{align}
\vec{I}_F(X,\eps) = N_F(X) \cdot \vec{I}'_F(X,\eps)  \,,
\end{align}
where the transformation matrix is diagonal and satisfies the DEs
\begin{align} \label{eq:DEsNF}
\frac{\partial N_F(X)}{\partial x}  + N_F(X) \cdot A_{F,x}^{(0)}(X)  = 0 
\end{align}
for all $x \in X$.
We obtain the analytic expression of $N_F(X)$ by solving these DEs.
Alternatively, one may determine it by computing the leading singularities~\cite{Arkani-Hamed:2010pyv}, but we find this to be unnecessary in this case as the DEs in 
\cref{eq:DEsNF} are fairly simple to solve.
We recall that, by construction, $N_F(X)$ contains only the square-root normalisations. 
In other words, $N_F(X)^2$ is a rational function.
The resulting DEs for $\vec{I}_F(X, \eps)$ take the $\eps$-factorised form in \cref{eq:DEsCanonical1}, where the connection matrix ${A}_F(X)$ satisfies
\begin{align} \label{eq:AtildeDEs}
\frac{\partial {A}_F(X)}{\partial x}  = N_F(X) \cdot A_{F,x}^{(1)}(X) \cdot N_F^{-1}(X) \,,
\end{align}
for all $x\in X$.

In \cref{sec:BasisConstruction} we discuss how we constructed the pre-canonical integral bases.
In \cref{sec:analyticDEs} we describe the alphabet, how we obtained it, and how we express the connection matrices ${A}_F(X)$ in terms of logarithms of letters, as in \cref{eq:DEsCanonical2}.
\Cref{sec:IC} is devoted to the initial conditions necessary to solve the DEs, to the solution of the DEs using \textsc{DiffExp}, and to the validation of our results.
Finally, in \cref{sec:IteratedIntegrals} we discuss how the canonical DEs and the initial conditions allow us to straightforwardly obtain the associated 
symbols~\cite{Goncharov:2010jf} and write the integrals in terms of 
Chen iterated integrals~\cite{Chen:1977oja}.

\subsection{Construction of the Pure Bases}
\label{sec:BasisConstruction}

We organise the construction of the integral bases sector by sector, starting from the lowest, i.e., the one with the fewest propagators. 
We first analyse each sector on its maximal cut. 
This amounts to focusing on the diagonal block of the connection matrices which couples the MIs of the chosen sector.
We refer to this part of the DEs as the {homogeneous} part.
Once the diagonal block of a given sector is in pre-canonical form, we extend the analysis to include all its sub-sectors.
A basis which is canonical on the maximal cut may still require sub-sector corrections to be pure.
If the entry $ij$ in the in-homogeneous part of the connection matrix is not $\eps$-factorised, we modify the definition of the $i^{\text{th}}$ MI by a term proportional to the $j^{\text{th}}$ MI, and fix the coefficient by imposing the factorisation of $\eps$ in the entry under consideration.
This approach requires the analytic expression of the relevant entries of the DEs. 
As mentioned above, obtaining it is not a bottleneck in this case, provided that the integral basis is already (pre-)canonical on the maximal cut.
This in fact ensures that the analytic expression of the connection matrices is substantially simpler than with an arbitrary basis.
Once a sector is fixed, we proceed to all its super-sectors, until we reach the top sector.

We construct candidate pure bases for the sectors with fewer than five external legs by mapping pure integrals taken from the literature onto our definitions.
In particular, we make use of the results of ref.~\cite{Dlapa:2021qsl} for the planar 
two-loop four-point integrals with three external massive legs, 
and of ref.~\cite{Abreu:2020jxa} for the MIs that overlap with those appearing in
planar two-loop five-point integrals with  one external massive leg. 

For the genuinely five-point sectors, we build upon the knowledge about two-loop five-point integral families with other external mass configurations available in the literature.
More precisely, we draw from refs.~\cite{Abreu:2020jxa,Badger:2022hno,Badger:2024fgb}. 
We take their choices of MIs, na\"ively change the kinematics to ours, and make manual adjustments to achieve the factorisation of $\eps$. This process is particularly simple
for integrals written in terms of the loop-momentum scalar products $\mu_{ij}$ defined 
in \cref{eq:muij} since, unlike the same expressions after expansion
in terms of scalar integrals, they have trivial generalisations to different kinematic
configurations. We emphasise that, in order to analyse the dependence on $\eps$ of the pure candidates constructed in this way, it suffices to reconstruct the $\eps$ dependence of the connection matrices, that is, it suffices to work on a univariate slice where all kinematic variables $X$ are set to random values, which is computationally inexpensive.

The top sectors of all families but ${\rm PBzzz}$ have 3 MIs each. 
Following the pattern known in the literature~\cite{Abreu:2020jxa,Badger:2022hno,Badger:2024fgb}, we find that the following numerators give rise to pure integrals:
\begin{align}
\begin{aligned}
& \mathcal{N}_F^{(1)} = \eps^4 \, \sqrt{\Delta_5} \, (p_{F,i} + p_{F,j})^2 \, \mu_{12} \,, \\
& \mathcal{N}_F^{(2)} = \eps^4 \, \sqrt{\Delta_5} \, (p_{F,i} + p_{F,j})^2 \, \mu_{22} \,,
\end{aligned}
\end{align}
where $p_{F,i}$ and $p_{F,j}$ are the external momenta attached to the box sub-graph.
For the third MI in the top sector, the literature suggests to start from a numerator proportional to $(k_2 - q_F)^2$, where $q_F$ is the external momentum on the bottom left of the graphs in \cref{fig:pentaboxes}. 
In addition to an overall normalisation factor, we find that a simple sub-sector correction is required for some of the families.
Explicitly, we find
\begin{align}
\begin{aligned}
& \mathcal{N}_{\rm PBmmz}^{(3)} = \eps^4 \, s_{12} (s_4 s_{12} - s_{34} s_{45}) \,  (k_2 - p_1)^2 \,, \\ 
& \mathcal{N}_{\rm PBmzm}^{(3)} = \eps^4 \, s_{12} (s_4 s_5 - s_4 s_{34} - s_5 s_{34} - s_{12} s_{34} + s_{34}^2 + s_{34} s_{45}) \,  (k_2 - p_1)^2 \,, \\ 
& \mathcal{N}_{\rm PBmzz}^{(3)} = \eps^4 \, s_{12} s_{15} \left[ s_{34} \, (k_2 - p_3)^2 - s_4 \, k_2^2 \right] \,, \\
& \mathcal{N}_{\rm PBzmz}^{(3)} = \eps^4 \, (s_4 s_{15} - s_4 s_5 + s_{12} s_{15} + s_5 s_{23} + s_5 s_{34} - s_{15} s_{34} -  s_{15} s_{45}) 
  \left[s_{24} \, (k_2 - p_2)^2 -s_4 \, k_2^2 \right] \,, \\
& \mathcal{N}_{\rm PBzzm}^{(3)} = \eps^4 \, s_{23} (s_4+s_5+s_{12}-s_{34}-s_{45}) \left[s_{14} \, (k_2 - p_1)^2 - s_4 \, k_2^2 \right] \,.
\end{aligned}
\end{align}

Unlike the cases above, the top sector of the family ${\rm PBzzz}$ has 4 MIs. 
It was already studied in ref.~\cite{Jiang:2024eaj}, but we provide here
a new representation of the pure basis in terms of $\mu$-insertions 
for a subset of the MIs.
Indeed, three of them can be chosen as above:
\begin{align}
\begin{aligned}
& \mathcal{N}_{\rm PBzzz}^{(1)} = \eps^4 \, \sqrt{\Delta_5} \, s_{45} \, \mu_{12} \,, \\
& \mathcal{N}_{\rm PBzzz}^{(2)} = \eps^4 \, \sqrt{\Delta_5} \, s_{45} \, \mu_{22} \,, \\
& \mathcal{N}_{\rm PBzzz}^{(3)} =  \eps^4 \, s_{45} \left[s_{12} s_{23} \, (k_2 - p_4)^2 - s_{12} s_{15} \, k_2^2 - s_{23} s_{34} \, (k_2 - p_4 - p_5)^2\right] \,.
\end{aligned}
\end{align}
For the fourth MI we could not find a simple representation, 
and we adopted the definition from ref.~\cite{Jiang:2024eaj}. 
The expression is lengthy and we thus omit it here.
We limit ourselves to highlighting that its normalisation involves two square roots 
(of $\Delta_5$ and of $\lambda(s_4,s_5,s_{45})$), and that a simple numerator which yields $\eps$-factorised DEs on the maximal cut is
\begin{align}
 \mathcal{N}_{\rm PBzzz}^{(4)'} = \eps^4 \, s_{45} \, \sqrt{\Delta_5} \, \sqrt{\lambda\left(s_4,s_5,s_{45}\right)} \, \frac{ 2 \, (k_2-p_4)^2 \, \mu_{12} + (s_{45}-s_4-s_5) (\mu_{12}+\mu_{22})}{\lambda\left(s_4,s_5,s_{45}\right)} + (\text{sub-sectors}) \,.
\end{align}

We present our pure bases for all the other five-point sectors in \cref{app:5ptSectors}. 
The complete bases can be found in the ancillary files~\cite{ancillary}.

\subsection{Analytic Differential Equations and Alphabet}
\label{sec:analyticDEs}

Having determined pure bases for each family, we now turn to 
obtaining analytic DEs in the form of \cref{eq:DEsCanonical1,eq:DEsCanonical2}.
The main missing ingredient are the set of 
letters of the alphabet corresponding to each family, i.e., the 
logarithms in \cref{eq:DEsCanonical2}. Despite having analytic DEs, casting them
in the form of \cref{eq:DEsCanonical1,eq:DEsCanonical2} still requires to 
integrate the entries of the DE matrices to identify the letters. 
In practice, we find it more convenient to follow the approach of ref.~\cite{Abreu:2018rcw,Abreu:2021smk}, 
where the letters are not obtained directly from the DEs,
and the analytic differential equations are obtained by numerically fitting
the matrices $a_i^{(F)}$ in \cref{eq:DEsCanonical2} once the alphabet is known.

In order to determine the letters for each of the families in \cref{fig:pentaboxes} 
we rely on recent developments in constructing symbol alphabets~\cite{Fevola:2023kaw,
Fevola:2023fzn,Jiang:2023qnl,Dlapa:2023cvx,He:2023umf,Jiang:2024eaj,Caron-Huot:2024brh}, 
in particular on the implementation of the ideas of ref.~\cite{Jiang:2024eaj} in the 
\texttt{Mathematica} library \texttt{Baikovletter}.
In this section, we discuss which letters could be determined using 
ref.~\cite{Jiang:2024eaj}, 
and which letters we had to construct ourselves.\footnote{Following the completion of this work, a new version of \texttt{Baikovletter} was released, capable of identifying the missing letters, with the exception of the one in \cref{eq:missedRatL}. The latter requires analyzing a next-to-minimal Baikov representation, rather than the minimal representation used by \texttt{Baikovletter}. We thank Xuhang Jiang for correspondence on this matter.}

The first question we can ask about the alphabets $\mathcal{A}_F$ associated with each
of the integral families we consider in this paper is their dimension,
that is the number of linearly independent (combinations of) dlogs that appear
in the associated differential equations. Answering this question
does not require knowledge of the analytic form of the dlogs, and we
collect the alphabet dimensions in \cref{tab:masterCount}. 

Once the dimension of the alphabet is known, we have a target for the 
number of letters we must construct for each family. Based on previous 
experience \cite{Abreu:2021smk}, 
we distinguish several types of letters. First, we have {even letters}
that are polynomials in the Mandelstam variables. 
Second, we have {odd letters} which change their sign 
together with the sign of the square roots that were introduced
to construct the pure basis. Odd letters can depend on either
a single square root $\sqrt{\Lambda}$ or two square roots 
$\sqrt{\Lambda_1}$ and $\sqrt{\Lambda_2}$. In the first case we assume they take
the form
\begin{equation}\label{eq:let1root}
\frac{p(X)+q(X)\sqrt{\Lambda}}{p(X)-q(X)\sqrt{\Lambda}}\,,
\end{equation}
and in the second case they take the form
\begin{equation}\label{eq:let2root}
\frac{p(X)+q(X)\sqrt{\Lambda_1}\sqrt{\Lambda_2}}
{p(X)-q(X)\sqrt{\Lambda_1}\sqrt{\Lambda_2}}\,,
\end{equation}
where $p(X)$ and $q(X)$ are polynomials in $X$.
Odd letters have the property that they are singular at places where
the even letters vanish~\cite{Heller:2019gkq,Abreu:2021smk,Zoia:2021zmb}. 
Given that the square roots are known,
this observation can be used to constrain the polynomials
$p(X)$ and $q(X)$ and thus construct candidate odd letters.
Finally, we note that in the alphabet corresponding to five-point
one-mass kinematics at two loops we found that we could always
set $q(X)=1$, but this is not possible for the two-mass case.

We used the \texttt{Mathematica} package \texttt{Baikovletter} \cite{Jiang:2024eaj}
to identify most of the alphabet for each of the families  in 
\cref{fig:pentaboxes} (the alphabet of the one-loop families in
\cref{fig:pentagons} is a subset of the two-loop one). 
For PBmmz, PBmzz and PBzzz, the letters identified
in this way exactly span the space corresponding to the alphabet
we find in the differential equations.
For PBmzm, we find that \texttt{Baikovletter} identifies one odd letter
that is in fact not required.
For PBzzm, the code does not identify
one of the square roots, which appears as the leading singularity of
one of our pure integrals (a permutation of the one
given in \cref{eq:r3} below, see also \cref{fig:r3}), and thus also misses
the associated odd letters. We constructed four letters of the type
given in \cref{eq:let1root} that involve
only that square root, as well as two letters of the type
given in \cref{eq:let2root}, one involving the new root
and $\sqrt{\lambda(s_4,s_{23},s_{15})}$, and one involving the new root and
$\sqrt{\Delta_5}$. One of the odd letters identified by the code is not
required.
Finally, \texttt{Baikovletter} misses sixteen letters for PBzmz. Fourteen out of 
these sixteen letters are related to the letters that are missed for
PBzzm. These are two permutations
of the square root given by \cref{eq:r3} below, 
and the associated twelve odd letters (six for each root) described above. 
The rational letter
\begin{equation}\label{eq:missedRatL}
	s_4 s_{12} s_{15} + s_5 s_{23} s_{34} - s_{15} s_{34} s_{45} \,
\end{equation}
is also not identified,
even though it is a permutation of a rational letter that is identified.
The last missed letter is odd in $\sqrt{\Delta_5}$, and the
derivative of its logarithm 
is singular when the rational letter in \cref{eq:missedRatL} vanishes, 
which presumably explains why it is not identified.

The representative families of \cref{fig:pentaboxes} correspond
to a particular ordering of the external momenta. When computing
a physical process, all permutations of massless
and massive legs may appear. 
In order to obtain the full alphabet required for planar five-point two-mass processes 
at two loops, we must thus consider the closure of
the alphabets discussed above under all such permutations.

\begin{figure}[t]
    \centering
    \begin{subfigure}[c]{0.32\linewidth}
        \centering
        \includegraphics[width=\textwidth]{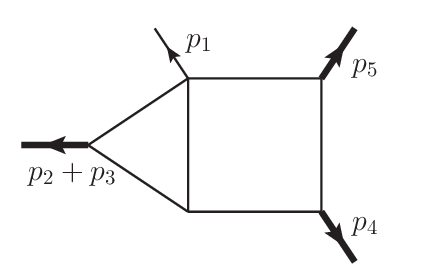}
        \caption{$r_1^{(1)}$ of \cref{eq:r1}.}
        \label{fig:r1}
    \end{subfigure}
    \begin{subfigure}[c]{0.32\linewidth}
        \centering
        \includegraphics[width=\textwidth]{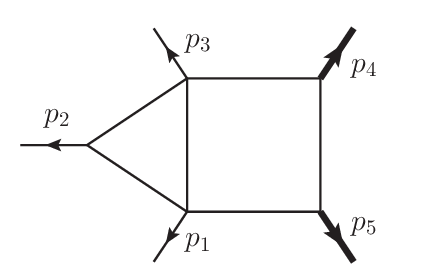}
        \caption{$r_2^{(1)}$ of \cref{eq:r2}.}
        \label{fig:r2}
    \end{subfigure}
    \begin{subfigure}[c]{0.32\linewidth}
        \centering
        \includegraphics[width=\textwidth]{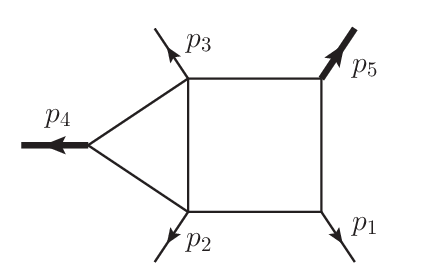}
        \caption{$r_3^{(1)}$ of \cref{eq:r3}.}
        \label{fig:r3}
    \end{subfigure}
    \caption{Representative diagrams that introduce square roots.}
    \label{fig:roots}
\end{figure}

As already highlighted, square roots play a distinguished role in building
the alphabet, allowing us to classify the letters into even and odd
letters depending on their charge under the change of the square-root sign.
In order to organise our alphabet, we start by
noticing that all square roots can be grouped into five 
permutation orbits.%
\footnote{Note that this requires viewing five-point 
two-mass kinematics
as a seven-point massless process, and then considering all permutations
of the seven-point process that are consistent with five-point two-mass kinematics.
Considering only the permutations of the massless momenta $\{p_1,p_2,p_3\}$
and the massive momenta $\{p_4,p_5\}$ would miss some relations.}

The first is the (square-root of) K\"all\'en function, and we
take as a representative
\begin{equation}
	\Delta_3^{(1)}=\lambda(s_4,s_5,s_{45})\,.
\end{equation}
It can appear in 7 permutations $\Delta_3^{(i)}$, $i=1,\ldots,7$
(note the minus sign difference between the definition of $\Delta_3$
in \cref{eq:grams} and $\Delta_3^{(i)}$).
The second root appears as the leading singularity of the integral
in \cref{fig:r1} with unit numerator, and its argument is
\begin{equation}\label{eq:r1}
	r_1^{(1)}=s_4^2 s_{23}^2 - 
 2 s_4 s_{23} (2 s_5 - s_{15} + s_{23}) s_{45} + (s_{15} - s_{23})^2 s_{45}^2\,,
\end{equation}
and it can appear in 18 permutations $r_1^{(i)}$, $i=1,\ldots,18$.
This root is associated with four-point three-mass kinematics, and was already
identified in ref.~\cite{Dlapa:2021qsl}.
The third root appears as the leading singularity of the integral
in \cref{fig:r2} with unit numerator, its argument is
\begin{equation}\label{eq:r2}
	r_2^{(1)}=s_4^2 s_{12}^2 + 
 2 s_4 s_{12} (s_5 s_{23} + (s_{15} - s_{34}) s_{45}) + (s_5 s_{23} + (
 s_{34}-s_{15}) s_{45})^2\,,
\end{equation}
and it can appear in 6 permutations $r_2^{(i)}$, $i=1,\ldots,6$.
The fourth root appears as the leading singularity of the integral
in \cref{fig:r3} with unit numerator, its argument is
\begin{equation}\label{eq:r3}
	r_3^{(1)}=4 s_4 s_{12} (s_5 - s_{15}) s_{15} + (s_5 (s_{23} + s_{34}) - s_{15} (s_{34} + s_{45}))^2\,,
\end{equation}
and it can appear in 12 permutations $r_3^{(i)}$, $i=1,\ldots,12$.
This square root can be computed in a very
similar way as the $\Sigma_5$ square root was computed in \cite{Abreu:2023rco}.
As mentioned previously, it is missed by the \texttt{Baikovletter} code.
It is however captured by the recursive Landau approach of \cite{Caron-Huot:2024brh}. 
The package \texttt{PLD.jl} \cite{Fevola:2023kaw} also detects it when computing Euler 
discriminants, but fails to detect it when computing principal Landau 
discriminants.\footnote{We thank Mathieu Giroux and Sebastian Mizera 
for assistance in these checks. The failure of the principal Landau 
discriminant approach in identifying this singularity
 is related to the fact that one of the subloops
is a triangle, whose leading singularity corresponds to taking the loop
momentum to infinity and is a more subtle case to handle within this approach.}
Finally, we also find the square-root of the five-point Gram determinant 
$\Delta_5$,
\begin{align}\begin{split}
	\Delta_5=&\,(s_4 s_{12}+s_{12} (s_{15} - s_{23}) + s_5 s_{23} - 
	s_{23} s_{34} -s_{15} s_{45}+s_{34} s_{45})^2
	-4 s_5 s_{12} s_{23} s_{34}\\
	&\,-4 s_4 s_{12} (s_{12} s_{15} + (s_{15} - s_{34}) (s_{23} - s_{45}))
	-4 s_{12} s_{23} s_{34} (s_{23}-s_{15} - s_{45})\,,
\end{split}\end{align}
which is invariant under permutations.
In total, this means that there are 44 roots for planar five-point two-mass scattering
at two loops.

\begin{table}[]
\newcolumntype{C}[1]{>{\centering\arraybackslash}m{#1}}
\centering
\begin{subtable}{.45\textwidth}
  \centering
  \begin{tabular}{l*{2}{C{6ex}}}
    \toprule
    Root & Degree & Orbit length \\
    \midrule
    $\Delta_3^{(i)}$ &  2 & 7   \\
    $\Delta_5$ &  4 & 1   \\
    $r_1^{(i)}$ &       4 & 18  \\
    $r_2^{(i)}$ &       4 & 6   \\
    $r_3^{(i)}$ &       4 & 12  \\
    \bottomrule
  \end{tabular}
  \caption{Polynomial degrees and permutation orbit lengths of the square roots appearing in the alphabet.}
  \label{tab:roots}
\end{subtable}%
\hfill
\begin{subtable}{.45\textwidth}
  \renewcommand{\arraystretch}{1.5}
  \[
  \begin{array}{l|ccccc}
                  & ~\Delta_3^{(i)}~ & ~\Delta_5~ & ~r_1^{(i)}~ & r_2^{(i)} &  ~r_3^{(i)}~\\ \hline
  \Delta_3^{(i)}  &       38       &          &           &           &   \\ 
  \Delta_5        &       8        &    69    &           &           &   \\ 
  r_1^{(i)}       &       36       &    30    &    66     &           &   \\ 
  r_2^{(i)}       &        6       &    6     &     0     &    24     &   \\ 
  r_3^{(i)}       &        12      &   12     &     0     &     0     &  48  \\ 
  \end{array}
\]
\caption{The number of letters that are odd in the square roots given in each row and column.}
  \label{tab:algebraic-letters}
\end{subtable}
\caption{Summary of the algebraic part of the alphabet, which contains 44 square roots and 355 algebraic letters.}
\label{tab:algebraic-alphabet}
\end{table}

Let us now return to the closure of the letters in the alphabet under 
all permutations. We find that there is a total of 570 letters, out of which 
215 are even and 355 are odd. 
Out of the odd letters, 236 are of the form in \cref{eq:let1root} and
depend on a single square-root and 119 are of the form in \cref{eq:let2root}
and depend on two square-roots (see \cref{tab:algebraic-alphabet}).
The full alphabet can be found in our ancillary files.

\subsection{Initial Conditions and Checks}
\label{sec:IC}

In order to solve the differential equations we need the evaluation of
the master integrals at a point. The package \texttt{AMFlow} \cite{Liu:2017jxz,Liu:2022chg} makes
this a triviality, and as such the determination of the initial
conditions for the numerical solution of the differential equations
is now a simple problem.

For completeness, we include in our ancillary files the numerical
evaluation of our bases of integrals at a point in the
Euclidean region (where integrals are either purely real or 
imaginary) and a point in what we call the `$s_{12}$-channel',
corresponding to a process where $p_1$ and $p_2$ are in the initial
state and $p_3$, $p_4$ and $p_5$ are in the final state (e.g.,
the production of two massive vector bosons together with a jet
at a hadron collider). We refer the reader to \cref{sec:app_kin}
for more details on how these kinematic regions are defined, and
here simply quote the point we chose in the Euclidean region,
\begin{equation}\label{eq:euclP}
X_{\rm eu} = \left( -\frac{3}{2}, -3, -\frac{57}{8}, -\frac{23}{4}, -\frac{5}{8}, -11, -1 \right) \,,
\end{equation}
and the point in the $s_{12}$-channel,
\begin{equation}\label{eq:X0}
X_0 = \bigl( 7, -1, 2, 5, -2, 1, 1 \bigr) \,.
\end{equation}
We note that the point $X_{\rm eu}$ is randomly chosen, but we
verified that it does not correspond to a singular point of the
differential equations. On the other hand, $X_0$ is chosen
to be a good initial condition for the construction of pentagon
functions according to the criteria of ref.~\cite{Chicherin:2021dyp}. In particular,
aside from being in the $s_{12}$-channel kinematic region,
it is symmetric under the external-momenta index swaps $1\leftrightarrow 2$
and $4\leftrightarrow 5$, which implies that it lies on the surface
where $s_4=s_5$.

The evaluations at these two points were obtained with 80-digit precision using
\texttt{AMFlow}. We verified that, starting from $X_{\rm eu}$ and evolving
the differential equations to $X_0$ with \texttt{DiffExp}~\cite{Hidding:2020ytt} 
we obtain the same results.
Given that $X_0$ is a very constrained kinematic point, we also evaluated the
functions at a generic point $X_1$ in the $s_{12}$-channel,
\begin{equation}\label{eq:X1}
X_1=\left(\frac{7}{2}, -\frac{15}{53}, \frac{11}{8}, 
\frac{15}{17}, -\frac{7}{30}, \frac{1}{15}, 
\frac{4}{31}\right)\,.
\end{equation}
Once again, we find complete agreement with the \texttt{AMFlow} evaluations and
the \texttt{DiffExp} results obtained by using either $X_0$ or $X_{\rm eu}$ as
an initial condition.\footnote{
	We note that when evolving the solution from $X_{\rm eu}$ to $X_1$ with 
	\texttt{DiffExp} for Pb, PBmzm and PBzmz we encounter a logarithmic 
	singularity that cannot be analytically continued through by simply 
	providing a positive imaginary part to $s_{12}$, $s_{23}$, $s_{34}$, 
	$s_{45}$, $s_{15}$, $s_4$ and $s_5$ as one might have expect. The singularity
	is associated with the letter 
	$s_4 (s_5 - s_{34}) + s_{34} (s_{34} + s_{45}-s_5 - s_{12}) $
	and occurs outside of the physical $s_{12}$ channel. 
  To bypass this issue we can first evolve from $X_{\rm eu}$ to $X_0$
  and then from $X_{0}$ to $X_1$, finding agreement with the
  \texttt{AMFlow} result. Within \texttt{DiffExp}, this result can be 
  reproduced by giving the letter above a small negative imaginary part.
}

Finally, solving the differential equations up to order $\eps$ is particularly
simple (all integrals are normalised to start at order $\eps^0$). 
One must simply require that the solutions to the differential equations
only have logarithmic singularities at the physical thresholds, which amounts
to imposing the so-called first entry condition~\cite{Gaiotto:2011dt}. In our case,
the physical thresholds are at $s_4=0$, $s_5=0$ and $s_{ij}=0$ if and only if the indices
$i$ and $j$ correspond to external momenta appearing next to each other in 
the graph representing each family (see \cref{fig:pentagons,fig:pentaboxes}).
This fixes the solutions at order $\eps^0$, which are just rational numbers.
At order $\eps$, the solutions are obtained from the differential equation as
linear combinations of 
$\log(-s_4-\i 0^+)$, 
$\log(-s_5-\i 0^+)$ and 
$\log(-s_{ij}-\i 0^+)$, where $0^+$ is a positive infinitesimal, and $i$ and $j$ satisfy the conditions above.
We compared the analytic solutions determined in this way to the numerical 
evaluations described above and found complete agreement.

\subsection{Iterated Integral Solution}
\label{sec:IteratedIntegrals}

While in the previous section we solved the differential equations numerically, their canonical form also allows us to write the solution analytically in terms of {Chen iterated integrals}~\cite{Chen:1977oja}.
Order by order in $\eps$, we write the expansion of the master integrals as
\begin{align}
\vec{I}_F(X,\eps) = \sum_{w \ge 0} \eps^w \, \vec{I}_F^{(w)}(X) \,.
\end{align}
At each order, the solution of the  DE is then given by
\begin{align}
\vec{I}_F^{(w)}(X) = \sum_{w'=0}^w \sum_{i_1,i_2,\ldots,i_{w'}} a_{i_1}^{(F)} \cdot a_{i_2}^{(F)} \cdots \, a_{i_{w'}}^{(F)} \cdot \vec{I}_F^{(w-w')}(X_0) \, \bigl[W_{i_{w'}}, \ldots, W_{i_2}, W_{i_1} \bigr]_{X_0}(X) \,,
\end{align}
where the sum in $i_1$, $i_2$, ..., $i_{w'}$ runs over the indices of all letters of the alphabet which are relevant for family $F$, $\vec{I}_F^{(w-w')}(X_0)$ are the initial values, and $\bigl[W_{i}, \ldots \bigr]_{X_0}(X)$ are the iterated integrals.
The latter are defined iteratively as
\begin{align} \label{eq:IIdef}
\bigl[ W_{i_1}, \ldots, W_{i_n} \bigr]_{X_0}(X) & \coloneqq \int_0^1 \d t \, \frac{\partial \log\left[W_{i_n}(\gamma(t)) \right]}{\partial t} 
  \, \bigl[ W_{i_1}, \ldots, W_{i_{n-1}} \bigr]_{X_0}\left(\gamma(t)\right) \,,
\end{align}
starting from $\bigl[ \bigr]_{X_0}(X) \coloneqq 1$.
Here, $\gamma$ is an arbitrary path in the space of kinematic variables $X$ connecting the initial and the target point, i.e., $\gamma(0)=X_0$ and $\gamma(1)=X$. 
The number of iterated integrations, $n$ in \cref{eq:IIdef}, is called transcendental weight.
Setting to zero all initial values with weight greater than 0 in \cref{eq:IIdef} ($\vec{I}_F^{(w)}(X_0) = 0$ for $w>0$) removes the dependence on the initial point $X_0$ and yields the {symbol} of the solution~\cite{Goncharov:2010jf}.
This formalism is the starting point in the construction of a basis of special functions to express the solution~\cite{Gehrmann:2018yef,Chicherin:2020oor,Badger:2021nhg,Chicherin:2021dyp,Abreu:2023rco,Badger:2023xtl} and enables the study of its analytic properties.
We refer to the review~\cite{Abreu:2022mfk} for a thorough discussion, and limit ourselves to mention two properties which we will make use of in the next section.
First, the algebraic independence of the letters $W_i$ implies that iterated integrals with different sets of letters are $\mathbb{Q}$-linearly independent. 
This enables explicit cancellations and simplifications in analytic expressions, such as the cancellation of the poles at $\eps = 0$ in the double Lagrangian insertion computed in the next section.
Second, the right-most entry of an iterated integral encodes the information about its derivatives.
In particular, the definition in \cref{eq:IIdef} implies that
\begin{align}
\d \bigl[ W_{i_1}, \ldots, W_{i_n} \bigr]_{X_0}(X) = \d \log W_{i_n}(X) \ \bigl[ W_{i_1}, \ldots, W_{i_{n-1}} \bigr]_{X_0}(X) \,.
\end{align}
We can then construct differential equations directly for the ($\mathbb{Q}$-linear combinations of) iterated integrals appearing in the result ---~say, for the double Lagrangian insertion computed in the next section~--- and solve them numerically, e.g., with \texttt{DiffExp}.
This allows us to sidestep the more expensive evaluation of the master integrals when evaluating a result obtained from them~\cite{Badger:2021nhg}.

\section{Wilson Loop with Two Lagrangian Insertions}
\label{sec:WilsonLoop}

The families of two-loop integrals studied in this paper are indispensable for calculating 
many physically relevant quantities, such as QCD corrections to electroweak production processes. Here we provide a more modest application of one of the two-loop families (namely family PBmzz, see fig.~\ref{fig:pentaboxes}) in the world of maximally super-symmetric Yang-Mills (sYM) theory. As compared to QCD amplitudes, the analytic structure of amplitudes and correlation functions in the ${\cal N} = 4$ sYM theory is usually restricted, which makes this theory a perfect testing ground for new results, allowing us
to check if the integrals we have computed reproduce the expected properties of 
the theory.

The finite gauge-invariant quantity we are going to calculate is not an amplitude, and is naturally defined in coordinate space. Let us consider a Wilson loop $W_{\rm F}$ in the fundamental representation of the colour group $SU(N_c)$, defined as
\begin{align}
W_{\rm F} = \frac{1}{N_c} {\rm tr} \, {\rm P} \exp \left( \i g_{\rm YM}\oint\limits_C A_\mu(x) \d x^\mu \right) \,. \label{eq:WL}
\end{align}
Here, $A_\mu(x) = A^a _\mu(x)\, t^a$ is a gauge field, where $t^a$ are the generators of $SU(N_c)$ in the fundamental representation, 
and $P$ stands for the path ordering of the colour indices.
We take the simplest non-trivial contour $C$, that is a quadrilateral formed by four cusps with coordinates $x_1,x_2,x_3,x_4$, with all edges lying on the light cone, i.e.
\begin{align}
x_{12}^2 = x_{23}^2 = x_{34}^2 = x_{14}^2  = 0\,, \label{contour}
\end{align}
where $x_{ij}^\mu \coloneqq (x_i -x_j)^\mu$.
We refer to \cref{eq:WL} with the light-like geometry of the contour as a 
{\it null Wilson loop}.
The simplest nontrivial gauge-invariant quantity involving $W_{\rm F}$ one could consider is its vacuum expectation value, denoted $\vev{W_{\rm F}}$, and throughout this paper we will consider it in the planar limit where $N_c \to \infty$.
Despite the ultra-violet finiteness of the ${\cal N} = 4$ sYM theory,
$\vev{W_{\rm F}}$ is divergent owing to short-distance integrations in the vicinities of the cusps which require a regulator~\cite{Korchemskaya:1992je,Bassetto:1993xd,Korchemsky:1992xv}.
 
The dimensionally-regulated $\vev{W_{\rm F}}$ is well-known to coincide with the four-gluon Maximally Helicity Violating (MHV) amplitude, both at weak~\cite{Drummond:2007aua} and strong coupling~\cite{Alday:2007hr}, upon identification of the light-like momenta of the gluons ($q_i^2 = 0$) with the edges of the Wilson loop,
\begin{align}
q_1 = x_{12} \,,\qquad \quad q_2 = x_{23} \,, \qquad \quad q_3 = x_{34} \,, \qquad \quad q_4 = x_{41} \,.  \label{xp}
\end{align}
The coordinates $x_i$ are then called dual momenta (or region momenta) of the amplitude. The null Wilson loops are known to capture infrared divergences of amplitudes \cite{Korchemsky:1985xj,Korchemsky:1987wg} in a gauge theory. In the case of ${\cal N}=4$ sYM, the duality not only maps between infrared divergences of the amplitude and cusp divergences of the null Wilson loop, but also identifies their finite parts.

\begin{figure}
\begin{center}
\includegraphics[width=6cm]{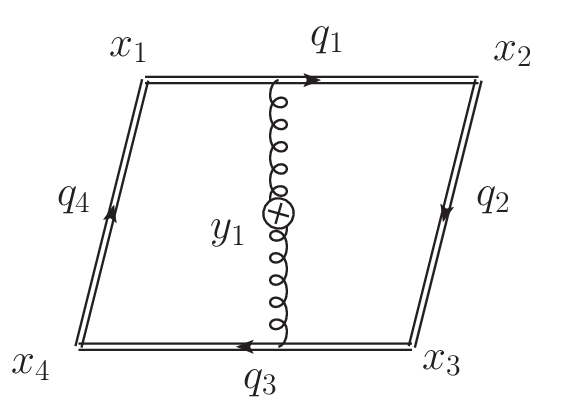} \qquad
\raisebox{4mm}{\includegraphics[width=4.8cm]{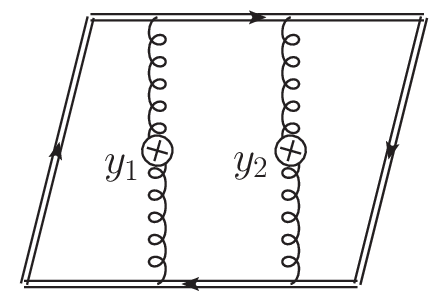}}
\end{center}
\caption{Typical Feynman diagrams representing the leading order contributions of the single ($l=1$) and double ($l=2$) Lagrangian insertions in the Wilson loop (see \cref{integrandW}) are depicted on the left and right, respectively.}\label{fig-WL}
\end{figure}

The equivalence between null Wilson loops and  MHV amplitudes also holds at the level 
of their four-dimensional integrands~\cite{Eden:2010zz,Eden:2010ce,Mason:2010yk}, 
which do not require a regulator. The Lagrangian-insertion procedure~\cite{Eden:2000mv}
provides a consistent definition of the four-dimensional Wilson loop integrand. 
It relies on the observation that, upon a suitable rescaling of the fields (e.g.~$A_\mu \to 1/g_{\rm YM} A_\mu$), differentiation of the correlation function 
$\vev{W_{\rm F}}$ with respect to the coupling constant results in a new 
correlation function involving the Lagrangian of the 
theory~\cite{Eden:2010zz}.
In other words, the $l$-loop integrand of $\vev{W_{\rm F}}$, which we denote by $M^{(l)}(y_1,\ldots,y_l)$,
is given by the correlation function of $W_{\rm F}$ and ${\cal N} = 4$ sYM Lagrangians ${\cal L}$ located at $y_{1},\ldots, y_{l}$,
which is to be calculated at the lowest perturbative order, i.e.\ $(g^2_{\rm YM})^{l}$,
\begin{align}
M^{(l)}(y_1,\ldots,y_l)
\coloneqq \vev{W_{\rm F} \,  {\cal L}(y_1) \ldots {\cal L}(y_l) }_{(g^2_{\rm YM})^{l}} \,. \label{integrandW}
\end{align}
Strictly speaking, ${\cal L}$ is the so-called {\it chiral on-shell} form of the Lagrangian,
whose classical dimension is protected from quantum corrections by the superconformal symmetry. The expression of ${\cal L}$ in terms of fields of the theory and more details can be found in~\cite{Eden:2011yp}.
In \cref{fig-WL} we present examples of Feynman diagrams contributing to \cref{integrandW}
in the cases of one and two Lagrangian insertions.
We stress that the correlator in the RHS of \cref{integrandW} is finite in four space-time dimensions only at leading order in the coupling. 
The higher order corrections, which are not relevant for \cref{integrandW}, would require a regulator.

Thanks to the duality above, $M^{(l)}$ is a four-dimensional $l$-loop integrand of both $\vev{W_{\rm F}}$ and the MHV amplitude.\footnote{More precisely, we are talking about colour-ordered MHV amplitudes normalised by their tree-level approximation, so that the integrand does not carry any colour nor helicity.}
From the amplitude point of view, the integrand in \cref{integrandW} is written in terms of dual momenta. 
The conformal symmetry of the Wilson-loop integrand thus translates into the dual-conformal symmetry of the amplitude's integrand. Integrating out the coordinates of one or several Lagrangian operators on the right-hand side of \cref{integrandW} produces cusp divergences.
Equivalently, performing the loop integrations in the corresponding integrand of the MHV amplitude leads to infrared divergences. 

This motivates us to extend \cref{integrandW} beyond the leading order of perturbation theory, and to define the following ratio of correlation functions,
\begin{align}
F_{l}(x_1,\ldots,x_4;y_{1},\ldots,y_{l}) \coloneqq \frac{\pi^{2l}}{\vev{W_{\rm F}}} \vev{W_{\rm F} \, {\cal L}(y_{1}) \ldots {\cal L}(y_{l})} \,. \label{Fn}
\end{align}
As compared to \cref{integrandW}, where the perturbative expansion is truncated at the leading order, we are interested in higher orders in the expansion of \cref{Fn} in the coupling. The correlators in the numerator and denominator of \cref{Fn} are both divergent. These divergences originate from gluon exchanges in the vicinities of the Wilson loop cusps, but they cancel out in the ratio. 
Therefore, $F_l$ is a finite quantity, well-defined in four space-time dimensions.

$F_{l}$ has the same kinematics as the $l$-loop integrand $M^{(l)}$ defined in \cref{integrandW}, but we do not aim to integrate over any of the $y_{i}$ in \cref{Fn}, since such integrations are not well-defined in four space-time dimensions. As we motivate below, this gauge-invariant quantity, depending on both on-shell and off-shell variables, is of interest on its own. The perturbative expansion of $F_l$ at weak coupling $g^2 \coloneqq g_{\rm YM}^2 N_c/(16 \pi^2)$ starts at order $(g^2)^l$,
\begin{align}
F_l = \sum_{L\geq 0} (g^2)^{l+L} F_{l}^{(L)} \,.
\end{align}
The lowest order term is the $l$-loop integrand of the Wilson loop, $F_{l}^{(0)} = M^{(l)}$, see \cref{integrandW}. In what follows, when talking about $L$-loop corrections of $F_{l}$, we mean that $L$ loop integrations are carried out.

The kinematic dependence of $F_l$ is restricted by the conformal symmetry acting in the coordinate space. 
Indeed, the light-like contour of $W_{\rm F}$ is covariant upon conformal transformations, the scaling dimension of the Lagrangian does not receive quantum corrections, and the beta-function of the theory vanishes. 
Given the cancellation of cusp divergences in the ratio on the right-hand side of 
\cref{Fn}, it follows that $F_l$ is exactly conformal in four dimensions 
with respect to the cusp coordinates. However,
$F_{l}$ carries conformal weight $(+4)$ at the Lagrangian points. It is thus
convenient to extract a factor which carries the nonzero conformal weights of $F_{l}$
and multiplies a nontrivial function of conformal cross-ratios. We choose this prefactor to be $(x_{13}^2 x_{24}^2)^l\prod_{i=1}^{4} \prod_{j=1}^{l} (x_{i}- y_{j})^{-2}$. 
Up to this normalization, $F_l$ depends on the kinematics non-trivially only 
through $1+(l-1)(l+6)/2$ conformal cross-ratios built from Lagrangian coordinates 
and cusps of the light-like contour.

In addition to conformal symmetry, $F_l$ exhibits also a discrete symmetry:
the dihedral transformations of the Wilson-loop contour. 
This means that $F_l$ is invariant upon the cyclic shift of all coordinates ($x_i \to x_{i+1}$ for all $i=1,\ldots,4$, with $x_{i+4}\equiv x_i$) and the inversion of their order ($x_i \to x_{5-i}$ for all $i=1,\ldots,4$).  

Another way to understand the finiteness of $F_l$ is to invoke the logarithm of the Wilson loop, $\log(\vev{W_{\rm F}})$, which has improved short-distance behaviour as compared to $\vev{W_{\rm F}}$: the cusp divergences of $\vev{W_{\rm F}}$ exponentiate, and $\log(\vev{W_{\rm F}})$ features only a one-loop divergence governed by the cusp $\Gamma_{\rm cusp}$ and collinear anomalous dimensions. For example, for the leading pole we have that $\log(\vev{W_{\rm F}}) \propto \Gamma_{\rm cusp}/{\eps^2}$. Differentiating $l$ times 
$\log(\vev{W_{\rm F}})$ with respect to the coupling $g^2$ results in the insertion of $l$ Lagrangians, e.g.
\begin{align}
\begin{aligned}
& g^2 \pa_{g^2} \log(\vev{W_{\rm F}}) =  \int \frac{\d^d y}{\i \pi^{\frac{d}{2}}} \, F_{l=1} (y) \,, \\
& g^4 \left( \pa_{g^2} \right)^2 \log(\vev{W_{\rm F}}) = \int \frac{\d^d y_1 }{\i \pi^{\frac{d}{2}}}  \frac{\d^d y_2}{\i \pi^{\frac{d}{2}}}  \, \left[ F_{l=2} (y_1,y_2) - F_{l=1} (y_1) F_{l=1} (y_2) \right] \,, 
\end{aligned}
\label{diffLogW}
\end{align}
et cetera.
In other words, we can think of $F_l$ as the integrand of $\log(\vev{W_{\rm F}})$  (up to products of $F_{m}$ with $m < l$) where all but $l$ loop integrations are carried out. More explicitly, $F^{(L)}_{l}$ results from $L$ loop integrations in an $(L+l)$-loop rational four-dimensional integrand. The integrands in \cref{diffLogW} are finite, and the divergence in $\log(\vev{W_{\rm F}})$ arises only upon carrying out the remaining loop integrations.

The case of a single Lagrangian insertion, $l=1$, in the four-cusp Wilson loop has been extensively studied in the literature, both at strong~\cite{Alday:2011ga} and at weak~\cite{Alday:2012hy,Alday:2013ip,Henn:2019swt} coupling, and using the negative geometry decomposition~\cite{Arkani-Hamed:2021iya,Brown:2023mqi} of the loop  corrections. Here we initiate the study of double Lagrangian insertion, $l=2$, and calculate $F_{l=2}$ for the first time in the two-loop approximation $F_{l=2}^{(2)}$.
We obtain the functional form of the one-loop result in terms of familiar one-loop polylogarithmic functions, whereas we provide an iterated integral expression at two loops.
Before we proceed to the case $l=2$, we briefly recall the available perturbative results for $F_{l=1}$ and some of its remarkable properties.

\subsection{Single Lagrangian Insertion}

Let us briefly review the structure of $F_{l=1}$ in perturbation theory.
In the case of a single Lagrangian insertion in the four-cusp Wilson loop, the kinematics is especially simple. Due to the conformal symmetry, it depends non-trivially on the single conformal cross-ratio $z$ that can be built from the Lagrangian coordinate $x_0$ and cusp coordinates $x_1,\ldots,x_4$,
\begin{align}
F_{l=1}(x_1,\ldots,x_4;x_0) = \frac{x^2_{13} x^2_{24}}{x_{10}^2 x_{20}^2 x^2_{30} x_{40}^2} \sum_{L \geq 0} (g^2)^{1+L} J^{(L)}\left(z \coloneqq \frac{x_{24}^2 x_{10}^2 x^2_{30}}{x_{13}^2 x_{20}^2 x_{40}^2}\right) \,. \label{Fx0}
\end{align}
Here, $J^{(L)}(z)$ are pure harmonic polylogarithms~\cite{Remiddi:1999ew} of weight $2L$, and the overall rational prefactor is the one-loop MHV amplitude integrand.
The first two orders \cite{Alday:2012hy} are given by
\begin{align}
J^{(0)} = -1 \,,\qquad \qquad
J^{(1)} = \log^2(z) + \pi^2 \,. \label{G1L}
\end{align}
The expressions of $J^{(2)}$ and $J^{(3)}$, of transcendental weights 4 and 6, respectively, can be found in refs.~\cite{Alday:2013ip,Henn:2019swt}. 
The dihedral symmetry implies that 
\begin{align}
J^{(L)}(z)=J^{(L)}\left(\frac{1}{z}\right) \,.
\end{align}

\begin{figure}
\begin{center}
\includegraphics[width=5cm]{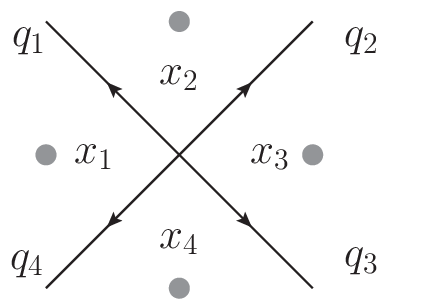} \qquad\qquad
\includegraphics[width=5cm]{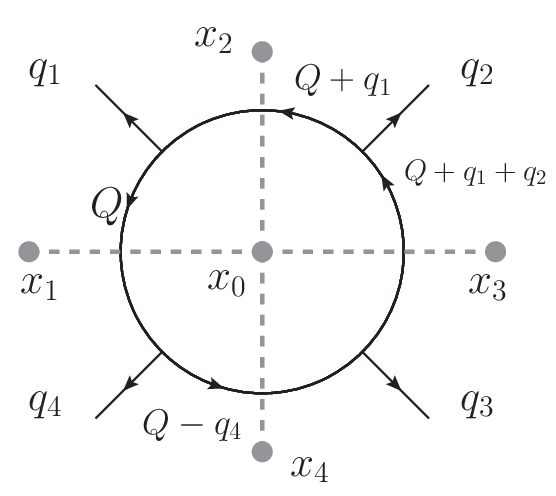}
\end{center}
\caption{Kinematics of the single (left) and double (right) Lagrangian insertions in the conformal frame, see eqs.~\p{st} and~\p{1loopvar}, where one of the Lagrangians is located at infinity. Momenta and dual-momentum variables (see \cref{xp}) correspond to dual graphs shown on the same picture. The arrows denote the directions of the momenta, the bullets denote the dual-momentum variables.}\label{kin}
\end{figure}

Without loss of generality, we can choose the conformal frame $x_0 \to \infty$ by means of a conformal transformation. Then, $z = x_{24}^2/x_{13}^2$ and, after identifying cusp coordinates with momenta according to \cref{xp}, we see that $z = t/s$ is the dimensionless ratio of the bi-particle Mandelstam variables, 
\begin{align}
s=(q_1+q_2)^2 \equiv x_{13}^2 \,,\qquad \qquad t=(q_2+q_3)^2 \equiv x_{24}^2 \,. \label{st}
\end{align}
In other words, the loop corrections $J^{(L)}$ have the same kinematics as four-particle massless amplitudes.
This correspondence is represented pictorially in \cref{kin}.
Moreover, according to the conjecture in ref.~\cite{Chicherin:2022bov}, $J^{(L)}$ coincides with the part of maximal transcendental weight of the planar $(L+1)$-loop all-plus helicity amplitude in pure Yang-Mills theory (up to an overall normalisation factor).

Another remarkable observation about available perturbative data made in 
\cite{Arkani-Hamed:2021iya} is that they do not change sign in the anti-Euclidean 
region $s,t > 0$, namely at $z > 0$, but their sign alternates with the loop order,
\begin{align}
(-1)^{L+1} J^{(L)} (z)> 0 \,,\qquad \text{at} \quad z > 0 \,. \label{eq:GLsign} 
\end{align}
We note that working in the anti-Euclidean region is conventional, and the same
result holds in the Euclidean region $s,t < 0$.

\subsection{Double Lagrangian Insertion}

The kinematics of the double Lagrangian insertion in the four-cusp Wilson loop $F_{l=2}$ is much richer as compared to the $l=1$ case shown in  \cref{Fx0}. 
The conformal symmetry implies that $F_{l=2}$ is a nontrivial function of five conformal cross ratios, ${\bf z} \coloneqq (z_1,\ldots,z_5)$, built from two Lagrangian coordinates ($x_{0}$, $x_{0'}$) and four cusp coordinates ($x_1,\ldots,x_4$), 
\begin{align}
F_{l=2}(x_1,\ldots,x_4;x_0,x_{0'}) = \frac{x^2_{13} x^2_{24}}{x_{10}^2 x_{20}^2 x^2_{30} x_{40}^2} \frac{x^2_{13} x^2_{24}}{x_{10'}^2 x_{20'}^2 x^2_{30'} x_{40'}^2} \sum_{L \geq 0} (g^2)^{2+L} G^{(L)}({\bf z}) \,. \label{FLL}
\end{align}

There is some freedom in choosing the five independent cross-ratios. In the following, we switch to the conformal frame $x_{0'} \to \infty$, and use the following set:
\begin{align}
{\bf z} := \left( \frac{x_{10}^2 x_{30'}^2}{x_{13}^2 x_{00'}^2},\frac{x_{20}^2 x_{10'}^2 x_{30'}^2}{x_{13}^2 x_{20'}^2 x_{00'}^2},\frac{x_{30}^2 x_{10'}^2}{x_{13}^2 x_{00'}^2},\frac{x_{40}^2 x_{10'}^2 x_{30'}^2}{x_{13}^2 x_{40'}^2 x_{00'}^2},\frac{x_{24}^2 x_{10'}^2 x_{30'}^2}{x_{13}^2 x_{20'}^2 x_{40'}^2}\right) \xrightarrow{x_{0'} \to \infty} \left(\frac{x_{10}^2}{x_{13}^2},\frac{x_{20}^2}{x_{13}^2},\frac{x_{30}^2}{x_{13}^2},\frac{x_{40}^2}{x_{13}^2},\frac{x_{24}^2}{x_{13}^2} \right) \,. \label{z15}
\end{align}
In order to translate the latter kinematic variables into amplitude language, we identify $x^\mu_0$ with an off-shell momentum $Q^\mu$, e.g.\ we define $Q^\mu \coloneqq x^\mu_{01}$ with $Q^2 \neq 0$.
Together with the definition of the dual momenta in \cref{xp}, this implies that
\begin{align}
x_{10}^2 = Q^2 \,, \quad x_{20}^2 = (Q+q_1)^2 \,, \quad x_{30}^2 = (Q+q_1+q_2)^2 \,,\quad x_{40}^2 = (Q-q_4)^2 \,. \label{1loopvar}
\end{align}
We represent this identification pictorially in \cref{kin}.
In other words, $G^{(L)}$ in \cref{FLL} has the same kinematic dependence as the one-loop integrand of a massless four-particle amplitude. However, contrary to the usual one-loop integrands, $G^{(L)}$ is not rational beyond the leading order, namely for $L \geq 1$.

In this work, we calculate $G^{(L)}$ at $L=0,1,2$. In order to achieve this goal, we need an efficient way to construct the loop integrands of $F^{(L)}_{l=2}$ in \cref{Fn}. The four-dimensional integrands are sufficient for our goals since the loop integrations do not introduce divergences. As the integrands of the Wilson loops are identical to those of the MHV amplitudes, they can be easily obtained from the literature~\cite{Bern:2005iz,Bern:2006ew}. Moreover, the $L$-loop integrand of the Wilson loop with $l$ Lagrangian insertions is the $(l+L)$-loop MHV amplitude integrand. Then, calculating the ratio in \cref{Fn}, we find that the loop integrand of $F^{(L)}_{l=2}$ is a combination of MHV amplitude loop integrands up to order $2+L$. 
Explicitly, we have that\footnote{See ref.~\cite{Chicherin:2022zxo} for an analogous derivation of the loop integrands in the case of the single Lagrangian insertion.}
\begin{align}
& F_{l=2}^{(0)}(x_0,x_{0'}) = M^{(2)}(x_0,x_{0'}) \,, \label{F0} \\
& F^{(1)}_{l=2}(x_0,x_{0'}) = \int \frac{\d^4 y}{\i\pi^2} \left[ M^{(3)}(x_0,x_{0'},y) - M^{(1)}(y) M^{(2)}(x_0,x_{0'}) \right] \,,  \label{F1}\\
& \begin{aligned} & F^{(2)}_{l=2}(x_0,x_{0'}) = \frac{1}{2}\int \frac{\d^4 y_1}{\i \pi^2} \frac{\d^4 y_2}{\i \pi^2} \Bigl[ M^{(4)}(x_0,x_{0'},y_1,y_2) - 2 M^{(1)}(y_1) M^{(3)}(x_0,x_{0'},y_2) \\
& \phantom{F^{(2)}_{l=2}(x_0,x_{0'}) =} - M^{(2)}(x_0,x_{0'}) M^{(2)}(y_1,y_2) +  2 M^{(1)}(y_1) M^{(1)}(y_2) M^{(2)}(x_0,x_{0'}) \Bigr]\,,
\end{aligned} \label{F2} 
\end{align}
where we recall that $M^{(L)}$ denotes the $L$-loop integrand of the MHV amplitude in \cref{integrandW}, and we omit the dependence on the cusp coordinates for the sake of compactness.

At the lowest order, $L=0$, we see from \cref{F0} that $F^{(0)}_{l=2}$ coincides with the two-loop integrand of the MHV amplitude. 
With the normalisation and expansion shown in \cref{FLL}, in the frame $x_{0'}\to \infty$, we obtain
\begin{align}
& G^{(0)} = \frac{1}{x_{13}^2  x_{24}^2}\left[ x_{13}^2 (x_{20}^2 + x_{40}^2) + x_{24}^2 (x_{10}^2+x_{30}^2)\right]\,,
\end{align}
which can be rewritten in terms of the cross-ratios defined in \cref{z15} as
\begin{align}
& G^{(0)}({\bf z}) = z_1 + z_3 + \frac{z_2}{z_5} + \frac{z_4}{z_5} \eqqcolon r_1 ({\bf z}) \,. \label{eq:G0}
\end{align}

\begin{figure}
\begin{center}
\includegraphics[width=4.5cm]{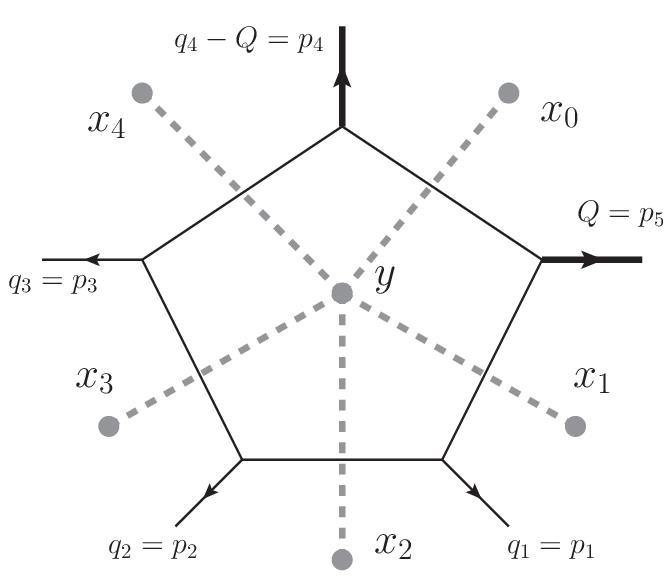} \qquad\qquad
\includegraphics[width=5.5cm]{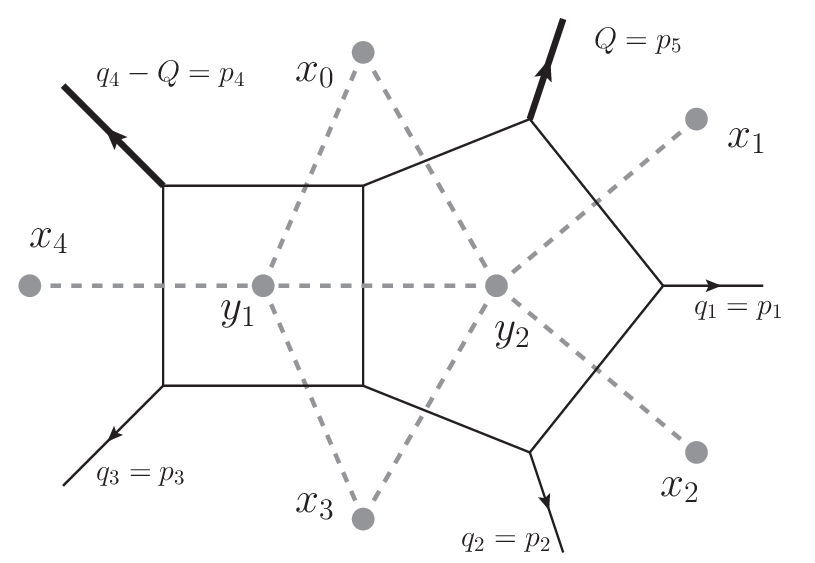}
\end{center}
\caption{Two-mass pentagon and two-mass pentabox topologies contributing to the one- and two-loop integrand of the double Lagrangian insertion $F_{l=2}$. We interpose on the same figure the Feynman diagrams drawn in momentum and dual momentum variables. The solid lines represent propagators in momentum space. The dashed lines, connecting bullets, represent propagators in the dual momentum notation. We also identify the momenta of the integrands $q_i$ with the momenta $p_i$ in figs.~\ref{fig:pentagons} and~\ref{fig:pentaboxes} (see \cref{eq:pq}). The kinematics is constrained since $(p_4+p_5)^2 = 0$.}\label{FD}
\end{figure}

Let us now move on to the loop corrections. First of all, in order to simplify the calculation, we eliminate one of the Lagrangian coordinates by choosing the conformal frame $x_{0'}\to \infty$ in the one-loop integrand in \cref{F1}. 
The loop integration in \cref{F1} is well-defined in four space-time dimensions provided we do not break the integrand into the sum of smaller pieces and integrate them separately. 
Nonetheless, we find it convenient to adopt dimensional regularisation, so we can employ the usual amplitude workflow: we IBP-reduce the appearing scalar Feynman integrals to a basis of master integrals, and express them in terms of special functions. 
Then, the cancellation of the $\eps$-poles of the individual master integrals is a strong self-consistency check of the calculation. 

The kinematics of these Feynman integrals are however more constrained than in the usual momentum-space computations.
Consider for example the two-mass pentagon Feynman integral with external momenta $p_1,\ldots,p_5$, whose kinematics is discussed in \cref{sec:kin}.
The map between the momenta of the integral ($p_i$) and those of the Lagrangian insertion ($q_i$) is given by
\begin{align} \label{eq:pq}
p_1 = x_{12} = q_1 \,, \quad 
p_2 = x_{23} = q_2 \,, \quad 
p_3 = x_{34} = q_3 \,, \quad 
p_4 = x_{40} = q_4 -Q \,, \quad
p_5 = x_{01} = Q\,,
\end{align}
as shown in \cref{FD}.
The constraints in \cref{contour} then imply that $p_1^2 = p_2^2 = p_3^2 = 0$, as in \cref{sec:kin}.
While $p_4$ and $p_5$ are off-shell, they are constrained by the fact that their sum must be light-like, since $p_4+p_5 = x_{41} = q_4$.
We then have an additional constraint on the Mandelstam variables $X$ defined in \cref{eq:sdef},
\begin{align} \label{eq:constraint}
s_{45} = (p_4+p_5)^2 = 0 \,.
\end{align}
As a result, instead of six dimensionless variables, only five are required ($z_1,\ldots,z_5$, defined in \cref{z15}), 
and we find that $F^{(1)}_{l=2}$ in \cref{F1} contains only the following one-loop Feynman integrals: the two-mass pentagon depicted in \cref{FD} (as well as its dihedral permutations), boxes (with zero and two external masses at adjacent legs), triangles (with
one, two, and three external masses), and bubbles.

By repeating this analysis for the two-loop integrand of $F^{(2)}_{l=2}$, we find that the most complicated Feynman integral topology that contributes is the two-mass pentabox PBmzz depicted in \cref{FD}, along with its dihedral permutations. As in the case of the two-mass pentagon, the kinematic dependence of the pentabox is restricted to five dimensionless variables by the constraint in \cref{eq:constraint}. The other contributing topologies are the double-boxes (with zero and two external massive legs attached to the same box)~\cite{Henn:2014lfa}, and products of one-loop topologies.

In the previous sections we have derived the canonical differential equations and identified the symbol alphabet for the PBmzz family.
We need to impose on them the additional constraint $s_{45} = 0$ required in the current calculation. 
This can be done straightforwardly through the differential equations. 
We write down the asymptotic solution of the canonical differential equations for PBmzz family in the $s_{45} \to 0$ limit by following the method of ref.~\cite{wasow1965asymptotic} (see ref.~\cite{Caron-Huot:2020vlo} for an application to Feynman integrals). In doing this, one must take care that only the so-called `hard region' of the asymptotic expansion is kept, which amounts to setting $s_{45} = 0$ at the integrand level (see ref.~\cite{Smirnov:2002pj} for a thorough discussion of the method of regions). In order to remove from the asymptotic expansion all the other regions, we drop all terms which go as $s_{45}^{-a \eps + b}$, where $a$ is a positive rational number and $b$ is a non-negative integer, and then set $s_{45} = 0$.
The resulting iterated integral representation of the PBmzz Feynman integrals at $s_{45} = 0$ involves a smaller alphabet.

The symbol of the two-mass pentabox PBmzz integrals (with the orientation of the external legs shown in \cref{fig:pentaboxes}) in the full $7$-variable kinematics of \cref{sec:kin} involves 80 letters of the 570-letter alphabet discussed in \cref{sec:analyticDEs}.
By taking the limit $s_{45} \to 0$, rewriting the resulting letters in the variables ${\bf z}$ defined in \cref{z15} and including all dihedral permutations, we obtain an alphabet of 86 letters. 
Among them, there are 43 even and 43 odd (algebraic) letters which involve 11 square roots. 
We emphasise that this is not a sub-alphabet of the 570-letter alphabet from \cref{sec:analyticDEs}.

We treat similarly the other contributing families. We use the canonical bases for the double-box families  
from~\cite{Henn:2014lfa} and re-derive their differential equations in terms of our 86-letter alphabet with the setup discussed in \cref{sec:bases}.  The identified 86-letter alphabet therefore captures the analytic structure of $F_{l=2}$ up to two loops. The one-loop approximation $F^{(1)}_{l=2}$ requires a 14-letter subalphabet involving only two square roots.

Finally, we find that the one- and two-loop coefficients of the double Lagrangian insertion in \cref{FLL} have the form
\begin{align}
G^{(1)}({\bf z}) = \sum_{i=1}^{7} r_i({\bf z})\, f^{(1)}_i({\bf z})\,, \qquad
G^{(2)}({\bf z}) = \sum_{i=1}^{64} r_i({\bf z})\, f^{(2)}_i({\bf z})\,, \label{G12}
\end{align}
where the $r_i({\bf z})$ are algebraic functions, while $f^{(1)}_i({\bf z})$ and $f^{(2)}_i({\bf z})$ are pure special functions of uniform transcendental weight two and four, respectively. Therefore, we observe that the double Lagrangian insertion $F_{l=2}$ satisfies the principle of maximal transcendentality~\cite{Kotikov:2002ab} as one
might have expected 
(we note that the same is true for the single Lagrangian insertion $F_{l=1}$). 
While we are only interested in the four-dimensional result here, we have also verified that the one- and two-loop corrections to the double Lagrangian insertion have uniform transcendental weight to all orders in $\eps$ by writing them as linear combinations of pure integrals multiplied by $\eps$-independent coefficients

Among the 86 letters appearing in the two-loop Feynman integrals, there is one dihedral-invariant square-root letter which originates from $\Delta_5$ in \cref{eq:delta5} by setting $s_{45} = 0$,
\begin{align}\label{eq:barDelta5}
\bar\Delta_5 := {\rm Gram}(x_{12},x_{23},x_{34},x_{10}) \,.
\end{align}
We observe that letter $\bar\Delta_5$ drops out of the finite quantity $G^{(2)}({\bf z})$. This is in line with previous observations about the cancellation of the analogous square-root letter $\Delta_5$ from the finite remainders of five-particle massless and one-mass amplitudes.
In the massless case, this phenomenon was linked to cluster algebras~\cite{Chicherin:2020umh} and Gr\"obner fans~\cite{Bossinger:2022eiy}.

As compared to the single Lagrangian insertion $F_{l=1}$, which requires just one rational prefactor (see \cref{Fx0}), the double Lagrangian insertion has a more complicated structure.
It involves 64 coefficients ($r_i({\bf z})$ in \cref{G12}), of which 28 are rational, and 36 are algebraic.
The algebraic coefficients are normalised by one of the 11 square roots in the alphabet, 
and are otherwise rational.
The double Lagrangian insertion is however independent of the choice of the branches of the square roots.
Indeed, each algebraic coefficient is accompanied in \cref{G12} by a pure function which is odd with respect to the sign of the corresponding square root, so that their product is even.
Furthermore, some coefficients contain spurious singularities, i.e., loci where some coefficients $r_i({\bf z})$ diverge while $G^{(1)}({\bf z})$ and $G^{(2)}({\bf z})$ should stay finite. 
The spurious singularities come from polynomials in the denominators of $r_i({\bf z})$ as well as from the square roots discussed above. 
We checked at symbol level that the spurious poles of $r_i({\bf z})$ are suppressed by zeros of the accompanying pure functions $f^{(L)}_i({\bf z})$, and expect this to hold at function level.

Despite this higher complexity, the analogy with the single Lagrangian insertion extends also to certain properties of the rational coefficients.
As for the rational coefficient of the single Lagrangian insertion in \cref{Fx0}, in fact, we find that all the coefficients of the one- and two-loop double Lagrangian insertion in \cref{G12} have unit leading singularities~\cite{Arkani-Hamed:2010pyv}.
The leading singularities of an integrand are the residues at the singularities of the highest codimension in the integration variables.
Denoting by $\underset{v}{\rm LS}\left[f \right]$ the leading singularities of an integrand $f$ with integration variables $v$, for the rational coefficient of the single Lagrangian insertion (see \cref{Fx0}) we have that
\begin{align}
\underset{x_0}{\rm LS}  \left[ \frac{x^2_{13} x^2_{24}}{x_{10}^2 x_{20}^2 x^2_{30} x_{40}^2} \right] = 1 \,. \label{LS1}
\end{align}
This is nothing but the translation into dual momenta of the well-known fact that the leading singularity of the massless one-loop box is $1/(s t)$.
We have verified by means of the package \texttt{DlogBasis}~\cite{Henn:2020lye} that the same holds for the coefficients of the double Lagrangian insertion,\footnote{Strictly speaking, these leading singularities are non-zero rational constants. The normalisation of the coefficients is however arbitrary, as these constants can be absorbed into the definition of the pure functions they multiply.}
\begin{align}
\underset{x_0, x_{0'}}{\rm LS} \left[ r_i({\bf z}) \frac{x^2_{13} x^2_{24}}{x_{10}^2 x_{20}^2 x^2_{30} x_{40}^2} \frac{x^2_{13} x^2_{24}}{x_{10'}^2 x_{20'}^2 x^2_{30'} x_{40'}^2} \right] = \underset{x_0}{\rm LS} \left[ r_i({\bf z}|_{x_{0'} \to \infty}) \frac{x^2_{13} x^2_{24}}{x_{10}^2 x_{20}^2 x^2_{30} x_{40}^2} \right] = 1 \,. \label{LS2}
\end{align}
We stress that, for this property to hold, the overall rational normalisation factor carrying the conformal weights of the double Lagrangian insertion in \cref{FLL} needs to be kept into account as well.
Conformal symmetry then allows us to simplify the computation by a transformation $x_0' \to \infty$ to the conformal frame.
This property can be explained heuristically as follows.
The coefficients $r_i({\bf z})$ in \cref{G12} can be obtained from $G^{(1)}({\bf z})$ and $G^{(2)}({\bf z})$ by taking discontinuities iteratively until all pure functions $f_i^{(L)}$ are replaced by powers of $\i \pi$.
These discontinuities correspond to suitable residues of the integrands of $F_{l=2}^{(1)}$ and $F_{l=2}^{(2)}$. 
The latter are built from loop integrands of MHV amplitudes $M^{(L)}$ (see eqs.~\p{F0}~--~\p{F2}), which are known to admit a $\d\log$ representation and have unit leading singularities~\cite{Arkani-Hamed:2012zlh,Bern:2014kca}. 
Since all residues of a $\d\log$ form with unit leading singularities are themselves $\d\log$ forms with unit leading singularities, we can expect that this property should hold for the coefficients $r_i({\bf z})$.

Finally, we organised $G^{(1)}$ and $G^{(2)}$ in such a way that they are manifestly invariant under dihedral transformations of the kinematic variables. The latter in fact act by permuting the addends $r_i({\bf z}) f^{(L)}_i({\bf z})$ in the expressions of $G^{(1)}$ and $G^{(2)}$ given in \cref{G12}, thus leaving the sums invariant.

We provide analytic expressions for $G^{(1)}$ and $G^{(2)}$ in the ancillary files, together with the definition of the corresponding alphabet letters and of the algebraic coefficients $r_i({\bf z})$. In the one-loop case we provide an explicit polylogarithmic representation for $\{ f_i^{(1)} \}_{i=1}^{7}$. They are given by the zero-mass-box, two-mass-hard-box, and three-mass-triangle functions. 
In the two-loop case, we provide an iterated integral expression for $\{ f_{i}^{(2)} \}_{i=1}^{64}$ with iterated integrals defined with respect to the base point 
\begin{align}
\label{eq:z0base}
{\bf z}_0 \, : \, \bigl(x_{10}^2 = 1,\, x_{20}^2 = 1 ,\, x_{30}^2 = 3, \, x_{40}^2 = 3,\, x_{13}^2 = 1,\, x_{24}^2 = 1\bigr)\,.
\end{align}
In order to evaluate them numerically, we follow the approach of ref.~\cite{Badger:2021nhg}: we derive the system of differential equations they satisfy, and solve them with the method of generalised power series expansions~\cite{Moriello:2019yhu}.
In other words, we construct the canonical system of differential equations 
\begin{align} \label{eq:Fcanon}
\d {\vec{F}}({\bf z}) = \d A({\bf z}) \cdot {\vec{F}}({\bf z})\,,\qquad \d = \sum_{i=1}^{5} \d z_i \frac{\pa}{\pa z_i} \,, 
\end{align}
for 187 uniform-weight iterated integrals $\{F_i\}_{i=1}^{187}$. 
The first 64 are the weight-four iterated integrals appearing in $G^{(2)}$, i.e., $F_i = f^{(2)}_i$ for $i=1,\ldots,64$.
The remaining ones ($F_i$ for $i=65,\ldots,187$) are lower-weight iterated integrals which result from the iterative differentiation of $\{ f^{(2)}_i \}_{i=1}^{64}$. 
The absence of the letter $\bar{\Delta}_5$ \p{eq:barDelta5} is manifest in the connection matrix $\d A({\bf z})$, which is written in terms of the other 85 letters.
Finally, we obtained numerical boundary values ${\vec{F}}({\bf z}_0)$ at the base point ${\bf z}_{0}$ with (at least) 55-digit precision by means of \texttt{AMFlow}~\cite{Liu:2017jxz,Liu:2022chg}. 
The canonical differential equations in \cref{eq:Fcanon}, supplemented by the boundary values, can then be integrated numerically.
We make use of \texttt{DiffExp}~\cite{Hidding:2020ytt} to evaluate numerically the two-loop double Lagrangian insertion, allowing us to investigate its positivity properties.

\subsection{Uniform Sign of the Loop Corrections}
\label{sec:AMPLpos}

The remarkable positivity property of the loop corrections to the single Lagrangian insertion (see \cref{eq:GLsign}) begs for a generalisation to the double Lagrangian insertion. 
We conjecture that the loop corrections to $F_{l=2}$ also have a uniform sign which alternates with the loop order, as
\begin{align}
(-1)^{L} G^{(L)}({\bf z}) > 0  \,,\qquad {\bf z} \in \mathbb{A} \,,
\label{eq:signFl2}
\end{align}
in a special kinematic region $\mathbb{A}$ which is defined by the amplituhedron~\cite{Arkani-Hamed:2013jha}. 

The double Lagrangian insertion suggests that we consider the two-loop four-particle MHV amplituhedron. 
This geometry is carved out by inequalities on the four-brackets of momentum twistors~\cite{Arkani-Hamed:2017vfh}, which are translated to our space-time variables in the frame $x_{0'} \to \infty$ as follows:
\begin{empheq} [left=\mathbb{A}\;:\quad {\empheqlbrace}]{alignat=2}
& x_{10}^2 > 0 \,, \quad x_{20}^2 >0 \,, \quad x_{30}^2 > 0 \,, \quad x_{40}^2 > 0 \,, \quad x_{13}^2 > 0 \,, \quad x_{24}^2 > 0 \, , \label{eq:12AB} \\[0.2cm]
& x_{13}^2 (x_{20}^2 + x_{40}^2) + x_{24}^2 (x_{10}^2+x_{30}^2) - x_{13}^2 x_{24}^2 > 0\,, \quad
\bar\Delta_5 > 0 \,. \label{eq:13AB} 
\end{empheq}
The region $\mathbb{A}$ is thus a subregion of the anti-Euclidean region (see the comment below \cref{eq:GLsign}), which is defined by the inequalities in \cref{eq:12AB}, restricted by the constraints in \cref{eq:13AB}. We provide more details on this region 
in~\cref{sec:app_kin}.

One can easily see that the lowest order correction in \cref{eq:G0} is positive, $G^{(0)}({\bf z}) >0$, in the whole Euclidean region. 
The amplituhedron constraints become relevant starting from one loop. 
In other words, while $G^{(1)}({\bf z})$ does change sign within the Euclidean region, we observe that $G^{(1)}({\bf z})<0$ inside the subregion $\mathbb{A}$. 
This is based on the evaluation of the polylogarithmic function $G^{(1)}({\bf z})$ in $O(10^{7})$ random points in $\mathbb{A}$.
The negativity of $G^{(1)}({\bf z})$ in $\mathbb{A}$, namely that a 5-variable polylogarithmic function does not change sign inside a curvy region $\mathbb{A}$, appears to be very nontrivial. Let us note that seven terms $\{ r_i f_i \}_{i=1}^{7}$ in the expression for $G^{(1)}({\bf z})$ given by \cref{G12} do not have fixed sign inside $\mathbb{A}$, yet they conspire to guarantee that $G^{(1)}({\bf z}) < 0$. In our numerical study, we detected $45 < 2^7-1$ different sign patterns $\{ {\rm sign}(r_i f_i) \}_{i=1}^{7}$. Of course, this counting holds for our particular representation of the answer, and a rearrangement of terms in \cref{G12} could potentially decrease the number of sign patterns.

The iterated integral expression for the two-loop correction allows us to evaluate $G^{(2)}$ numerically as well. 
Here we provide the benchmark values
\begin{align}
G^{(0)}({\bf z}_1) \approx 24.261630456 \,, \quad G^{(1)}({\bf z}_1) \approx -988.27502992 \,, \quad
G^{(2)}({\bf z}_1) \approx 27222.154196 \,,
\end{align} 
at the random point
\begin{align}
{\bf z}_1 \in \mathbb{A}: \quad  \left( x_{10}^2 = \frac{271}{13},\, x_{20}^2 = \frac{463}{29} ,\, x_{30}^2 = \frac{499}{79}, \, x_{40}^2 = \frac{299}{83},\, x_{13}^2 = \frac{73}{53},\, x_{24}^2 = \frac{367}{89} \right)\,.
\end{align}
We evaluated $G^{(2)}({\bf z}_1)$ in two different ways. 
On the one hand, we evaluated numerically all Feynman integrals contributing to $G^{(2)}$ at ${\bf z}={\bf z_1}$ using \texttt{AMFlow}, and once again observed the cancellation of the $\eps$-poles coming from individual integrals. 
On the other hand, using \texttt{DiffExp}, we integrated numerically the canonical differential equation in \cref{eq:Fcanon} satisfied by $\{f^{(2)}_i\}_{i=1}^{64}$, this way transporting their values from the base point ${\bf z} = {\bf z}_0 \in \mathbb{A}$ given in eq.~\p{eq:z0base} to ${\bf z} = {\bf z}_1$. 
We find agreement between the two evaluations within the expected numerical accuracy, which is an additional cross-check of our calculation. 
The second approach is advantageous, since it requires fewer computational resources. 
Indeed, it does not require dimensional regularisation, and directly provides the values of the transcendental functions $\{f^{(2)}_i\}_{i=1}^{64}$.

We would like to stress that \emph{all} the amplituhedron constraints defining the region $\mathbb{A}$ in eqs.~\eqref{eq:12AB} and~\eqref{eq:13AB} are essential for the uniform sign conjecture to hold. 
Indeed, the sign is not uniform in the whole anti-Euclidean region defined by \cref{eq:12AB} alone. 
To see this, let us consider the following ray parametrised by $t > 0$,
\begin{align*}
{\bf z}_{\rm E}(t) : \quad  \left( x_{10}^2 = \frac{31}{4046},\, x_{20}^2 = \frac{86}{663} ,\, x_{30}^2 = \frac{3824}{2329}, \, x_{40}^2 = \frac{2858}{4159},\, x_{13}^2 = t\frac{4741}{85},\, x_{24}^2 = t\frac{4262}{79} \right)\,,
\end{align*}
which belongs to the anti-Euclidean region.
This ray punctures the amplituhedron region $\mathbb{A}$ defined in \cref{eq:13AB}, as ${\bf z}_{\rm E}(t) \notin \mathbb{A}$ at $t \gtrapprox 0.02817$ and ${\bf z}_{\rm E}(t) \in \mathbb{A}$ otherwise. 
We verify that the uniform sign conjecture of \cref{eq:signFl2} holds on the segment of the ray which is inside $\mathbb{A}$. 
However, the one-loop correction evaluated on the ray, $G^{(1)}\left({\bf z}_{\rm E}(t) \right)$, changes sign at $t \approx 1.34435$, while the change of sign for the two-loop correction $G^{(2)}({\bf z}_{\rm E}(t) )$ happens at $t \approx 0.45154$. 
In other words, the sign of the loop corrections is not uniform outside of the amplituhedron region $\mathbb{A}$.

We evaluated $G^{(2)}({\bf z})$ at a number of points ${\bf z} \in \mathbb{A}$ with \texttt{DiffExp}, and found agreement with our positivity conjecture (see \cref{eq:signFl2}). 
More explicitly, we evaluated it along $\mathcal{O}(10)$ one-dimensional slices of the kinematics emanating from ${\bf z}_0$.
Still, the current approach to the numerical evaluation of $G^{(2)}$ is not efficient enough to test the conjecture on a sample of the same size as we used in the one-loop case.
It would be extremely interesting to undertake a more detailed study of the positivity of $G^{(2)}({\bf z})$. 
Given that the analytic structure of $G^{(2)}({\bf z})$ is much more complicated than that of $G^{(1)}({\bf z})$, the expected positivity of $G^{(2)}({\bf z})$ seems to be even more miraculous.

The four-dimensional amplitude integrands are differential forms with positive coefficients inside the amplituhedron geometry~\cite{Arkani-Hamed:2014dca}, which belong to an anti-Euclidean region. At the same, the alternating sign property of the loop corrections arises after integration over a Minkowski contour. Thus, it is not clear whether the positivity of the integrands can explain the uniform sign of the loop corrections. Our conjecture for the double Lagrangian insertion adds to a list of similar observations about positivity properties of the integrated loop corrections in the amplituhedron geometry, e.g.\ the finite ratio function of the six-particle amplitude~\cite{Dixon:2016apl}, and the single Lagrangian insertion in the four-cusp~\cite{Arkani-Hamed:2021iya} and five-cusp~\cite{Chicherin:2022zxo} null Wilson loops.

\section{Conclusions}
\label{sec:conclusions}

In this paper we constructed analytic differential equations for a 
complete set of planar two-loop five-point Feynman integrals with two off-shell
external legs. 
There are six different two-loop integral families that do not factorise
into products of one-loop integrals, and for each of them we have determined
a pure basis, satisfying a differential equation in canonical form.
The corresponding logarithmic forms were mostly obtained from newly developed
tools that allow one to construct symbol letters. 
The analytic differential equations were then obtained from finite-field samples
with techniques that are by now standard. We observed that, despite the 
large number of scales involved in these integrals, modern 
IBP-reduction tools were able to handle these calculations.
The derived analytic differential equations can be readily solved through generalised power series expansions.
This allowed us to perform consistency checks by
transporting numerical solutions between the Euclidean region 
and a physical region, finding agreement with an independent evaluation through the auxiliary mass flow method.

The families we considered need to be closed under permutations to cover all
integrals that appear in physical quantities such as amplitudes. 
With that observation in mind, we completed the alphabet 
with these transformations, finding a total of 570 letters.
This is substantially
larger than the alphabet for planar five-point scattering with
a single off-shell external leg at two-loops (which has 58 letters). 
In particular,
we observe a large increase in the number of square roots appearing in these
differential equations compared to the one-mass case. 
Nevertheless, the analytic structure of the letters is similar to that observed
in other five-point integrals involving several roots.
This observation gives us great confidence that the strategy to build
pentagon functions established in ref.~\cite{Abreu:2023rco} will be 
directly applicable for this set of integrals. This is however left
for future work.

In the second part of this paper, we
used our integrals to initiate the study of the double Lagrangian insertion 
in the null Wilson loop in ${\cal N} = 4$ super-symmetric Yang-Mills. 
Our motivation was twofold. First, the analytic properties of 
quantities in ${\cal N} = 4$ sYM are very constrained, so recovering
those expected properties with our integrals is a non-trivial check of their
correctness. Second, we would like to understand 
which of the beautiful properties observed for the single Lagrangian insertions 
extend to the double Lagrangian insertions.
Compared to the former, the kinematic space of the latter is multidimensional already for the simplest null-polygonal contour, 
the quadrilateral, which we considered in this paper.
We confirmed the expectations that the double Lagrangian insertion is finite, 
conformally invariant in four dimensions, and has uniform transcendentality. 
The analytic structure is described by a 85-letter alphabet involving 11 square 
roots. The rational coefficients accompanying the pure functions are rather 
special as well, as they have unit leading singularities.
We took the necessary steps required to numerically evaluate the
double Lagrangian insertions up to two-loops, which allowed
us to formulate and test a new conjecture on the positivity of these 
quantities inside a kinematic region defined by the amplituhedron. 

There are however many open questions that remain to be answered about
double Lagrangian insertions in a Wilson loop.
For instance, with a view to a three-loop bootstrap, 
it would be interesting to understand whether the alphabet stabilises at 
two loops, and whether new rational coefficients can appear at 
higher loop orders. 
Another question we have not investigated is whether the hidden momentum-space 
conformal symmetry of the single Lagrangian insertion \cite{Chicherin:2022bov} 
has a counterpart for the double Lagrangian insertion. 
Also, we would like to understand whether the double Lagrangian insertion can 
be identified with any amplitude in non-supersymmetric theories, as is the 
case for the single Lagrangian insertion and the all-plus helicity amplitude 
in pure Yang-Mills theory~\cite{Chicherin:2022bov}. 
The perturbative data we provide are a good starting point for these 
investigations, which we leave to future work.

\section*{Acknowledgments}
The authors thank Antonela Matijašić and Julian Miczajka for useful discussions about the construction of algebraic letters,
and Xuhang Jiang for helpful correspondence about the package \texttt{Baikovletter}.
D.C.\ is grateful to the Max-Planck Institute for Physics, where part of the work has been done.
This project has received funding from the European Union's Horizon Europe research and innovation programme under the Marie Skłodowska-Curie grant agreement No.~101105486. 
V.S.\ has received funding from the European Research Council (ERC) under the
European Union's Horizon 2020 research and innovation programme grant agreement
101019620 (ERC Advanced Grant TOPUP). D.C.\ is supported by ANR-24-CE31-7996.

\appendix

\section{Pure Master Integrals for the Five-Point Sectors}
\label{app:5ptSectors}

In this appendix, we present our pure bases for all independent five-point sub sectors (modulo permutations of the external massless legs, and exchanges $p_4 \leftrightarrow p_5$). 
We omit those sectors whose integrals are products of one-loop integrals.
Explicit, machine-readable expressions for all master integrals can be found in our ancillary files~\cite{ancillary}.

\subsubsection*{PBmmz, $\{1,1,1,1,1,1,1,1,0,0,0\}$, 3 MIs}

\hspace{\parindent}
\begin{minipage}{0.4\textwidth}
\begin{figure}[H]
\centering
\vspace{-0.4cm}
\includegraphics[width=\textwidth]{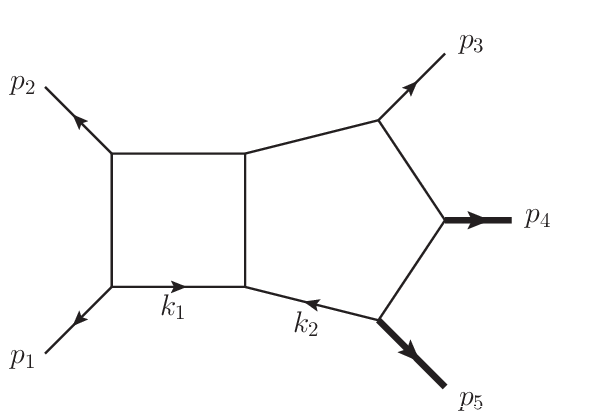}
\end{figure}
 \end{minipage}
\begin{minipage}{0.65\textwidth}
\begin{align}
\begin{split}
  & \mathcal{N}_{\rm PBmmz}^{(1)} = \eps^4 \, \sqrt{\Delta_5} \, (p_{1} + p_{2})^2 \, \mu_{12} \,, \\
  & \mathcal{N}_{\rm PBmmz}^{(2)} = \eps^4 \, \sqrt{\Delta_5} \, (p_{1} + p_{2})^2 \, \mu_{22} \,, \\
  & \mathcal{N}_{\rm PBmmz}^{(3)} = \eps^4 \, s_{12} (s_4 s_{12} - s_{34} s_{45}) \,  (k_2 - p_1)^2  \,. \\ 
\end{split}
\end{align}
\end{minipage}

\medskip

\subsubsection*{PBmzm, $\{1,1,1,1,1,1,1,1,0,0,0\}$, 3 MIs}

\hspace{\parindent}
\begin{minipage}{0.4\textwidth}
\begin{figure}[H]
\centering
\vspace{-0.4cm}
\includegraphics[width=\textwidth]{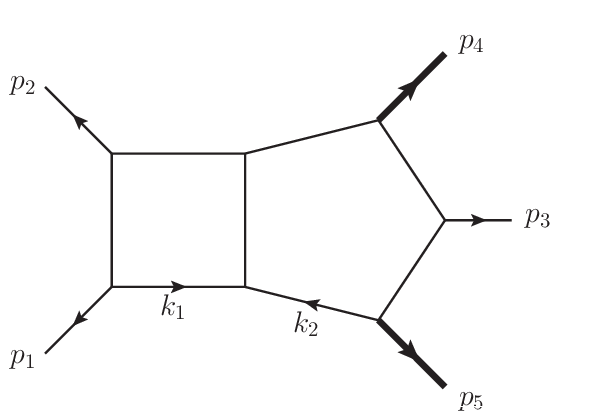}
\end{figure}
 \end{minipage}
\begin{minipage}{0.65\textwidth}
\begin{align}
\begin{split}
  & \mathcal{N}_{\rm PBmzm}^{(1)} = \eps^4 \, \sqrt{\Delta_5} \, (p_{1} + p_{2})^2 \, \mu_{12} \,, \\
  & \mathcal{N}_{\rm PBmzm}^{(2)} = \eps^4 \, \sqrt{\Delta_5} \, (p_{1} + p_{2})^2 \, \mu_{22} \,, \\
  & \mathcal{N}_{\rm PBmzm}^{(3)} = \eps^4 \, s_{12} (s_4 s_5 - s_4 s_{34} - s_5 s_{34} - s_{12} s_{34} + s_{34}^2 + s_{34} s_{45})  \\
  & \phantom{ \mathcal{N}_{\rm PBmzm}^{(3)} = } \times (k_2 - p_1)^2   \,. \\ 
\end{split}
\end{align}
\end{minipage}

\medskip

\subsubsection*{PBmzz, $\{1,1,1,1,1,1,1,1,0,0,0\}$, 3 MIs}

\hspace{\parindent}
\begin{minipage}{0.4\textwidth}
\begin{figure}[H]
\centering
\vspace{-0.4cm}
\includegraphics[width=\textwidth]{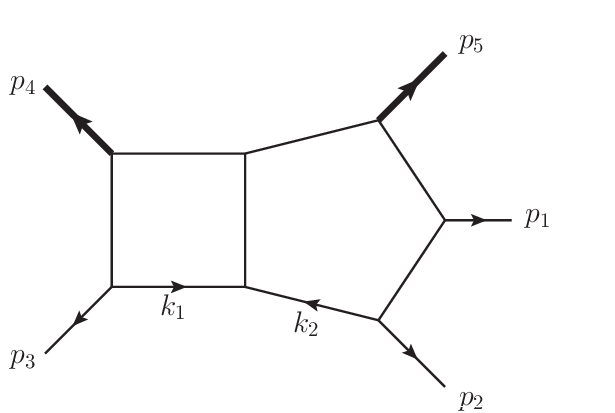}
\end{figure}
 \end{minipage}
\begin{minipage}{0.65\textwidth}
\begin{align}
\begin{split}
  & \mathcal{N}_{\rm PBmzz}^{(1)} = \eps^4 \, \sqrt{\Delta_5} \, (p_{3} + p_{4})^2 \, \mu_{12} \,, \\
  & \mathcal{N}_{\rm PBmzz}^{(2)} = \eps^4 \, \sqrt{\Delta_5} \, (p_{3} + p_{4})^2 \, \mu_{22} \,, \\
  & \mathcal{N}_{\rm PBmzz}^{(3)} = \eps^4 \, s_{12} s_{15} \left[ s_{34} \, (k_2 - p_3)^2 - s_4 \, k_2^2 \right]  \,. \\
\end{split}
\end{align}
\end{minipage}

\medskip

\subsubsection*{PBzmz, $\{1,1,1,1,1,1,1,1,0,0,0\}$, 3 MIs}

\hspace{\parindent}
\begin{minipage}{0.4\textwidth}
\begin{figure}[H]
\centering
\vspace{-0.4cm}
\includegraphics[width=\textwidth]{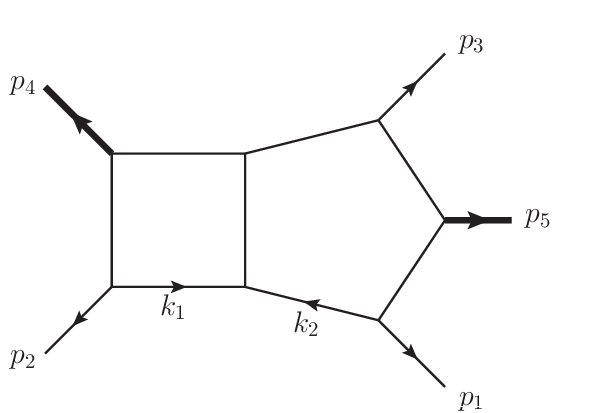}
\end{figure}
 \end{minipage}
\begin{minipage}{0.65\textwidth}
\begin{align}
\begin{split}
  & \mathcal{N}_{\rm PBzmz}^{(1)} = \eps^4 \, \sqrt{\Delta_5} \, (p_{2} + p_{4})^2 \, \mu_{12} \,, \\
  & \mathcal{N}_{\rm PBzmz}^{(2)} = \eps^4 \, \sqrt{\Delta_5} \, (p_{2} + p_{4})^2 \, \mu_{22} \,, \\
  & \mathcal{N}_{\rm PBzmz}^{(3)} = \eps^4 \, (s_4 s_{15} - s_4 s_5 + s_{12} s_{15} + s_5 s_{23} + s_5 s_{34} \\
   & \phantom{ \mathcal{N}_{\rm PBzmz}^{(3)} = } - s_{15} s_{34} -  s_{15} s_{45}) \left[s_{24} \, (k_2 - p_2)^2 -s_4 \, k_2^2 \right] \\
\end{split}
\end{align}
\end{minipage}

\medskip

\subsubsection*{PBzzm, $\{1,1,1,1,1,1,1,1,0,0,0\}$, 3 MIs}

\hspace{\parindent}
\begin{minipage}{0.4\textwidth}
\begin{figure}[H]
\centering
\vspace{-0.4cm}
\includegraphics[width=\textwidth]{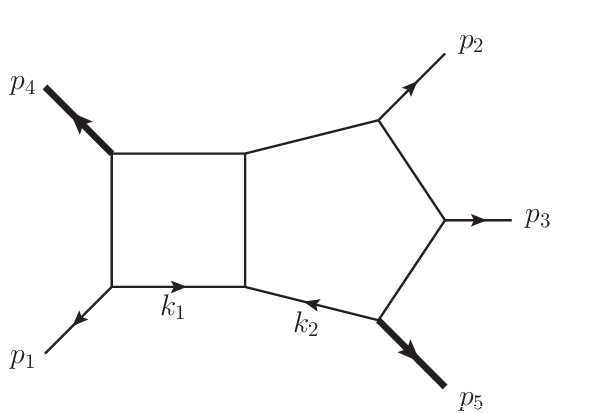}
\end{figure}
 \end{minipage}
\begin{minipage}{0.65\textwidth}
\begin{align}
\begin{split}
  & \mathcal{N}_{\rm PBzzm}^{(1)} = \eps^4 \, \sqrt{\Delta_5} \, (p_{1} + p_{4})^2 \, \mu_{12} \,, \\
  & \mathcal{N}_{\rm PBzzm}^{(2)} = \eps^4 \, \sqrt{\Delta_5} \, (p_{1} + p_{4})^2 \, \mu_{22} \,, \\
  & \mathcal{N}_{\rm PBzzm}^{(3)} = \eps^4 \, s_{23} (s_4+s_5+s_{12}-s_{34}-s_{45}) \\
  & \phantom{ \mathcal{N}_{\rm PBzzm}^{(3)} = } \times \left[s_{14} \, (k_2 - p_1)^2 - s_4 \, k_2^2 \right]  \,. \\
\end{split}
\end{align}
\end{minipage}

\medskip

\subsubsection*{PBzzz, $\{1,1,1,1,1,1,1,1,0,0,0\}$, 4 MIs}

\hspace{\parindent}
\begin{minipage}{0.4\textwidth}
\begin{figure}[H]
\centering
\vspace{-0.4cm}
\includegraphics[width=\textwidth]{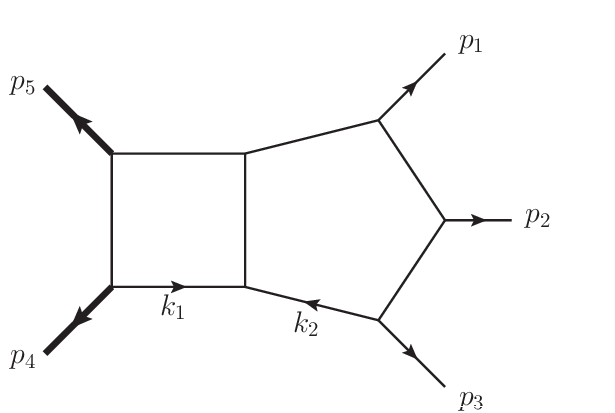}
\end{figure}
 \end{minipage}
\begin{minipage}{0.65\textwidth}
\begin{align}
\begin{split}
  & \mathcal{N}_{\rm PBzzz}^{(1)} = \eps^4 \, \sqrt{\Delta_5} \, (p_{4} + p_{5})^2 \, \mu_{12} \,, \\
  & \mathcal{N}_{\rm PBzzz}^{(2)} = \eps^4 \, \sqrt{\Delta_5} \, (p_{4} + p_{5})^2 \, \mu_{22} \,, \\
  & \mathcal{N}_{\rm PBzzz}^{(3)} = \eps^4 \, s_{45} \bigl[s_{12} s_{23} \, (k_2 - p_4)^2 - s_{12} s_{15} \, k_2^2 \\
  & \phantom{ \mathcal{N}_{\rm PBzzz}^{(3)} = } - s_{23} s_{34} \, (k_2 - p_4 - p_5)^2 \bigr] \,. \\
\end{split}
\end{align}
\end{minipage}
We took the definition of the forth MI from ref.~\cite{Jiang:2024eaj}.
Its normalisation involves the product of two square roots 
(of $\Delta_5$ and $\lambda(s_4,s_5,s_{45})$).
The expression is however lengthy and we thus omit it here.

\medskip

\subsubsection*{PBmmz, $\{1,1,1,1,1,1,0,1,0,0,0\}$, 4 MIs}

\hspace{\parindent}
\begin{minipage}{0.35\textwidth}
\begin{figure}[H]
\centering
\vspace{-0.4cm}
\includegraphics[width=\textwidth]{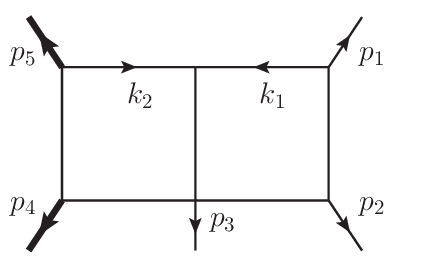}
\end{figure}
 \end{minipage}
\begin{minipage}{0.65\textwidth}
\begin{align}
\begin{split}
  & \mathcal{N}_{\rm PBmmz}^{(4)} = \eps^4 \, \sqrt{\Delta_5} \, \mu_{12} \,, \\
  & \mathcal{N}_{\rm PBmmz}^{(5)} = \eps^4 \, \sqrt{\lambda(s_4,s_5,s_{45})} \, s_{12} \, (k_2 - p_1)^2 \,, \\
  & \mathcal{N}_{\rm PBmmz}^{(6)} = \eps^4 \, s_{12} (s_{23} s_5 - s_{15} s_{45}) \,, \\ 
  & \mathcal{N}_{\rm PBmmz}^{(7)} = \eps^4 \, N_{\rm PBmmz}^{(7)}(X) \biggl[ (k_1 - p_5)^2 + R_{\rm PBmmz}^{(7,a)}(X) \, \mu_{12} \\
  & \phantom{ \mathcal{N}_{\rm PBmmz}^{(7)} = } +   R_{\rm PBmmz}^{(7,b)}(X) \biggr] + (\text{sub-sectors}) \,, \\
\end{split}
\end{align}
\end{minipage}
where $N_{\rm PBmmz}^{(7)}(X)$, $R_{\rm PBmmz}^{(7,a)}(X)$ and $R_{\rm PBmmz}^{(7,b)}(X)$ are rational functions of $X$.
To construct the fourth numerator, we started from $\eps^4 \, (k_1 - p_5)^2$, which leads to DEs linear in $\eps$, and added terms to achieve the factorisation of $\eps$ as described in \cref{sec:BasisConstruction}.

\medskip

\subsubsection*{PBmmz, $\{1,1,1,0,1,1,1,1,0,0,0\}$, 3 MIs}

\hspace{\parindent}
\begin{minipage}{0.35\textwidth}
\begin{figure}[H]
\centering
\includegraphics[width=\textwidth]{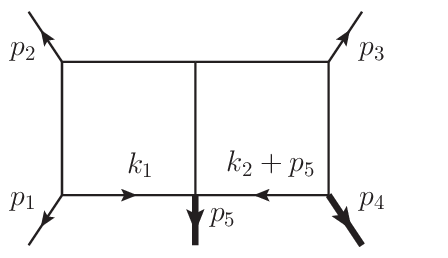}
\end{figure}
 \end{minipage}
\begin{minipage}{0.65\textwidth}
\begin{align}
\begin{split}
 & \mathcal{N}_{\rm PBmmz}^{(13)} = \eps^4 \, \sqrt{\Delta_5} \, \mu_{12} \,, \\
& \mathcal{N}_{\rm PBmmz}^{(14)} = \eps^4 \, s_{12} \left[s_{34} \, (k_1 - p_4 - p_5)^2 - s_4 \, (k_1 + p_1 + p_2)^2 \right] \,, \\
& \mathcal{N}_{\rm PBmmz}^{(15)} = \eps^4 \, s_{12} s_{23} s_{34} \,.
\end{split}
\end{align}
\end{minipage}

\medskip

\subsubsection*{PBmzm, $\{1,1,1,1,1,1,0,1,0,0,0\}$, 3 MIs}

\hspace{\parindent}
\begin{minipage}{0.35\textwidth}
\begin{figure}[H]
\centering
\includegraphics[width=\textwidth]{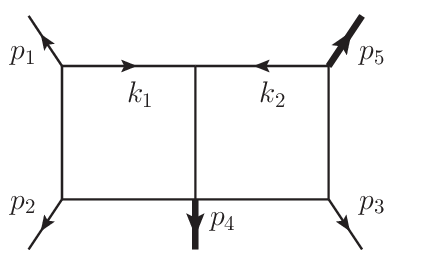}
\end{figure}
 \end{minipage}
\begin{minipage}{0.65\textwidth}
\begin{align}
\begin{split}
& \mathcal{N}_{\rm PBmzm}^{(4)} = \eps^4 \, \sqrt{\Delta_5} \, \mu_{12} \,, \\
& \mathcal{N}_{\rm PBmzm}^{(5)} = \eps^4 \, s_{12} (s_{34} + s_{45} - s_4 - s_{12}) \, (k_2 - p_1)^2 \,, \\
& \mathcal{N}_{\rm PBmzm}^{(6)} = \eps^4 \, s_{12} (s_4 s_5 - s_4 s_{15} - s_{12} s_{15} - s_5 s_{23} \\
& \phantom{ \mathcal{N}_{\rm PBmzm}^{(6)} = }  - s_5 s_{34} + s_{15} s_{34} + s_{15} s_{45}) \,.
\end{split}
\end{align}
\end{minipage}

\medskip

\subsubsection*{PBmzz, $\{1,1,1,0,1,1,1,1,0,0,0\}$, 3 MIs}

\hspace{\parindent}
\begin{minipage}{0.35\textwidth}
\begin{figure}[H]
\vspace{-0.5cm}
\centering
\includegraphics[width=\textwidth]{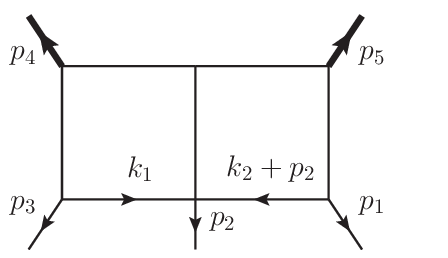}
\end{figure}
 \end{minipage}
\begin{minipage}{0.65\textwidth}
\begin{align}
\begin{split}
 & \mathcal{N}_{\rm PBmzz}^{(12)} = \eps^4 \, \sqrt{\Delta_5} \, \mu_{12} \,, \\
& \mathcal{N}_{\rm PBmzz}^{(13)} =  \eps^4 \, (s_{12} s_4 s_{15} + s_5 s_{23} s_{34} - s_{15} s_{34} s_{45} ) \,, \\
& \mathcal{N}_{\rm PBmzz}^{(14)} = \eps^4 \, N_{\rm PBmzz}^{(14)}(X) \biggl[ (k_1 - p_2)^2 + R_{\rm PBmzz}^{(14,a)}(X) \, \mu_{12}  \\
& \phantom{ \mathcal{N}_{\rm PBmzz}^{(14)} = } + R_{\rm PBmzz}^{(14,b)}(X) \biggr] + (\text{sub-sectors}) \,, \\
\end{split}
\end{align}
\end{minipage}
where $N_{\rm PBmzz}^{(14)}(X)$, $R_{\rm PBmzz}^{(14,a)}(X)$ and $R_{\rm PBmzz}^{(14,b)}(X)$ are rational functions of $X$.

\medskip

\subsubsection*{PBzmz, $\{1,1,1,1,1,1,0,1,0,0,0\}$, 3 MIs}

\hspace{\parindent}
\begin{minipage}{0.35\textwidth}
\begin{figure}[H]
\centering
\includegraphics[width=\textwidth]{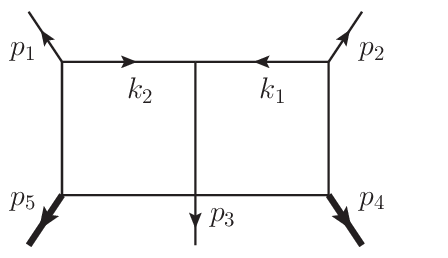}
\end{figure}
 \end{minipage}
\begin{minipage}{0.65\textwidth}
\begin{align}
\begin{split}
& \mathcal{N}_{\rm PBzmz}^{(4)} = \eps^4 \, \sqrt{\Delta_5} \, \mu_{12} \,, \\
& \mathcal{N}_{\rm PBzmz}^{(5)} =  \eps^4 \, (s_{23}+s_{34}-s_{15}) \left[ s_{15} \, (k_1 + p_1)^2 + (s_5 - 2 s_{15}) \, k_1^2 \right] \,, \\
& \mathcal{N}_{\rm PBzmz}^{(6)} = \eps^4 \, s_{12} s_{15} (s_{23} + s_{34} - s_4 - s_{15} ) \,.
\end{split}
\end{align}
\end{minipage}

\medskip

\subsubsection*{PBzmz, $\{1,1,1,0,1,1,1,1,0,0,0\}$, 3 MIs}

\hspace{\parindent}
\begin{minipage}{0.35\textwidth}
\begin{figure}[H]
\centering
\includegraphics[width=\textwidth]{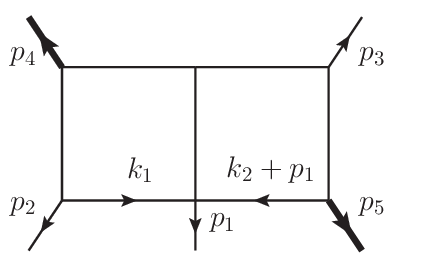}
\end{figure}
 \end{minipage}
\begin{minipage}{0.65\textwidth}
\begin{align}
\begin{split}
& \mathcal{N}_{\rm PBzmz}^{(11)} = \eps^4 \, \sqrt{\Delta_5} \, \mu_{12} \,, \\
& \mathcal{N}_{\rm PBzmz}^{(12)} =  \eps^4 \, (s_{15}-s_{23}-s_{34}) \bigl[ s_5 \, (k_1 + p_2 + p_4)^2  \\ 
& \phantom{ \mathcal{N}_{\rm PBzmz}^{(12)} = } - (s_4 + s_5 + s_{12} - s_{34} - s_{45}) (k_1 - p_1 - p_5)^2 \bigr] \,, \\
& \mathcal{N}_{\rm PBzmz}^{(13)} = \eps^4 \, (s_4 s_{15} - s_4 s_{34} - s_{15} s_{34} + s_{23} s_{34} + s_{34}^2) \\
& \phantom{ \mathcal{N}_{\rm PBzmz}^{(13)} =  } \times (s_4 + s_5 + s_{12} -
    s_{34} - s_{45}) \,. 
\end{split}
\end{align}
\end{minipage}

\medskip

\subsubsection*{PBmmz, $\{1, 1, 1, 0, 1, 1, 0, 1, 0, 0, 0\}$, 6 MIs}

\hspace{\parindent}
\begin{minipage}{0.35\textwidth}
\begin{figure}[H]
\vspace{-0.5cm}
\centering
\includegraphics[width=\textwidth]{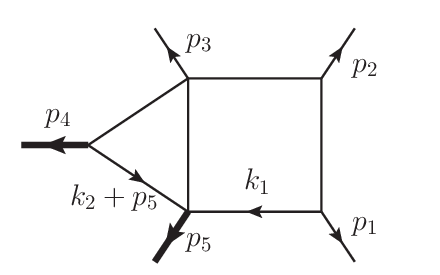}
\end{figure}
 \end{minipage}
\begin{minipage}{0.65\textwidth}
\begin{align}
\begin{split}
& \mathcal{N}_{\rm PBmmz}^{(18)} = \eps^3 \, \sqrt{\Delta_5} \, \dfrac{\mu_{12}}{(k_1+k_2)^2} \,, \\
& \mathcal{N}_{\rm PBmmz}^{(19)} = \eps^3 \, \sqrt{\Delta_5} \, \dfrac{\mu_{11}}{(k_1+k_2)^2} \,, \\
& \mathcal{N}_{\rm PBmmz}^{(20)} = \eps^4 \, s_{12} \sqrt{\lambda(s_{23},s_4,s_{15})} \,, \\
& \mathcal{N}_{\rm PBmmz}^{(21)} = \eps^4 \, s_{12} \left[ \dfrac{s_{15} s_4}{\eps \, (k_2+p_5)^2} - (s_4 + s_{15} - s_{23}) \right] \,, \\
& \mathcal{N}_{\rm PBmmz}^{(22)} = \eps^4 \, s_{12} \left[ \dfrac{s_{23} s_4}{\eps \, (k_2+p_4+p_5)^2} - (s_4 - s_{15} + s_{23}) \right] \,, \\
& \mathcal{N}_{\rm PBmmz}^{(23)} = \eps^4 \, s_{12} \left[ \dfrac{ s_4 (k_1 - p_5)^2}{\eps \, (k_2+p_5)} - (s_4 + s_{15} - s_{23}) \right] \,.
\end{split}
\end{align}
\end{minipage}

\medskip

\subsubsection*{PBmmz, $\{1, 1, 0, 0, 1, 1, 1, 1, 0, 0, 0\}$, 2 MIs}

\hspace{\parindent}
\begin{minipage}{0.35\textwidth}
\begin{figure}[H]
\centering
\includegraphics[width=\textwidth]{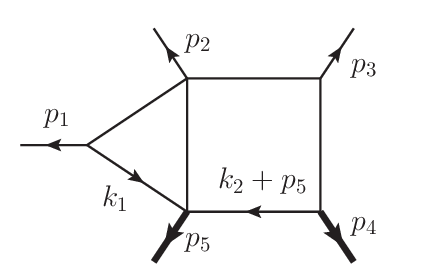}
\end{figure}
 \end{minipage}
\begin{minipage}{0.65\textwidth}
\begin{align}
\begin{split}
& \mathcal{N}_{\rm PBmmz}^{(25)} = \eps^3 \, \sqrt{\Delta_5} \, \dfrac{\mu_{22}}{(k_1+k_2)^2} \,, \\
& \mathcal{N}_{\rm PBmmz}^{(26)} = \eps^4 \, (s_{12} s_4 + s_{23} s_{34} - s_{34} s_{45}) \,.
\end{split}
\end{align}
\end{minipage}

\medskip

\subsubsection*{PBmmz, 30, $\{0, 1, 1, 1, 1, 1, 0, 1, 0, 0, 0\}$, 2 MIs}

\hspace{\parindent}
\begin{minipage}{0.35\textwidth}
\begin{figure}[H]
\centering
\includegraphics[width=\textwidth]{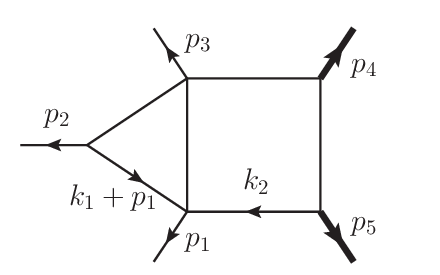}
\end{figure}
 \end{minipage}
\begin{minipage}{0.65\textwidth}
\begin{align}
\begin{split}
& \mathcal{N}_{\rm PBmmz}^{(30)} = \eps^3 \, \sqrt{\Delta_5} \, \dfrac{\mu_{22}}{(k_1+k_2)^2} \,, \\
& \mathcal{N}_{\rm PBmmz}^{(31)} = \eps^4 \, \sqrt{r_2^{(1)}} \,, 
\end{split}
\end{align}
\end{minipage}
where $r_2^{(1)}$ is defined in \cref{eq:r2}.

\medskip

\subsubsection*{PBmzm, $\{1, 1, 1, 0, 1, 1, 0, 1, 0, 0, 0\}$, 2 MIs}

\hspace{\parindent}
\begin{minipage}{0.35\textwidth}
\begin{figure}[H]
\centering
\includegraphics[width=\textwidth]{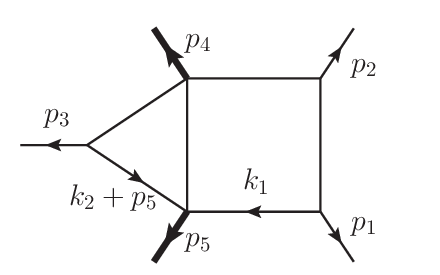}
\end{figure}
 \end{minipage}
\begin{minipage}{0.65\textwidth}
\begin{align}
\begin{split}
& \mathcal{N}_{\rm PBmzm}^{(18)} = \eps^3 \, \sqrt{\Delta_5} \, \dfrac{\mu_{11}}{(k_1+k_2)^2} \,, \\
& \mathcal{N}_{\rm PBmzm}^{(19)} = \eps^4 \, s_{12} (s_{23}+s_{34}-s_4) \,.
\end{split}
\end{align}
\end{minipage}

\medskip

\subsubsection*{PBmzm, $\{1, 1, 0, 0, 1, 1, 1, 1, 0, 0, 0\}$, 2 MIs}

\hspace{\parindent}
\begin{minipage}{0.35\textwidth}
\begin{figure}[H]
\centering
\includegraphics[width=\textwidth]{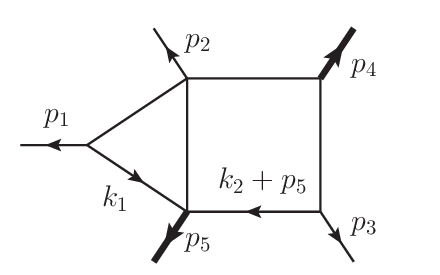}
\end{figure}
 \end{minipage}
\begin{minipage}{0.65\textwidth}
\begin{align}
\begin{split}
& \mathcal{N}_{\rm PBmzm}^{(22)} = \eps^3 \, \sqrt{\Delta_5} \, \dfrac{\mu_{22}}{(k_1+k_2)^2} \,, \\
& \mathcal{N}_{\rm PBmzm}^{(23)} = \eps^4 \, \left[ s_4 (s_5-s_{15})-s_{34} (s_5+s_{12}-s_{15}+s_{23}-s_{45}) \right] \,. 
\end{split}
\end{align}
\end{minipage}

\medskip

\subsubsection*{PBmzz, $\{1, 1, 1, 0, 1, 1, 0, 1, 0, 0, 0\}$, 2 MIs}

\hspace{\parindent}
\begin{minipage}{0.35\textwidth}
\begin{figure}[H]
\centering
\includegraphics[width=\textwidth]{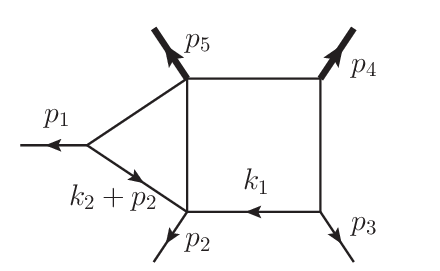}
\end{figure}
 \end{minipage}
\begin{minipage}{0.65\textwidth}
\begin{align}
\begin{split}
& \mathcal{N}_{\rm PBmzz}^{(17)} = \eps^3 \, \sqrt{\Delta_5} \, \dfrac{\mu_{11}}{(k_1+k_2)^2} \,, \\
& \mathcal{N}_{\rm PBmzz}^{(18)} = \eps^4 \, \left( s_4 s_{12}+s_{23} s_{34}-s_{34} s_{45} \right) \,. 
\end{split}
\end{align}
\end{minipage}

\medskip

\subsubsection*{PBzmz, $\{1, 1, 1, 0, 1, 1, 0, 1, 0, 0, 0\}$, 6 MIs}

\hspace{\parindent}
\begin{minipage}{0.35\textwidth}
\begin{figure}[H]
\vspace{-1cm}
\centering
\includegraphics[width=\textwidth]{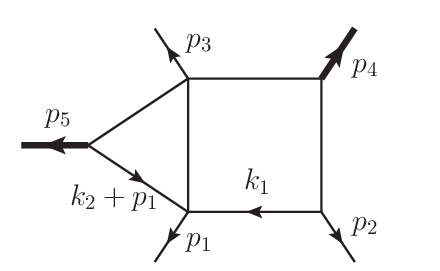}
\end{figure}
 \end{minipage}
\begin{minipage}{0.65\textwidth}
\begin{align}
\begin{split}
& \mathcal{N}_{\rm PBzmz}^{(16)} = \eps^3 \, \sqrt{\Delta_5} \, \dfrac{\mu_{12}}{(k_1+k_2)^2} \,, \\
& \mathcal{N}_{\rm PBzmz}^{(17)} = \eps^3 \, \sqrt{\Delta_5} \, \dfrac{\mu_{11}}{(k_1+k_2)^2} \,, \\
& \mathcal{N}_{\rm PBzmz}^{(18)} = \eps^4 \, \sqrt{r_3^{(3)}} \,, \\
& \mathcal{N}_{\rm PBzmz}^{(19)} =  \eps^4 \, \biggl[ \dfrac{s_5 s_{12} (s_{23}+s_{34}-s_4-s_{15})}{\eps \, (k_2+p_1)^2} \\
& \phantom{ \mathcal{N} }  + s_4 (s_{12}+s_{15}-s_{34}) + (s_{12}-s_{34}+s_5) (s_{15}-s_{23}-s_{34})  \biggr] \,, \\
& \mathcal{N}_{\rm PBzmz}^{(20)} = \eps^4 \, \biggl[ \dfrac{s_5 \bigl(s_4 (s_{15}-s_{34})+s_{34} (s_{23}+s_{34}-s_{15}) \bigr)}{ \eps \, (k_2+p_1+p_5)^2} \\
& \phantom{ \mathcal{N} } -s_4 (s_{12}+s_{15}-s_{34})  - (s_{12}-s_{34}-s_5) (s_{15}-s_{23}-s_{34}) \biggr] \,, \\
& \mathcal{N}_{\rm PBzmz}^{(21)} = \eps^4 \, \biggl[ \dfrac{s_5 (s_{23}+s_{34}-s_{15}) \, (k_1-p_1)^2}{\eps \, (k_2+p_1)^2} \\
& \phantom{ \mathcal{N} } + s_4 (s_{12} + s_{15} - s_{34}) + (s_{12} - s_{34} + s_5) (s_{15}-s_{23} - s_{34}) \biggr]  \,,
\end{split}
\end{align}
\end{minipage}
where $r_3^{(3)}$ is obtained by swapping $p_4 \leftrightarrow p_5$ and $p_1 \leftrightarrow p_2$ in $r_3^{(1)}$, defined in \cref{eq:r3}.

\medskip

\subsubsection*{PBmzz, $\{0,1,1,1,1,1,1,1,0,0,0\}$, 1 MI}

\hspace{\parindent}
\begin{minipage}{0.35\textwidth}
\begin{figure}[H]
\centering
\includegraphics[width=\textwidth]{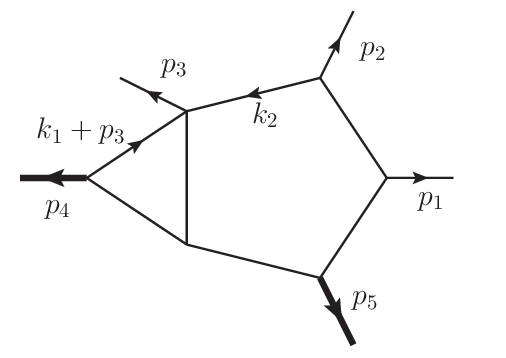}
\end{figure}
 \end{minipage}
\begin{minipage}{0.65\textwidth}
\begin{align}
\mathcal{N}_{\rm PBmzz}^{(15)} = \eps^4 \, \sqrt{\Delta_5} \, \mu_{22} \,.
\end{align}
\end{minipage}

\medskip

\subsubsection*{PBzmz, $\{0,1,1,1,1,1,1,1,0,0,0\}$, 1 MI}

\hspace{\parindent}
\begin{minipage}{0.35\textwidth}
\begin{figure}[H]
\centering
\includegraphics[width=\textwidth]{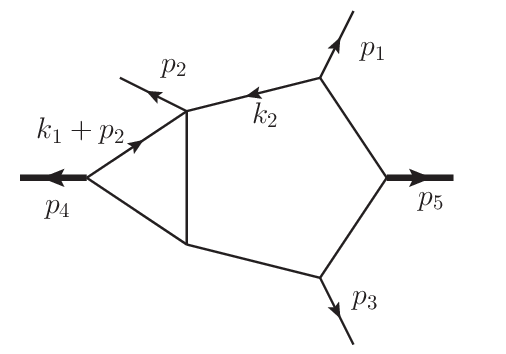}
\end{figure}
 \end{minipage}
\begin{minipage}{0.65\textwidth}
\begin{align}
\mathcal{N}_{\rm PBzmz}^{(14)} = \eps^4 \, \sqrt{\Delta_5} \, \mu_{22} \,.
\end{align}
\end{minipage}

\medskip

\subsubsection*{PBzzm, $\{0,1,1,1,1,1,1,1,0,0,0\}$, 1 MI}

\hspace{\parindent}
\begin{minipage}{0.35\textwidth}
\begin{figure}[H]
\centering
\includegraphics[width=\textwidth]{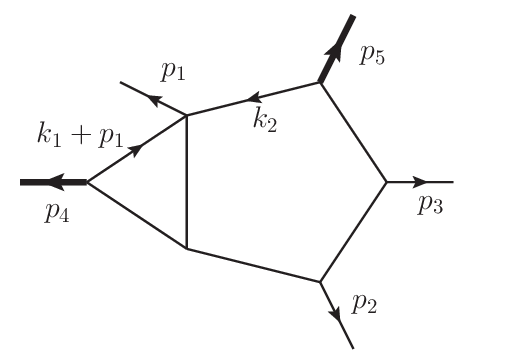}
\end{figure}
 \end{minipage}
\begin{minipage}{0.65\textwidth}
\begin{align}
\mathcal{N}_{\rm PBzzm}^{(15)} = \eps^4 \, \sqrt{\Delta_5} \, \mu_{22} \,.
\end{align}
\end{minipage}

\medskip

\subsubsection*{PBzzz, $\{1,1,0,1,1,1,1,1,0,0,0\}$, 1 MI}

\hspace{\parindent}
\begin{minipage}{0.35\textwidth}
\begin{figure}[H]
\centering
\includegraphics[width=\textwidth]{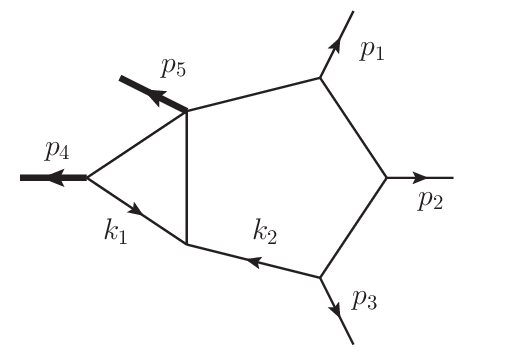}
\end{figure}
 \end{minipage}
\begin{minipage}{0.65\textwidth}
\begin{align}
\begin{split}
& \mathcal{N}_{\rm PBzzz}^{(20)} = \eps^4 \, \sqrt{\Delta_5} \, \frac{s_{12} s_{23} (s_4 s_{12} s_{15} + s_5 s_{23} s_{34} - s_{15} s_{34} s_{45})}{\Delta_5} \\ 
& \phantom{ \mathcal{N}_{\rm PBzzz}^{(20)} = } + (\text{sub-sectors}) \,.
\end{split}
\end{align}
\end{minipage}

\medskip

\subsubsection*{PBmmz, $\{0,1,0,1,1,1,1,1,0,0,0\}$, 2 MIs}

\hspace{\parindent}
\begin{minipage}{0.35\textwidth}
\begin{figure}[H]
\centering
\includegraphics[width=\textwidth]{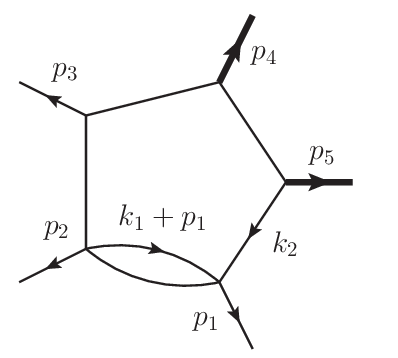}
\end{figure}
 \end{minipage}
\begin{minipage}{0.65\textwidth}
\begin{align}
\begin{split}
& \mathcal{N}_{\rm PBmmz}^{(32)} = \eps^3 \, \sqrt{\Delta_5} \dfrac{\mu_{22}}{(k_1 + k_2)^2} \,, \\
& \mathcal{N}_{\rm PBmmz}^{(33)} = \eps^3 (1 - 2 \eps) (s_{34} s_{45} - s_4 s_{12}) \,. 
\end{split}
\end{align}
\end{minipage}

\medskip

\subsubsection*{PBmzm, $\{0,1,0,1,1,1,1,1,0,0,0\}$, 2 MIs}

\hspace{\parindent}
\begin{minipage}{0.35\textwidth}
\begin{figure}[H]
\centering
\includegraphics[width=\textwidth]{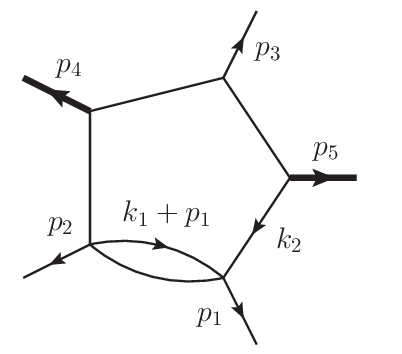}
\end{figure}
 \end{minipage}
\begin{minipage}{0.65\textwidth}
\begin{align}
\begin{split}
& \mathcal{N}_{\rm PBmzm}^{(29)} = \eps^3 \, \sqrt{\Delta_5} \dfrac{\mu_{22}}{(k_1 + k_2)^2} \,, \\
& \mathcal{N}_{\rm PBmzm}^{(30)} = \eps^3 (1 - 2 \eps) (s_4 s_5 - s_4 s_{34} - s_5 s_{34} - s_{12} s_{34} \\
& \phantom{ \mathcal{N}_{\rm PBmzm}^{(30)} = } + s_{34}^2 + s_{34} s_{45}) \,. 
\end{split}
\end{align}
\end{minipage}

\medskip

\subsubsection*{PBmzz, $\{0,1,0,1,1,1,1,1,0,0,0\}$, 2 MIs}

\hspace{\parindent}
\begin{minipage}{0.35\textwidth}
\begin{figure}[H]
\centering
\includegraphics[width=\textwidth]{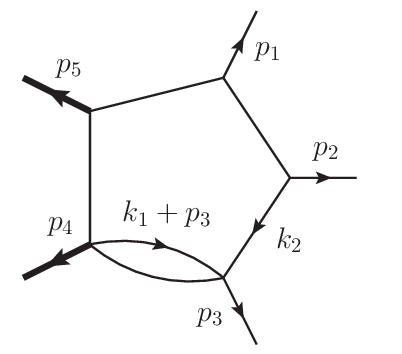}
\end{figure}
 \end{minipage}
\begin{minipage}{0.65\textwidth}
\begin{align}
\begin{split}
& \mathcal{N}_{\rm PBmzz}^{(36)} = \eps^3 \, \sqrt{\Delta_5} \dfrac{\mu_{22}}{(k_1 + k_2)^2} \,, \\
& \mathcal{N}_{\rm PBmzz}^{(37)} = \eps^3 (1 - 2 \eps) \, s_{12} \, s_{15} \,. 
\end{split}
\end{align}
\end{minipage}

\medskip

\subsubsection*{PBzmz, $\{0,1,0,1,1,1,1,1,0,0,0\}$, 2 MIs}

\hspace{\parindent}
\begin{minipage}{0.35\textwidth}
\begin{figure}[H]
\centering
\includegraphics[width=\textwidth]{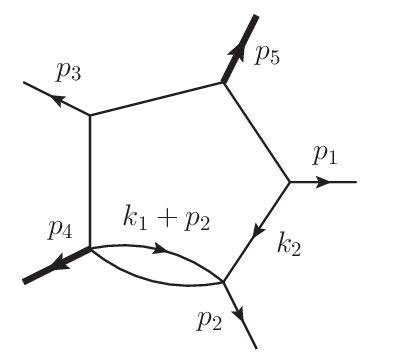}
\end{figure}
 \end{minipage}
\begin{minipage}{0.65\textwidth}
\begin{align}
\begin{split}
& \mathcal{N}_{\rm PBzmz}^{(38)} = \eps^3 \, \sqrt{\Delta_5} \dfrac{\mu_{22}}{(k_1 + k_2)^2} \,, \\
& \mathcal{N}_{\rm PBzmz}^{(39)} = \eps^3 (1 - 2 \eps) (s_4 s_{15} - s_4 s_5 + s_{12} s_{15} \\
& \phantom{ \mathcal{N}_{\rm PBzmz}^{(39)} = } + s_5 s_{23} + s_5 s_{34} - s_{15} s_{34} - 
 s_{15} s_{45}) \,. 
\end{split}
\end{align}
\end{minipage}

\medskip

\subsubsection*{PBzzm, $\{0,1,0,1,1,1,1,1,0,0,0\}$, 2 MIs}

\hspace{\parindent}
\begin{minipage}{0.35\textwidth}
\begin{figure}[H]
\centering
\includegraphics[width=\textwidth]{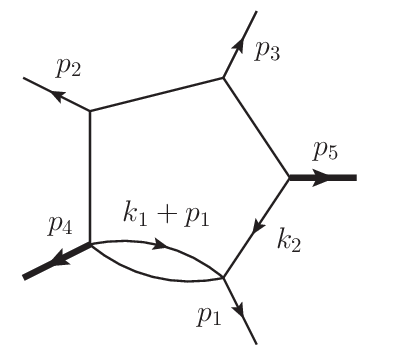}
\end{figure}
 \end{minipage}
\begin{minipage}{0.65\textwidth}
\begin{align}
\begin{split}
& \mathcal{N}_{\rm PBzzm}^{(36)} = \eps^3 \, \sqrt{\Delta_5} \dfrac{\mu_{22}}{(k_1 + k_2)^2} \,, \\
& \mathcal{N}_{\rm PBzzm}^{(37)} = \eps^3 (1 - 2 \eps) \, s_{23} (s_4 + s_5 + s_{12} - s_{34} - s_{45}) \,. 
\end{split}
\end{align}
\end{minipage}

\medskip

\subsubsection*{PBzzz, $\{0,1,0,1,1,1,1,1,0,0,0\}$, 2 MIs}

\hspace{\parindent}
\begin{minipage}{0.35\textwidth}
\begin{figure}[H]
\centering
\includegraphics[width=\textwidth]{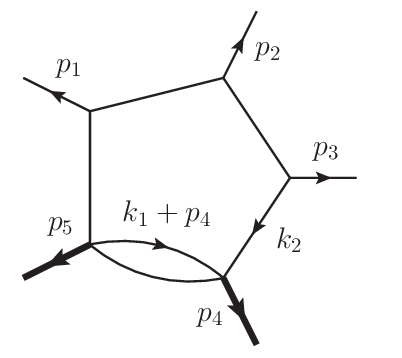}
\end{figure}
 \end{minipage}
\begin{minipage}{0.65\textwidth}
\begin{align}
\begin{split}
& \mathcal{N}_{\rm PBzzz}^{(49)} = \eps^3 \, \sqrt{\Delta_5} \dfrac{\mu_{22}}{(k_1 + k_2)^2} \,, \\
& \mathcal{N}_{\rm PBzzz}^{(50)} = \eps^3 (1 - 2 \eps) \, s_{12} \, s_{23} \,. 
\end{split}
\end{align}
\end{minipage}

\section{Kinematic Regions}
\label{sec:app_kin}

In \cref{sec:IC} we introduced two regions of phase space
where we obtained numerical evaluations of our integrals, namely
the Euclidean region and the $s_{12}$-channel. In section \cref{sec:AMPLpos} we further introduced the amplituhedron region $\mathbb{A}$. In this appendix
we define these regions explicitly.

Let us first discuss the Euclidean region.
Given the non-cyclic indexing of the external legs in the
integral families in \cref{fig:pentagons,fig:pentaboxes}, the Euclidean
region associated with each one of those families is not necessarily the same. 
We define our Euclidean region as the intersection of the Euclidean
regions associated with any permutation of the representative families given
in \cref{fig:pentagons,fig:pentaboxes}, i.e.,
\begin{equation}
	s_{ij}<0 \,, \qquad s_4 < 0 \,, \qquad s_5 < 0 \,, \label{eq:s-Eucl}
\end{equation}
for $i\neq j = 1,\ldots,5$.
Note that this region is not guaranteed to be non-empty, 
but it is in our case and the point in \cref{eq:euclP} lies within it.

The second kinematic region we consider is what we call the `$s_{12}$-channel', 
corresponding to the process where legs $p_1$ and $p_2$ are in the initial state, 
and legs $p_3$, $p_4$ and $p_5$ are in the final state, that is
$-p_1-p_2\to p_3+p_4+p_5$. For instance, such a process could
describe the production of two vector bosons in association with
a jet at an hadron collider, and as such is of phenomenological
interest. This channel is defined by the following set of constraints:
\begin{align}
\begin{gathered}\label{eq:s12Channel}
p_1\cdot p_2 > 0  \,, \quad p_3\cdot p_4 > 0 \,, \quad p_4\cdot p_5 > 0 \,, 
\quad p_3\cdot p_5> 0 \,, \\
p_1\cdot p_3 < 0  \,, \quad p_1\cdot p_4 < 0 \,, \quad p_1\cdot p_5 < 0 \,, 
\quad
p_2\cdot p_3 < 0  \,, \quad p_2\cdot p_4 < 0 \,, \quad p_2\cdot p_5 < 0 \,, \\
G(p_4) = 2 \, s_4 > 0 \,, \quad G(p_5) = 2 \, s_5> 0   \,, \\
G(p_i,p_j)  < 0 \,, \quad G(p_i,p_j,p_k)  > 0 \,, 
\quad G(p_1,p_2,p_3,p_4)  < 0 \,,
\end{gathered}
\end{align}
where $i$, $j$ and $k$ take distinct values in $\{1,2,3,4,5\}$.
The point $X_1$ in \cref{eq:X1} lies within the kinematic region
defined by these constraints, and is furthermore `generic', in the sense
that none of the letters of the alphabet for
integrals for five-point two-loop integrals with two external masses
vanish or diverge there.
Having in mind the construction of pentagon functions for this
set of integrals, we choose another point within the $s_{12}$-channel that also
has the symmetries of the region. In this case, this means that 
it is invariant under the exchanges $1\leftrightarrow2$
and $4\leftrightarrow5$. The point $X_0$ in \cref{eq:X0} satisfies these
conditions.

Finally, let us explain the origin of the inequalities in eqs.~\p{eq:12AB} and \p{eq:13AB} which specify the amplituhedron region.
We represent the Lagrangian coordinates $x_{0}$, $x_{0'´}$ by bi-twistors $Z_A Z_B$ and $Z_C Z_D$, and choose $x_{0'} \to \infty$ by taking $Z_C Z_D$ to be the infinity bi-twistor. The momentum twistors $Z_1,\ldots,Z_4$ represent the quadrilateral light-like contour. 
The two-loop four-particle MHV amplituhedron is specified by the following inequalities for the four-brackets of momentum twistors~\cite{Arkani-Hamed:2017vfh}, 
\begin{align}
\vev{AB12}, \vev{AB23}, \vev{AB34}, \vev{AB14},\vev{CD12}, \vev{CD23}, \vev{CD34}, \vev{CD14}, \vev{ABCD} >0 , \label{eq:amplh1} \\
\vev{CD13},\vev{CD24} < 0 \,, \label{eq:amplh2} \\
\vev{AB13}, \vev{AB24}<0 \,. \label{eq:amplh3}
\end{align}
The inequalities \p{eq:amplh1} and \p{eq:amplh2} are equivalent to those in eq.~\p{eq:12AB}
when written in dual momenta variables, and as such they impose the condition of being in the anti-Euclidean region.
As already noted below \cref{eq:GLsign}, working in the anti-Euclidean region instead of the
Euclidean region is purely conventional, and the same non-trivial positivity conclusions 
hold in the Euclidean region.
The inequalities \p{eq:amplh3} imply that 
\begin{align}
x_{13}^2 (x_{20}^2 + x_{40}^2) + x_{24}^2 (x_{10}^2+x_{30}^2) - x_{13}^2 x_{24}^2 \pm \epsilon_5 > 0
\end{align} 
where the pseudo-scalar invariant $\eps_5 \coloneqq {\rm tr}( \slashed{x}_{12} \slashed{x}_{23} \slashed{x}_{34} \slashed{x}_{10} \gamma_5)$ is related to the 
$\bar\Delta_5$ defined in \cref{eq:barDelta5} through $(\eps_5)^2 = \bar\Delta_5$. Thus, the space-time coordinates are complex valued in such a way that the Mandelstam invariants and $\eps_5$ are real-valued. The constraints in \cref{eq:13AB} then follow.

\bibliography{main.bib}

\providecommand{\href}[2]{#2}\begingroup\raggedright\begin{thebibliography}{100}

\bibitem{Bourjaily:2022bwx}
J.L.~Bourjaily et~al., \emph{{Functions Beyond Multiple Polylogarithms for
  Precision Collider Physics}},  in \emph{{Snowmass 2021}}, 3, 2022
  [\href{https://arxiv.org/abs/2203.07088}{{\ttfamily 2203.07088}}].

\bibitem{Chen:1977oja}
K.-T.~Chen, \emph{{Iterated path integrals}},
  \href{https://doi.org/10.1090/S0002-9904-1977-14320-6}{\emph{Bull. Am. Math.
  Soc.} {\bfseries 83} (1977) 831}.

\bibitem{Arkani-Hamed:2010pyv}
N.~Arkani-Hamed, J.L.~Bourjaily, F.~Cachazo and J.~Trnka, \emph{{Local
  Integrals for Planar Scattering Amplitudes}},
  \href{https://doi.org/10.1007/JHEP06(2012)125}{\emph{JHEP} {\bfseries 06}
  (2012) 125} [\href{https://arxiv.org/abs/1012.6032}{{\ttfamily 1012.6032}}].

\bibitem{Henn:2013pwa}
J.M.~Henn, \emph{{Multiloop integrals in dimensional regularization made
  simple}}, \href{https://doi.org/10.1103/PhysRevLett.110.251601}{\emph{Phys.
  Rev. Lett.} {\bfseries 110} (2013) 251601}
  [\href{https://arxiv.org/abs/1304.1806}{{\ttfamily 1304.1806}}].

\bibitem{Abreu:2022mfk}
S.~Abreu, R.~Britto and C.~Duhr, \emph{{The SAGEX review on scattering
  amplitudes Chapter 3: Mathematical structures in Feynman integrals}},
  \href{https://doi.org/10.1088/1751-8121/ac87de}{\emph{J. Phys. A} {\bfseries
  55} (2022) 443004} [\href{https://arxiv.org/abs/2203.13014}{{\ttfamily
  2203.13014}}].

\bibitem{Abreu:2017mtm}
S.~Abreu, R.~Britto, C.~Duhr and E.~Gardi, \emph{{Diagrammatic Hopf algebra of
  cut Feynman integrals: the one-loop case}},
  \href{https://doi.org/10.1007/JHEP12(2017)090}{\emph{JHEP} {\bfseries 12}
  (2017) 090} [\href{https://arxiv.org/abs/1704.07931}{{\ttfamily
  1704.07931}}].

\bibitem{Arkani-Hamed:2017ahv}
N.~Arkani-Hamed and E.Y.~Yuan, \emph{{One-Loop Integrals from Spherical
  Projections of Planes and Quadrics}},
  \href{https://arxiv.org/abs/1712.09991}{{\ttfamily 1712.09991}}.

\bibitem{Chen:2023kgw}
J.~Chen, B.~Feng and L.L.~Yang, \emph{Intersection theory rules symbology},
  \href{https://doi.org/10.1007/s11433-023-2239-8}{\emph{Sci. China Phys. Mech.
  Astron.} {\bfseries 67} (2024) 221011}
  [\href{https://arxiv.org/abs/2305.01283}{{\ttfamily 2305.01283}}].

\bibitem{Fevola:2023kaw}
C.~Fevola, S.~Mizera and S.~Telen, \emph{{Landau Singularities Revisited:
  Computational Algebraic Geometry for Feynman Integrals}},
  \href{https://doi.org/10.1103/PhysRevLett.132.101601}{\emph{Phys. Rev. Lett.}
  {\bfseries 132} (2024) 101601}
  [\href{https://arxiv.org/abs/2311.14669}{{\ttfamily 2311.14669}}].

\bibitem{Fevola:2023fzn}
C.~Fevola, S.~Mizera and S.~Telen, \emph{Principal {{Landau Determinants}}},
  \href{https://doi.org/10.1016/j.cpc.2024.109278}{\emph{Comput. Phys. Commun.}
  {\bfseries 303} (2024) 109278}
  [\href{https://arxiv.org/abs/2311.16219}{{\ttfamily 2311.16219}}].

\bibitem{Jiang:2023qnl}
X.~Jiang and L.L.~Yang, \emph{The recursive structure of {{Baikov}}
  representations {{I}}: Generics and application to symbology},
  \href{https://doi.org/10.1103/PhysRevD.108.076004}{\emph{Phys. Rev. D}
  {\bfseries 108} (2023) 076004}
  [\href{https://arxiv.org/abs/2303.11657}{{\ttfamily 2303.11657}}].

\bibitem{Dlapa:2023cvx}
C.~Dlapa, M.~Helmer, G.~Papathanasiou and F.~Tellander, \emph{{Symbol alphabets
  from the Landau singular locus}},
  \href{https://doi.org/10.1007/JHEP10(2023)161}{\emph{JHEP} {\bfseries 10}
  (2023) 161} [\href{https://arxiv.org/abs/2304.02629}{{\ttfamily
  2304.02629}}].

\bibitem{He:2023umf}
S.~He, X.~Jiang, J.~Liu and Q.~Yang, \emph{{On symbology and differential
  equations of Feynman integrals from Schubert analysis}},
  \href{https://doi.org/10.1007/JHEP12(2023)140}{\emph{JHEP} {\bfseries 12}
  (2023) 140} [\href{https://arxiv.org/abs/2309.16441}{{\ttfamily
  2309.16441}}].

\bibitem{Jiang:2024eaj}
X.~Jiang, J.~Liu, X.~Xu and L.L.~Yang, \emph{{Symbol letters of Feynman
  integrals from Gram determinants}},
  \href{https://arxiv.org/abs/2401.07632}{{\ttfamily 2401.07632}}.

\bibitem{Helmer:2024wax}
M.~Helmer, G.~Papathanasiou and F.~Tellander, \emph{Landau {{Singularities}}
  from {{Whitney Stratifications}}},
  \href{https://arxiv.org/abs/2402.14787}{{\ttfamily 2402.14787}}.

\bibitem{Caron-Huot:2024brh}
S.~Caron-Huot, M.~Correia and M.~Giroux, \emph{{Recursive Landau Analysis}},
  \href{https://arxiv.org/abs/2406.05241}{{\ttfamily 2406.05241}}.

\bibitem{Hannesdottir:2021kpd}
H.S.~Hannesdottir, A.J.~McLeod, M.D.~Schwartz and C.~Vergu, \emph{{Implications
  of the Landau equations for iterated integrals}},
  \href{https://doi.org/10.1103/PhysRevD.105.L061701}{\emph{Phys. Rev. D}
  {\bfseries 105} (2022) L061701}
  [\href{https://arxiv.org/abs/2109.09744}{{\ttfamily 2109.09744}}].

\bibitem{Hannesdottir:2024cnn}
H.S.~Hannesdottir, L.~Lippstreu, A.J.~McLeod and M.~Polackova, \emph{{Minimal
  Cuts and Genealogical Constraints on Feynman Integrals}},
  \href{https://arxiv.org/abs/2406.05943}{{\ttfamily 2406.05943}}.

\bibitem{Lee:2014ioa}
R.N.~Lee, \emph{{Reducing differential equations for multiloop master
  integrals}}, \href{https://doi.org/10.1007/JHEP04(2015)108}{\emph{JHEP}
  {\bfseries 04} (2015) 108} [\href{https://arxiv.org/abs/1411.0911}{{\ttfamily
  1411.0911}}].

\bibitem{Meyer:2017joq}
C.~Meyer, \emph{Algorithmic transformation of multi-loop master integrals to a
  canonical basis with {{CANONICA}}},
  \href{https://doi.org/10.1016/j.cpc.2017.09.014}{\emph{Comput. Phys. Commun.}
  {\bfseries 222} (2018) 295}
  [\href{https://arxiv.org/abs/1705.06252}{{\ttfamily 1705.06252}}].

\bibitem{Prausa:2017ltv}
M.~Prausa, \emph{{epsilon: A tool to find a canonical basis of master
  integrals}}, \href{https://doi.org/10.1016/j.cpc.2017.05.026}{\emph{Comput.
  Phys. Commun.} {\bfseries 219} (2017) 361}
  [\href{https://arxiv.org/abs/1701.00725}{{\ttfamily 1701.00725}}].

\bibitem{Gituliar:2017vzm}
O.~Gituliar and V.~Magerya, \emph{{Fuchsia: a tool for reducing differential
  equations for Feynman master integrals to epsilon form}},
  \href{https://doi.org/10.1016/j.cpc.2017.05.004}{\emph{Comput. Phys. Commun.}
  {\bfseries 219} (2017) 329}
  [\href{https://arxiv.org/abs/1701.04269}{{\ttfamily 1701.04269}}].

\bibitem{Lee:2020zfb}
R.N.~Lee, \emph{{Libra: A package for transformation of differential systems
  for multiloop integrals}},
  \href{https://doi.org/10.1016/j.cpc.2021.108058}{\emph{Comput. Phys. Commun.}
  {\bfseries 267} (2021) 108058}
  [\href{https://arxiv.org/abs/2012.00279}{{\ttfamily 2012.00279}}].

\bibitem{Henn:2020lye}
J.~Henn, B.~Mistlberger, V.A.~Smirnov and P.~Wasser, \emph{{Constructing d-log
  integrands and computing master integrals for three-loop four-particle
  scattering}}, \href{https://doi.org/10.1007/JHEP04(2020)167}{\emph{JHEP}
  {\bfseries 04} (2020) 167}
  [\href{https://arxiv.org/abs/2002.09492}{{\ttfamily 2002.09492}}].

\bibitem{Chen:2020uyk}
J.~Chen, X.~Jiang, X.~Xu and L.L.~Yang, \emph{Constructing canonical
  {{Feynman}} integrals with intersection theory},
  \href{https://doi.org/10.1016/j.physletb.2021.136085}{\emph{arXiv [hep-th]}
  {\bfseries 814} (2021) 136085}
  [\href{https://arxiv.org/abs/2008.03045}{{\ttfamily 2008.03045}}].

\bibitem{Chen:2022lzr}
J.~Chen, X.~Jiang, C.~Ma, X.~Xu and L.L.~Yang, \emph{Baikov representations,
  intersection theory, and canonical {{Feynman}} integrals},
  \href{https://doi.org/10.1007/JHEP07(2022)066}{\emph{arXiv [hep-th]}
  {\bfseries 07} (2022) 066}
  [\href{https://arxiv.org/abs/2202.08127}{{\ttfamily 2202.08127}}].

\bibitem{Dlapa:2020cwj}
C.~Dlapa, J.~Henn and K.~Yan, \emph{{Deriving canonical differential equations
  for Feynman integrals from a single uniform weight integral}},
  \href{https://doi.org/10.1007/JHEP05(2020)025}{\emph{JHEP} {\bfseries 05}
  (2020) 025} [\href{https://arxiv.org/abs/2002.02340}{{\ttfamily
  2002.02340}}].

\bibitem{Dlapa:2021qsl}
C.~Dlapa, X.~Li and Y.~Zhang, \emph{{Leading singularities in Baikov
  representation and Feynman integrals with uniform transcendental weight}},
  \href{https://doi.org/10.1007/JHEP07(2021)227}{\emph{JHEP} {\bfseries 07}
  (2021) 227} [\href{https://arxiv.org/abs/2103.04638}{{\ttfamily
  2103.04638}}].

\bibitem{Abreu:2020jxa}
S.~Abreu, H.~Ita, F.~Moriello, B.~Page, W.~Tschernow and M.~Zeng,
  \emph{{Two-Loop Integrals for Planar Five-Point One-Mass Processes}},
  \href{https://doi.org/10.1007/JHEP11(2020)117}{\emph{JHEP} {\bfseries 11}
  (2020) 117} [\href{https://arxiv.org/abs/2005.04195}{{\ttfamily
  2005.04195}}].

\bibitem{Abreu:2021smk}
S.~Abreu, H.~Ita, B.~Page and W.~Tschernow, \emph{{Two-loop hexa-box integrals
  for non-planar five-point one-mass processes}},
  \href{https://doi.org/10.1007/JHEP03(2022)182}{\emph{JHEP} {\bfseries 03}
  (2022) 182} [\href{https://arxiv.org/abs/2107.14180}{{\ttfamily
  2107.14180}}].

\bibitem{Abreu:2023rco}
S.~Abreu, D.~Chicherin, H.~Ita, B.~Page, V.~Sotnikov, W.~Tschernow et~al.,
  \emph{{All Two-Loop Feynman Integrals for Five-Point One-Mass Scattering}},
  \href{https://doi.org/10.1103/PhysRevLett.132.141601}{\emph{Phys. Rev. Lett.}
  {\bfseries 132} (2024) 141601}
  [\href{https://arxiv.org/abs/2306.15431}{{\ttfamily 2306.15431}}].

\bibitem{Henn:2024ngj}
J.M.~Henn, A.~Matija\v{s}i\'c, J.~Miczajka, T.~Peraro, Y.~Xu and Y.~Zhang,
  \emph{{A computation of two-loop six-point Feynman integrals in dimensional
  regularization}},  \href{https://arxiv.org/abs/2403.19742}{{\ttfamily
  2403.19742}}.

\bibitem{FebresCordero:2023pww}
F.~Febres~Cordero, G.~Figueiredo, M.~Kraus, B.~Page and L.~Reina,
  \emph{{Two-loop master integrals for leading-color $ pp\to t\overline{t}H $
  amplitudes with a light-quark loop}},
  \href{https://doi.org/10.1007/JHEP07(2024)084}{\emph{JHEP} {\bfseries 07}
  (2024) 084} [\href{https://arxiv.org/abs/2312.08131}{{\ttfamily
  2312.08131}}].

\bibitem{Badger:2022hno}
S.~Badger, M.~Becchetti, E.~Chaubey and R.~Marzucca, \emph{{Two-loop master
  integrals for a planar topology contributing to pp \textrightarrow{}$
  t\overline{t}j $}},
  \href{https://doi.org/10.1007/JHEP01(2023)156}{\emph{JHEP} {\bfseries 01}
  (2023) 156} [\href{https://arxiv.org/abs/2210.17477}{{\ttfamily
  2210.17477}}].

\bibitem{Badger:2024fgb}
S.~Badger, M.~Becchetti, N.~Giraudo and S.~Zoia, \emph{{Two-loop integrals for
  $ t\overline{t} $+jet production at hadron colliders in the leading colour
  approximation}}, \href{https://doi.org/10.1007/JHEP07(2024)073}{\emph{JHEP}
  {\bfseries 07} (2024) 073}
  [\href{https://arxiv.org/abs/2404.12325}{{\ttfamily 2404.12325}}].

\bibitem{Chicherin:2020oor}
D.~Chicherin and V.~Sotnikov, \emph{{Pentagon Functions for Scattering of Five
  Massless Particles}},
  \href{https://doi.org/10.1007/JHEP12(2020)167}{\emph{JHEP} {\bfseries 20}
  (2020) 167} [\href{https://arxiv.org/abs/2009.07803}{{\ttfamily
  2009.07803}}].

\bibitem{Chicherin:2021dyp}
D.~Chicherin, V.~Sotnikov and S.~Zoia, \emph{{Pentagon functions for one-mass
  planar scattering amplitudes}},
  \href{https://doi.org/10.1007/JHEP01(2022)096}{\emph{JHEP} {\bfseries 01}
  (2022) 096} [\href{https://arxiv.org/abs/2110.10111}{{\ttfamily
  2110.10111}}].

\bibitem{vonManteuffel:2014ixa}
A.~von Manteuffel and R.M.~Schabinger, \emph{{A novel approach to integration
  by parts reduction}},
  \href{https://doi.org/10.1016/j.physletb.2015.03.029}{\emph{Phys. Lett. B}
  {\bfseries 744} (2015) 101}
  [\href{https://arxiv.org/abs/1406.4513}{{\ttfamily 1406.4513}}].

\bibitem{Peraro:2016wsq}
T.~Peraro, \emph{{Scattering amplitudes over finite fields and multivariate
  functional reconstruction}},
  \href{https://doi.org/10.1007/JHEP12(2016)030}{\emph{JHEP} {\bfseries 12}
  (2016) 030} [\href{https://arxiv.org/abs/1608.01902}{{\ttfamily
  1608.01902}}].

\bibitem{Peraro:2019svx}
T.~Peraro, \emph{{FiniteFlow: multivariate functional reconstruction using
  finite fields and dataflow graphs}},
  \href{https://doi.org/10.1007/JHEP07(2019)031}{\emph{JHEP} {\bfseries 07}
  (2019) 031} [\href{https://arxiv.org/abs/1905.08019}{{\ttfamily
  1905.08019}}].

\bibitem{Alday:2007hr}
L.F.~Alday and J.M.~Maldacena, \emph{{Gluon scattering amplitudes at strong
  coupling}}, \href{https://doi.org/10.1088/1126-6708/2007/06/064}{\emph{JHEP}
  {\bfseries 06} (2007) 064} [\href{https://arxiv.org/abs/0705.0303}{{\ttfamily
  0705.0303}}].

\bibitem{Drummond:2007aua}
J.M.~Drummond, G.P.~Korchemsky and E.~Sokatchev, \emph{{Conformal properties of
  four-gluon planar amplitudes and Wilson loops}},
  \href{https://doi.org/10.1016/j.nuclphysb.2007.11.041}{\emph{Nucl. Phys. B}
  {\bfseries 795} (2008) 385}
  [\href{https://arxiv.org/abs/0707.0243}{{\ttfamily 0707.0243}}].

\bibitem{Brandhuber:2007yx}
A.~Brandhuber, P.~Heslop and G.~Travaglini, \emph{{MHV amplitudes in N=4 super
  Yang-Mills and Wilson loops}},
  \href{https://doi.org/10.1016/j.nuclphysb.2007.11.002}{\emph{Nucl. Phys. B}
  {\bfseries 794} (2008) 231}
  [\href{https://arxiv.org/abs/0707.1153}{{\ttfamily 0707.1153}}].

\bibitem{Arkani-Hamed:2012zlh}
N.~Arkani-Hamed, J.L.~Bourjaily, F.~Cachazo, A.B.~Goncharov, A.~Postnikov and
  J.~Trnka, \emph{{Grassmannian Geometry of Scattering Amplitudes}}, Cambridge
  University Press (4, 2016),
  \href{https://doi.org/10.1017/CBO9781316091548}{10.1017/CBO9781316091548},
  [\href{https://arxiv.org/abs/1212.5605}{{\ttfamily 1212.5605}}].

\bibitem{Alday:2011ga}
L.F.~Alday, E.I.~Buchbinder and A.A.~Tseytlin, \emph{{Correlation function of
  null polygonal Wilson loops with local operators}},
  \href{https://doi.org/10.1007/JHEP09(2011)034}{\emph{JHEP} {\bfseries 09}
  (2011) 034} [\href{https://arxiv.org/abs/1107.5702}{{\ttfamily 1107.5702}}].

\bibitem{Alday:2012hy}
L.F.~Alday, P.~Heslop and J.~Sikorowski, \emph{{Perturbative correlation
  functions of null Wilson loops and local operators}},
  \href{https://doi.org/10.1007/JHEP03(2013)074}{\emph{JHEP} {\bfseries 03}
  (2013) 074} [\href{https://arxiv.org/abs/1207.4316}{{\ttfamily 1207.4316}}].

\bibitem{Alday:2013ip}
L.F.~Alday, J.M.~Henn and J.~Sikorowski, \emph{{Higher loop mixed correlators
  in N=4 SYM}}, \href{https://doi.org/10.1007/JHEP03(2013)058}{\emph{JHEP}
  {\bfseries 03} (2013) 058} [\href{https://arxiv.org/abs/1301.0149}{{\ttfamily
  1301.0149}}].

\bibitem{Henn:2019swt}
J.M.~Henn, G.P.~Korchemsky and B.~Mistlberger, \emph{{The full four-loop cusp
  anomalous dimension in $\mathcal{N}=4$ super Yang-Mills and QCD}},
  \href{https://doi.org/10.1007/JHEP04(2020)018}{\emph{JHEP} {\bfseries 04}
  (2020) 018} [\href{https://arxiv.org/abs/1911.10174}{{\ttfamily
  1911.10174}}].

\bibitem{Chicherin:2022bov}
D.~Chicherin and J.M.~Henn, \emph{{Symmetry properties of Wilson loops with a
  Lagrangian insertion}},
  \href{https://doi.org/10.1007/JHEP07(2022)057}{\emph{JHEP} {\bfseries 07}
  (2022) 057} [\href{https://arxiv.org/abs/2202.05596}{{\ttfamily
  2202.05596}}].

\bibitem{Chicherin:2022zxo}
D.~Chicherin and J.~Henn, \emph{{Pentagon Wilson loop with Lagrangian insertion
  at two loops in $ \mathcal{N} $ = 4 super Yang-Mills theory}},
  \href{https://doi.org/10.1007/JHEP07(2022)038}{\emph{JHEP} {\bfseries 07}
  (2022) 038} [\href{https://arxiv.org/abs/2204.00329}{{\ttfamily
  2204.00329}}].

\bibitem{Arkani-Hamed:2013jha}
N.~Arkani-Hamed and J.~Trnka, \emph{{The Amplituhedron}},
  \href{https://doi.org/10.1007/JHEP10(2014)030}{\emph{JHEP} {\bfseries 10}
  (2014) 030} [\href{https://arxiv.org/abs/1312.2007}{{\ttfamily 1312.2007}}].

\bibitem{Arkani-Hamed:2021iya}
N.~Arkani-Hamed, J.~Henn and J.~Trnka, \emph{{Nonperturbative negative
  geometries: amplitudes at strong coupling and the amplituhedron}},
  \href{https://doi.org/10.1007/JHEP03(2022)108}{\emph{JHEP} {\bfseries 03}
  (2022) 108} [\href{https://arxiv.org/abs/2112.06956}{{\ttfamily
  2112.06956}}].

\bibitem{Brown:2023mqi}
T.V.~Brown, U.~Oktem, S.~Paranjape and J.~Trnka, \emph{{Loops of loops
  expansion in the amplituhedron}},
  \href{https://doi.org/10.1007/JHEP07(2024)025}{\emph{JHEP} {\bfseries 07}
  (2024) 025} [\href{https://arxiv.org/abs/2312.17736}{{\ttfamily
  2312.17736}}].

\bibitem{Henn:2023pkc}
J.M.~Henn, M.~Lagares and S.-Q.~Zhang, \emph{{Integrated negative geometries in
  ABJM}}, \href{https://doi.org/10.1007/JHEP05(2023)112}{\emph{JHEP} {\bfseries
  05} (2023) 112} [\href{https://arxiv.org/abs/2303.02996}{{\ttfamily
  2303.02996}}].

\bibitem{He:2023exb}
S.~He, C.-K.~Kuo, Z.~Li and Y.-Q.~Zhang, \emph{{Emergent unitarity, all-loop
  cuts and integrations from the ABJM amplituhedron}},
  \href{https://doi.org/10.1007/JHEP07(2023)212}{\emph{JHEP} {\bfseries 07}
  (2023) 212} [\href{https://arxiv.org/abs/2303.03035}{{\ttfamily
  2303.03035}}].

\bibitem{Lagares:2024epo}
M.~Lagares and S.-Q.~Zhang, \emph{{Higher-loop integrated negative geometries
  in ABJM}}, \href{https://doi.org/10.1007/JHEP05(2024)142}{\emph{JHEP}
  {\bfseries 05} (2024) 142}
  [\href{https://arxiv.org/abs/2402.17432}{{\ttfamily 2402.17432}}].

\bibitem{Henn:2024qwe}
J.~Henn and P.~Raman, \emph{Positivity properties of scattering amplitudes},
  \href{https://arxiv.org/abs/2407.05755}{{\ttfamily 2407.05755}}.

\bibitem{ancillary}
S.~Abreu, D.~Chicherin, V.~Sotnikov and S.~Zoia, \emph{{Ancillary files for
  ``Two-Loop Five-Point Two-Mass Planar Integrals and Double Lagrangian
  Insertions in a Wilson Loop''}},  8, 2024.
\newblock \href{https://zenodo.org/records/13254698}{10.5281/zenodo.13254698}.

\bibitem{Tkachov:1981wb}
F.V.~Tkachov, \emph{{A Theorem on Analytical Calculability of Four Loop
  Renormalization Group Functions}},
  \href{https://doi.org/10.1016/0370-2693(81)90288-4}{\emph{Phys. Lett.}
  {\bfseries 100B} (1981) 65}.

\bibitem{Chetyrkin:1981qh}
K.G.~Chetyrkin and F.V.~Tkachov, \emph{{Integration by Parts: The Algorithm to
  Calculate beta Functions in 4 Loops}},
  \href{https://doi.org/10.1016/0550-3213(81)90199-1}{\emph{Nucl. Phys. B}
  {\bfseries 192} (1981) 159}.

\bibitem{Lee:2012cn}
R.N.~Lee, \emph{{Presenting LiteRed: a tool for the Loop InTEgrals REDuction}},
   \href{https://arxiv.org/abs/1212.2685}{{\ttfamily 1212.2685}}.

\bibitem{Lee:2013mka}
R.N.~Lee, \emph{{LiteRed 1.4: a powerful tool for reduction of multiloop
  integrals}}, \href{https://doi.org/10.1088/1742-6596/523/1/012059}{\emph{J.
  Phys. Conf. Ser.} {\bfseries 523} (2014) 012059}
  [\href{https://arxiv.org/abs/1310.1145}{{\ttfamily 1310.1145}}].

\bibitem{Wu:2023upw}
Z.~Wu, J.~Boehm, R.~Ma, H.~Xu and Y.~Zhang, \emph{{NeatIBP 1.0, a package
  generating small-size integration-by-parts relations for Feynman integrals}},
  \href{https://doi.org/10.1016/j.cpc.2023.108999}{\emph{Comput. Phys. Commun.}
  {\bfseries 295} (2024) 108999}
  [\href{https://arxiv.org/abs/2305.08783}{{\ttfamily 2305.08783}}].

\bibitem{Laporta:2001dd}
S.~Laporta, \emph{{High precision calculation of multiloop Feynman integrals by
  difference equations}}, \href{https://doi.org/10.1016/S0217-751X(00)00215-7,
  10.1142/S0217751X00002157}{\emph{Int. J. Mod. Phys.} {\bfseries A15} (2000)
  5087} [\href{https://arxiv.org/abs/hep-ph/0102033}{{\ttfamily
  hep-ph/0102033}}].

\bibitem{Peraro:2019okx}
T.~Peraro, \emph{{Analytic multi-loop results using finite fields and dataflow
  graphs with FiniteFlow}},  in \emph{{14th International Symposium on
  Radiative Corrections}: {Application of Quantum Field Theory to
  Phenomenology}}, 12, 2019, \href{https://doi.org/10.22323/1.375.0077}{DOI}
  [\href{https://arxiv.org/abs/1912.03142}{{\ttfamily 1912.03142}}].

\bibitem{Gluza:2010ws}
J.~Gluza, K.~Kajda and D.A.~Kosower, \emph{{Towards a Basis for Planar Two-Loop
  Integrals}}, \href{https://doi.org/10.1103/PhysRevD.83.045012}{\emph{Phys.
  Rev. D} {\bfseries 83} (2011) 045012}
  [\href{https://arxiv.org/abs/1009.0472}{{\ttfamily 1009.0472}}].

\bibitem{Abreu:2018zmy}
S.~Abreu, J.~Dormans, F.~Febres~Cordero, H.~Ita and B.~Page, \emph{{Analytic
  Form of Planar Two-Loop Five-Gluon Scattering Amplitudes in QCD}},
  \href{https://doi.org/10.1103/PhysRevLett.122.082002}{\emph{Phys. Rev. Lett.}
  {\bfseries 122} (2019) 082002}
  [\href{https://arxiv.org/abs/1812.04586}{{\ttfamily 1812.04586}}].

\bibitem{Badger:2021imn}
S.~Badger, C.~Br\o{}nnum-Hansen, D.~Chicherin, T.~Gehrmann, H.B.~Hartanto,
  J.~Henn et~al., \emph{{Virtual QCD corrections to gluon-initiated diphoton
  plus jet production at hadron colliders}},
  \href{https://doi.org/10.1007/JHEP11(2021)083}{\emph{JHEP} {\bfseries 11}
  (2021) 083} [\href{https://arxiv.org/abs/2106.08664}{{\ttfamily
  2106.08664}}].

\bibitem{Goncharov:2010jf}
A.B.~Goncharov, M.~Spradlin, C.~Vergu and A.~Volovich, \emph{{Classical
  Polylogarithms for Amplitudes and Wilson Loops}},
  \href{https://doi.org/10.1103/PhysRevLett.105.151605}{\emph{Phys. Rev. Lett.}
  {\bfseries 105} (2010) 151605}
  [\href{https://arxiv.org/abs/1006.5703}{{\ttfamily 1006.5703}}].

\bibitem{Abreu:2018rcw}
S.~Abreu, B.~Page and M.~Zeng, \emph{{Differential equations from unitarity
  cuts: nonplanar hexa-box integrals}},
  \href{https://doi.org/10.1007/JHEP01(2019)006}{\emph{JHEP} {\bfseries 01}
  (2019) 006} [\href{https://arxiv.org/abs/1807.11522}{{\ttfamily
  1807.11522}}].

\bibitem{Heller:2019gkq}
M.~Heller, A.~von Manteuffel and R.M.~Schabinger, \emph{{Multiple
  polylogarithms with algebraic arguments and the two-loop EW-QCD Drell-Yan
  master integrals}},
  \href{https://doi.org/10.1103/PhysRevD.102.016025}{\emph{Phys. Rev. D}
  {\bfseries 102} (2020) 016025}
  [\href{https://arxiv.org/abs/1907.00491}{{\ttfamily 1907.00491}}].

\bibitem{Zoia:2021zmb}
S.~Zoia, \emph{Modern Analytic Methods for Computing Scattering Amplitudes:
  With Application to Two-Loop Five-Particle Processes}, Springer International
  Publishing (2022),
  \href{https://doi.org/10.1007/978-3-031-01945-6}{10.1007/978-3-031-01945-6}.

\bibitem{Liu:2017jxz}
X.~Liu, Y.-Q.~Ma and C.-Y.~Wang, \emph{{A Systematic and Efficient Method to
  Compute Multi-loop Master Integrals}},
  \href{https://doi.org/10.1016/j.physletb.2018.02.026}{\emph{Phys. Lett. B}
  {\bfseries 779} (2018) 353}
  [\href{https://arxiv.org/abs/1711.09572}{{\ttfamily 1711.09572}}].

\bibitem{Liu:2022chg}
X.~Liu and Y.-Q.~Ma, \emph{{AMFlow: A Mathematica package for Feynman integrals
  computation via auxiliary mass flow}},
  \href{https://doi.org/10.1016/j.cpc.2022.108565}{\emph{Comput. Phys. Commun.}
  {\bfseries 283} (2023) 108565}
  [\href{https://arxiv.org/abs/2201.11669}{{\ttfamily 2201.11669}}].

\bibitem{Hidding:2020ytt}
M.~Hidding, \emph{{DiffExp, a Mathematica package for computing Feynman
  integrals in terms of one-dimensional series expansions}},
  \href{https://doi.org/10.1016/j.cpc.2021.108125}{\emph{Comput. Phys. Commun.}
  {\bfseries 269} (2021) 108125}
  [\href{https://arxiv.org/abs/2006.05510}{{\ttfamily 2006.05510}}].

\bibitem{Gaiotto:2011dt}
D.~Gaiotto, J.~Maldacena, A.~Sever and P.~Vieira, \emph{{Pulling the straps of
  polygons}}, \href{https://doi.org/10.1007/JHEP12(2011)011}{\emph{JHEP}
  {\bfseries 12} (2011) 011} [\href{https://arxiv.org/abs/1102.0062}{{\ttfamily
  1102.0062}}].

\bibitem{Gehrmann:2018yef}
T.~Gehrmann, J.M.~Henn and N.A.~Lo~Presti, \emph{{Pentagon functions for
  massless planar scattering amplitudes}},
  \href{https://doi.org/10.1007/JHEP10(2018)103}{\emph{JHEP} {\bfseries 10}
  (2018) 103} [\href{https://arxiv.org/abs/1807.09812}{{\ttfamily
  1807.09812}}].

\bibitem{Badger:2021nhg}
S.~Badger, H.B.~Hartanto and S.~Zoia, \emph{{Two-Loop QCD Corrections to
  $Wb\bar{b}$ Production at Hadron Colliders}},
  \href{https://doi.org/10.1103/PhysRevLett.127.012001}{\emph{Phys. Rev. Lett.}
  {\bfseries 127} (2021) 012001}
  [\href{https://arxiv.org/abs/2102.02516}{{\ttfamily 2102.02516}}].

\bibitem{Badger:2023xtl}
S.~Badger, J.~Kry\'s, R.~Moodie and S.~Zoia, \emph{{Lepton-pair scattering with
  an off-shell and an on-shell photon at two loops in massless QED}},
  \href{https://doi.org/10.1007/JHEP11(2023)041}{\emph{JHEP} {\bfseries 11}
  (2023) 041} [\href{https://arxiv.org/abs/2307.03098}{{\ttfamily
  2307.03098}}].

\bibitem{Korchemskaya:1992je}
I.A.~Korchemskaya and G.P.~Korchemsky, \emph{{On lightlike Wilson loops}},
  \href{https://doi.org/10.1016/0370-2693(92)91895-G}{\emph{Phys. Lett. B}
  {\bfseries 287} (1992) 169}.

\bibitem{Bassetto:1993xd}
A.~Bassetto, I.A.~Korchemskaya, G.P.~Korchemsky and G.~Nardelli, \emph{{Gauge
  invariance and anomalous dimensions of a light cone Wilson loop in lightlike
  axial gauge}},
  \href{https://doi.org/10.1016/0550-3213(93)90133-A}{\emph{Nucl. Phys. B}
  {\bfseries 408} (1993) 62}
  [\href{https://arxiv.org/abs/hep-ph/9303314}{{\ttfamily hep-ph/9303314}}].

\bibitem{Korchemsky:1992xv}
G.P.~Korchemsky and G.~Marchesini, \emph{{Structure function for large x and
  renormalization of Wilson loop}},
  \href{https://doi.org/10.1016/0550-3213(93)90167-N}{\emph{Nucl. Phys. B}
  {\bfseries 406} (1993) 225}
  [\href{https://arxiv.org/abs/hep-ph/9210281}{{\ttfamily hep-ph/9210281}}].

\bibitem{Korchemsky:1985xj}
G.P.~Korchemsky and A.V.~Radyushkin, \emph{{Loop Space Formalism and
  Renormalization Group for the Infrared Asymptotics of {QCD}}},
  \href{https://doi.org/10.1016/0370-2693(86)91439-5}{\emph{Phys. Lett. B}
  {\bfseries 171} (1986) 459}.

\bibitem{Korchemsky:1987wg}
G.P.~Korchemsky and A.V.~Radyushkin, \emph{{Renormalization of the Wilson Loops
  Beyond the Leading Order}},
  \href{https://doi.org/10.1016/0550-3213(87)90277-X}{\emph{Nucl. Phys. B}
  {\bfseries 283} (1987) 342}.

\bibitem{Eden:2010zz}
B.~Eden, G.P.~Korchemsky and E.~Sokatchev, \emph{{From correlation functions to
  scattering amplitudes}},
  \href{https://doi.org/10.1007/JHEP12(2011)002}{\emph{JHEP} {\bfseries 12}
  (2011) 002} [\href{https://arxiv.org/abs/1007.3246}{{\ttfamily 1007.3246}}].

\bibitem{Eden:2010ce}
B.~Eden, G.P.~Korchemsky and E.~Sokatchev, \emph{{More on the duality
  correlators/amplitudes}},
  \href{https://doi.org/10.1016/j.physletb.2012.02.014}{\emph{Phys. Lett. B}
  {\bfseries 709} (2012) 247}
  [\href{https://arxiv.org/abs/1009.2488}{{\ttfamily 1009.2488}}].

\bibitem{Mason:2010yk}
L.J.~Mason and D.~Skinner, \emph{{The Complete Planar S-matrix of N=4 SYM as a
  Wilson Loop in Twistor Space}},
  \href{https://doi.org/10.1007/JHEP12(2010)018}{\emph{JHEP} {\bfseries 12}
  (2010) 018} [\href{https://arxiv.org/abs/1009.2225}{{\ttfamily 1009.2225}}].

\bibitem{Eden:2000mv}
B.~Eden, C.~Schubert and E.~Sokatchev, \emph{{Three loop four point correlator
  in N=4 SYM}},
  \href{https://doi.org/10.1016/S0370-2693(00)00515-3}{\emph{Phys. Lett. B}
  {\bfseries 482} (2000) 309}
  [\href{https://arxiv.org/abs/hep-th/0003096}{{\ttfamily hep-th/0003096}}].

\bibitem{Eden:2011yp}
B.~Eden, P.~Heslop, G.P.~Korchemsky and E.~Sokatchev, \emph{{The
  super-correlator/super-amplitude duality: Part I}},
  \href{https://doi.org/10.1016/j.nuclphysb.2012.12.015}{\emph{Nucl. Phys. B}
  {\bfseries 869} (2013) 329}
  [\href{https://arxiv.org/abs/1103.3714}{{\ttfamily 1103.3714}}].

\bibitem{Remiddi:1999ew}
E.~Remiddi and J.A.M.~Vermaseren, \emph{{Harmonic polylogarithms}},
  \href{https://doi.org/10.1142/S0217751X00000367}{\emph{Int. J. Mod. Phys. A}
  {\bfseries 15} (2000) 725}
  [\href{https://arxiv.org/abs/hep-ph/9905237}{{\ttfamily hep-ph/9905237}}].

\bibitem{Bern:2005iz}
Z.~Bern, L.J.~Dixon and V.A.~Smirnov, \emph{{Iteration of planar amplitudes in
  maximally supersymmetric Yang-Mills theory at three loops and beyond}},
  \href{https://doi.org/10.1103/PhysRevD.72.085001}{\emph{Phys. Rev. D}
  {\bfseries 72} (2005) 085001}
  [\href{https://arxiv.org/abs/hep-th/0505205}{{\ttfamily hep-th/0505205}}].

\bibitem{Bern:2006ew}
Z.~Bern, M.~Czakon, L.J.~Dixon, D.A.~Kosower and V.A.~Smirnov, \emph{{The
  Four-Loop Planar Amplitude and Cusp Anomalous Dimension in Maximally
  Supersymmetric Yang-Mills Theory}},
  \href{https://doi.org/10.1103/PhysRevD.75.085010}{\emph{Phys. Rev. D}
  {\bfseries 75} (2007) 085010}
  [\href{https://arxiv.org/abs/hep-th/0610248}{{\ttfamily hep-th/0610248}}].

\bibitem{Henn:2014lfa}
J.M.~Henn, K.~Melnikov and V.A.~Smirnov, \emph{{Two-loop planar master
  integrals for the production of off-shell vector bosons in hadron
  collisions}}, \href{https://doi.org/10.1007/JHEP05(2014)090}{\emph{JHEP}
  {\bfseries 05} (2014) 090} [\href{https://arxiv.org/abs/1402.7078}{{\ttfamily
  1402.7078}}].

\bibitem{wasow1965asymptotic}
W.~Wasow, \emph{Asymptotic expansions for ordinary differential equations},
  Pure and Applied Mathematics, Vol. XIV, Interscience Publishers John Wiley \&
  Sons, Inc., New York-London-Sydney (1965).

\bibitem{Caron-Huot:2020vlo}
S.~Caron-Huot, D.~Chicherin, J.~Henn, Y.~Zhang and S.~Zoia, \emph{{Multi-Regge
  Limit of the Two-Loop Five-Point Amplitudes in $\mathcal{N} = 4$ Super
  Yang-Mills and $\mathcal{N} = 8$ Supergravity}},
  \href{https://doi.org/10.1007/JHEP10(2020)188}{\emph{JHEP} {\bfseries 10}
  (2020) 188} [\href{https://arxiv.org/abs/2003.03120}{{\ttfamily
  2003.03120}}].

\bibitem{Smirnov:2002pj}
V.A.~Smirnov, \emph{{Applied asymptotic expansions in momenta and masses}},
  {\emph{Springer Tracts Mod. Phys.} {\bfseries 177} (2002) 1}.

\bibitem{Kotikov:2002ab}
A.V.~Kotikov and L.N.~Lipatov, \emph{{DGLAP and BFKL equations in the $N=4$
  supersymmetric gauge theory}},
  \href{https://doi.org/10.1016/S0550-3213(03)00264-5}{\emph{Nucl. Phys. B}
  {\bfseries 661} (2003) 19}
  [\href{https://arxiv.org/abs/hep-ph/0208220}{{\ttfamily hep-ph/0208220}}].

\bibitem{Chicherin:2020umh}
D.~Chicherin, J.M.~Henn and G.~Papathanasiou, \emph{{Cluster algebras for
  Feynman integrals}},
  \href{https://doi.org/10.1103/PhysRevLett.126.091603}{\emph{Phys. Rev. Lett.}
  {\bfseries 126} (2021) 091603}
  [\href{https://arxiv.org/abs/2012.12285}{{\ttfamily 2012.12285}}].

\bibitem{Bossinger:2022eiy}
L.~Bossinger, J.M.~Drummond and R.~Glew, \emph{{Adjacency for scattering
  amplitudes from the Gr\"obner fan}},
  \href{https://doi.org/10.1007/JHEP11(2023)002}{\emph{JHEP} {\bfseries 11}
  (2023) 002} [\href{https://arxiv.org/abs/2212.08931}{{\ttfamily
  2212.08931}}].

\bibitem{Bern:2014kca}
Z.~Bern, E.~Herrmann, S.~Litsey, J.~Stankowicz and J.~Trnka, \emph{{Logarithmic
  Singularities and Maximally Supersymmetric Amplitudes}},
  \href{https://doi.org/10.1007/JHEP06(2015)202}{\emph{JHEP} {\bfseries 06}
  (2015) 202} [\href{https://arxiv.org/abs/1412.8584}{{\ttfamily 1412.8584}}].

\bibitem{Moriello:2019yhu}
F.~Moriello, \emph{{Generalised power series expansions for the elliptic planar
  families of Higgs + jet production at two loops}},
  \href{https://doi.org/10.1007/JHEP01(2020)150}{\emph{JHEP} {\bfseries 01}
  (2020) 150} [\href{https://arxiv.org/abs/1907.13234}{{\ttfamily
  1907.13234}}].

\bibitem{Arkani-Hamed:2017vfh}
N.~Arkani-Hamed, H.~Thomas and J.~Trnka, \emph{{Unwinding the Amplituhedron in
  Binary}}, \href{https://doi.org/10.1007/JHEP01(2018)016}{\emph{JHEP}
  {\bfseries 01} (2018) 016}
  [\href{https://arxiv.org/abs/1704.05069}{{\ttfamily 1704.05069}}].

\bibitem{Arkani-Hamed:2014dca}
N.~Arkani-Hamed, A.~Hodges and J.~Trnka, \emph{{Positive Amplitudes In The
  Amplituhedron}}, \href{https://doi.org/10.1007/JHEP08(2015)030}{\emph{JHEP}
  {\bfseries 08} (2015) 030} [\href{https://arxiv.org/abs/1412.8478}{{\ttfamily
  1412.8478}}].

\bibitem{Dixon:2016apl}
L.J.~Dixon, M.~von Hippel, A.J.~McLeod and J.~Trnka, \emph{{Multi-loop
  positivity of the planar $ \mathcal{N} $ = 4 SYM six-point amplitude}},
  \href{https://doi.org/10.1007/JHEP02(2017)112}{\emph{JHEP} {\bfseries 02}
  (2017) 112} [\href{https://arxiv.org/abs/1611.08325}{{\ttfamily
  1611.08325}}].

\end{thebibliography}\endgroup
\end{document}